\documentclass[twoside,12pt]{article}
\usepackage{epsfig}
\usepackage{wrapfig}
\usepackage[linktocpage]{hyperref}

\newcommand{\be}{\begin{equation}}
\newcommand{\ee}{\end{equation}}
\newcommand{\bea}{\begin{eqnarray}}
\newcommand{\eea}{\end{eqnarray}}

\begin{document}

\title{ \vspace{1cm} 
Novel Phenomenology of Parton Distributions from  
the Drell-Yan Process}
\author{Jen-Chieh Peng$^{1}$ and Jian-Wei Qiu$^{2,3}$\\
\\
$^1$Department of Physics, University of Illinois at Urbana-Champaign\\
Urbana, Illinois 61801, USA\\
$^2$Physics Department, Brookhaven National Laboratory\\
Upton, NY 11973, USA\\
$^3$C.N. Yang Institute for Theoretical Physics\\
and Department of Physics and Astronomy\\
Stony Brook University, Stony Brook, NY 11794, USA
}
\maketitle

\begin{abstract} 
The Drell-Yan massive lepton-pair production in hadronic 
collisions provides 
a unique tool complementary to the Deep-Inelastic Scattering for probing the
partonic substructures in hadrons.  We review key concepts, 
approximations, and progress
for QCD factorization of the Drell-Yan process in terms of collinear or 
transverse momentum dependent (TMD) parton distribution functions.  
We present experimental results from recent fixed-target Drell-Yan as 
well as $W$ and $Z$ boson production at colliders,
focussing on the topics of flavor structure of the nucleon sea
as well as the extraction of novel Sivers and Boer-Mulders functions via 
single transverse spin asymmetries and azimuthal lepton angular distribution
of the Drell-Yan process.  
Prospects for future Drell-Yan experiments are also presented.
\end{abstract}
\eject
\tableofcontents
\eject

\section{Introduction}
\label{sec:intro}

The first direct evidence for point-like constituents in the nucleons
came from the discovery of scaling phenomenon in lepton-nucleon 
inclusive Deep-Inelastic Scattering (DIS) experiments at SLAC \cite{Bloom:1969kc}, 
which led to the introduction of the fundamental theory of strong interaction
known as Quantum Chromo-Dynamics (QCD) \cite{Fritzsch:1973pi}.
According to QCD, nucleons are composed of quarks and gluons 
(known as partons), bound together by color force through the exchange of gluons. 
However, no quarks and gluons have ever been observed directly in
the DIS or any other scattering experiments, 
a phenomenon that is believed to be a consequence of 
QCD color confinement.  
The structure as well as the dynamics of quarks and gluons inside the nucleons, 
their confined motion and spatial distribution, is one of the most 
intriguing aspects of QCD.  

Much of the predictive power of QCD is contained in factorization 
theorems 
\cite{Collins:1989gx}.  It is the QCD factorization that 
enables us to connect the quarks and gluons, 
the basic degrees of freedom of QCD, and their distributions inside hadrons to
physically measured high energy scattering cross sections with identified hadrons and leptons. 
QCD factorization separates long- from short-distance effects in hadronic collisions, 
provides systematic prescriptions and tools to calculate the 
short-distance dynamics perturbatively, and 
identifies the leading nonperturbative long-distance effects with 
universal hadron matrix elements of quark and gluon field operators and 
supplies them with physical content, which allows them to be extracted from 
experimental measurements or by numerical calculations in Lattice QCD (LQCD). 
Hadron parton distribution functions (PDFs), $\phi_f(x,\mu)$, probability distributions 
to find a parton of flavor $f$ (a quark or a gluon) to carry $x$ fraction of the 
momentum of a fast moving hadron, probed at a hard scale $\mu$,
are Fourier transforms of such universal hadron matrix elements.  
They are the most prominent nonperturbative quantities 
describing the relation between a hadron and the quarks and gluons within it. 
PDFs are universal and independent of the details of scattering processes,
from which they are extracted.  
They are fundamental and carry every secrets of QCD dynamics 
from the ``less-unknown'' QCD confinement to the ``well-known'' QCD asymptotic freedom.
Enormous experimental and theoretical efforts have been devoted to 
the extraction of the PDFs.  
On the experimental side, vast amount of data sensitive to the PDFs have 
been accumulated from low energy fixed-target experiments to 
the measurements performed at the LHC - the highest energy hadron 
collider in the world.  
On the theoretical side, many sets of PDFs have been extracted 
from QCD global analysis of all existing data 
\cite{CTEQ6.6,CT10,MSTW08,NNPDF2.0,NNPDF2.1,gluck08,Alekhin:2013nda}.
Our knowledge of PDFs has been much
improved by many surprises and discoveries from experimental measurements
throughout the years.
The excellent agreement between the theory and data on the scaling-violation behavior 
of the PDFs has provided one of the most stringent tests for QCD as the theory of
strong interaction.

\begin{wrapfigure}{R}{0.4\textwidth}
\begin{center}
\includegraphics[width=0.3\textwidth]{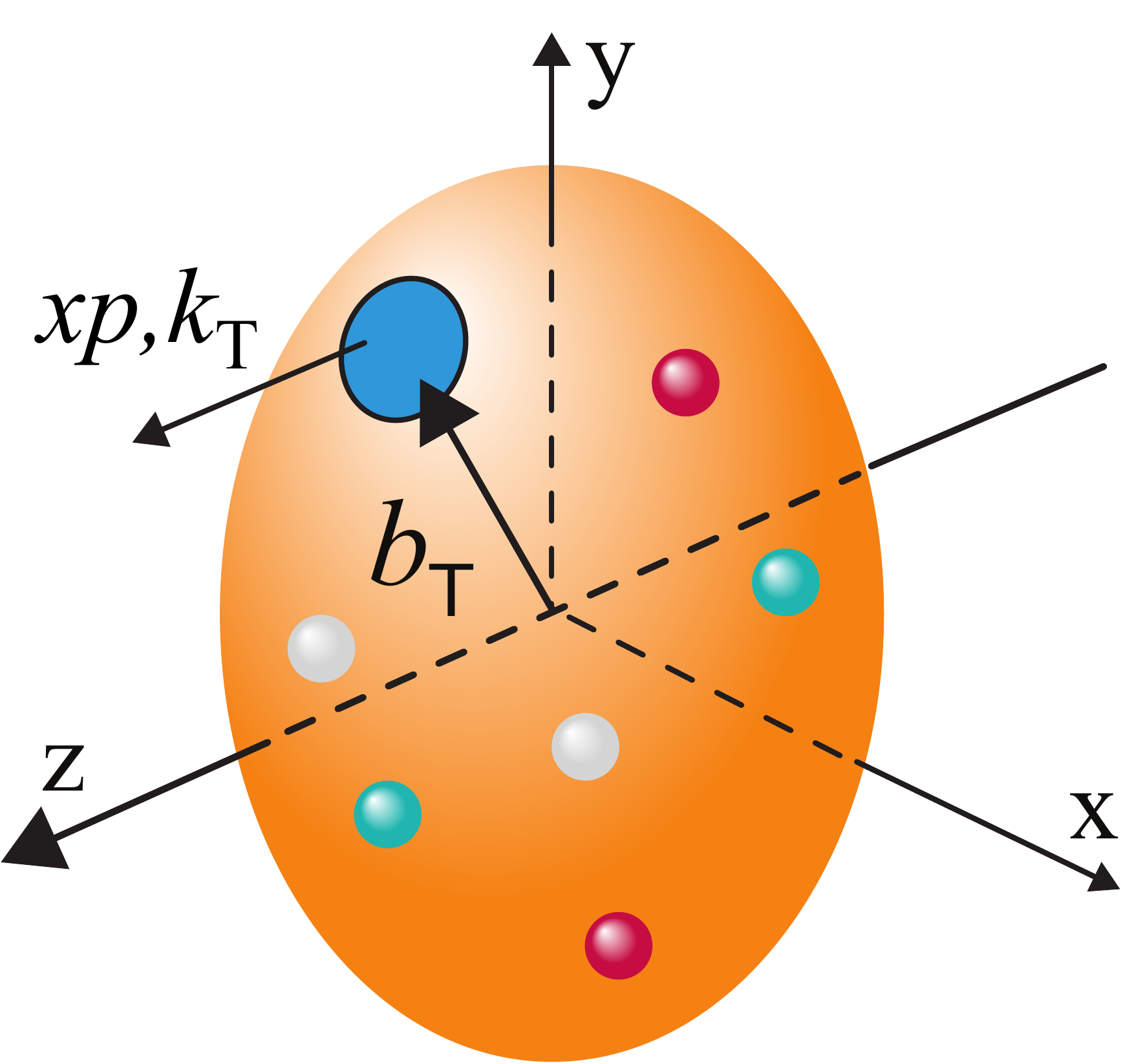}
\end{center}
\caption{Schematic view of a parton with longitudinal momentum fraction $x$, 
transverse momentum $k_T$, and transverse position $b_T$ in a hadron of 
momentum $p$ moving along z-direction.
}
\label{fig:pdf-5d}
\end{wrapfigure}

During the past two decades, many important new developments related to
the study of quark-gluon structure of the nucleons have taken place, 
which enable the systematic investigation of hadronic structure beyond the PDFs, 
the one-dimensional (1D) parton momentum distributions inside a fast moving hadron.
Theoretically, remarkable advances in QCD factorization were achieved, allowing 
the dynamical mapping in both 3D confined motion 
and transverse spatial distribution of quarks and gluons inside a fast moving hadron. 
The distributions of quarks and gluons in space are encoded in 
the Generalized Parton Distributions (GPDs) \cite{gpds,dvcs}.  
QCD collinear factorization for exclusive diffractive scatterings 
allows for systematic extractions of GPDs from these  
exclusive processes \cite{gpds-fac,dvcs-fac}.  
On the other hand, the information on the 3D confined motion 
of quarks and gluons is matched to novel 
Transverse Momentum Dependent PDFs (TMDs) 
\cite{brodsky02,collins02,tmds-gauge,tmds-boer-etal}.  
QCD transverse momentum dependent factorization formalisms 
\cite{collins-book,Ji:2004wu,Ji:2004xq,GarciaEchevarria:2011rb}
allow for systematic extraction of TMDs from semi-inclusive DIS (SIDIS)
- the DIS experiments with the detection of additional 
hadron(s) produced 
in coincidence with the scattered lepton 
\cite{Airapetian:2004tw,Alekseev:2008aa,Qian:2011py}.
Complementary to SIDIS, the massive lepton-pair production in hadronic collisions,
which is known as the Drell-Yan process \cite{Drell:1970wh}, 
is another ideal process to probe the TMDs 
when the transverse momentum of the lepton-pair 
is much smaller than the invariant mass of the pair.
Electroweak processes, such as DIS and the Drell-Yan processes,
bring to bear the unmatched precision of electroweak interaction 
as a probe to test QCD dynamics and to extract information on hadron structure. 
In addition, the advent of the LQCD in calculating moments of various 
PDFs, GPDs, and TMDs (even the $x$-distributions themselves according to 
the latest proposal \cite{Ji:2013dva}) allows direct comparison between 
such first-principle LQCD calculations and those extracted from the experimental data.

Pioneering measurements of exclusive Deep Virtual Compton Scattering
(DVCS) and J/$\psi$ photo-production at HERA
provided the first glimpse of spatial distributions of sea quarks 
and gluons, respectively.  
Recently, extensive inclusive and semi-inclusive DIS measurements 
using polarized lepton beams at HERMES of DESY \cite{Airapetian:2004tw}, 
COMPASS of CERN \cite{Alekseev:2008aa},
and several experiments at Jefferson Lab \cite{Qian:2011py}
have allowed the first observation and extraction of novel TMDs, 
such as the Sivers functions of nucleons, 
which provide direct quantum correlations between nucleon spin and 
the spatial preference of partons' confined transverse motion inside the nucleon \cite{sivers90}.
More recently, the prospect of using polarized hadronic beam colliding with
possibly polarized beam/target has attracted much attention. 
Measurements of the Drell-Yan process and other observables 
could be pursued in existing facilities such as the polarized
$p-p$ collider at RHIC, COMPASS at CERN, as well as other hadron machines
around the world. 
The novel TMDs obtained in hadron collisions are expected to provide 
unique and critical tests of TMD factorization in QCD, such as 
the sign change of the Sivers and Boer-Mulders functions measured 
in SIDIS and Drell-Yan.

Understanding the characteristics and physics content of the extracted PDFs, 
such as the shape and the flavor dependence of the distributions,
is the first necessary step in searching for answers to the ultimate question 
of QCD on how quarks and gluons are confined into hadrons.  Furthermore, 
going beyond the PDFs by extracting information on the parton's 
confined transverse motion (its transverse momentum $k_T$-distribution) 
and its spatial $b_T$ distribution of a fast moving hadron, 
as sketched in Fig.~\ref{fig:pdf-5d},   
necessarily yields a complementary picture of the hadron in both  
momentum and coordinate space, and 
pushes our investigation of hadron structure to a new frontier.

In this review article, we focus on the recent progress and future prospect
on using hadronic beams to explore the novel parton distributions in
the nucleons, while the tremendous and complementary potential 
to explore hadron structure by using a lepton-hadron collider
can be found in the newly released White Paper on future prospects 
of an Electron-Ion Collider \cite{Accardi:2012hwp}. 
We examine, in particular, the unique features of the massive 
lepton-pair production (the Drell-Yan process) in hadronic collision
in extracting the spin and flavor dependences of PDFs and TMDs.  
With the measurement of both invariant mass of the lepton pair $Q$ and 
its transverse momentum $q_T$, Drell-Yan massive lepton-pair  
production in hadronic collisions is an excellent laboratory for theoretical 
and experimental investigations of strong interaction dynamics, and 
has been a valuable and constant pursuit since 1970s.  
With a large invariant mass $Q$ to localize the probe to ``see'' a parton (a quark or a gluon), 
the natural small transverse momentum of the most lepton-pairs produced, $q_T \ll Q$, 
is an ideal scale to be sensitive to the parton's confined motion 
inside the hadron. The Drell-Yan process in this kinematic regime
is ideal for extracting the TMDs.  On the other hand, for events with
$q_T\sim Q$, or with $q_T$ integrated, 
the cross section of Drell-Yan lepton-pair production has
effectively one hard scale, and is the most suited for extracting PDFs.   
Since QCD factorization for both of these regimes are proved to be valid,
the Drell-Yan massive lepton-pair production is a unique and clean
observable to extract both TMDs and PDFs, and the transition between them
by varying the transverse momentum of the lepton pair, $q_T$.  
The same idea has been applied to production of any kind of lepton pair 
from decay of an electroweak gauge boson ($\gamma, W, Z$), 
as well as to production of Higgs boson and any color-neutral 
heavy bosons beyond the Standard Model. 
In addition, by measuring angular distribution of the lepton in the 
rest frame of the lepton pair, the Drell-Yan process is a
unique one to study quantum interference between two scattering 
amplitudes with the intermediate vector boson in different spin states.
The measurement of Drell-Yan massive lepton-pair production not only 
was performed in almost all high-energy hadronic facilities ever existed, 
but also is taking place now in an on-going fixed-target 
experiment (E906) at Fermilab
and all major experiments at the LHC.
Several future Drell-Yan experiments are also being planned at 
facilities around the world.

The rest of the article is organized as follows.  In the next section, we 
review the role of inclusive Drell-Yan measurements in probing PDFs
and hadron's partonic structure, and QCD factorization for Drell-Yan process,
which is necessary for connecting the measured massive lepton pair   
to the dynamics of quarks and gluons inside the colliding hadrons.
To illustrate the complementary nature of the Drell-Yan and
DIS in probing nucleon's parton structures, we
focus on the flavor structure
of the parton distributions in the nucleons and nuclei. The striking 
observation of the large up and down sea quark flavor asymmetry in the 
proton from the Drell-Yan and semi-inclusive DIS experiments continues
to motivate new theoretical interpretations, and further experimental
studies for testing the various theoretical models are currently
underway or being planned. The Drell-Yan process could also
probe the flavor dependence of parton distributions in nuclei, and
provide a sensitive test for theoretical models explaining the
famous EMC effect. We then discuss the strange quark and
gluon contents in the nucleons, as they could provide new
insight on the flavor structure of the nucleon sea. We examine the
$x$-dependence of the strange quark relative to those of the lighter
up and down quarks, as well as the possible difference between the
$s(x)$ and $\bar s(x)$ distributions. While $\bar u(x) (\bar d(x))$
in the proton is very different from $\bar d(x) (\bar u(x))$
in the neutron, we present the
experimental evidence that the gluons distributions in the proton 
and neutron are very similar.

In Sec.~3, we review the recent development
in probing TMDs by Drell-Yan process when $Q\gg q_T$.   
We focus on physics of two novel TMDs: Sivers function \cite{sivers90} 
and Boer-Mulders function \cite{boer98}, and the prospects for extracting them.
We review the theory developments of TMD factorization for Drell-Yan process, 
and the physics leading to the ultimate QCD prediction of the sign change 
of Sivers and and Boer-Mulders functions, when measured in SIDIS 
in comparison with that measured in the Drell-Yan process.  
We discuss the role of the Sivers functions in generating the 
novel single transverse-spin asymmetry of Drell-Yan
massive lepton pair production, and the physics behind the 
intriguing phenomena of single transverse-spin asymmetries.  
We summarize the current world effort in planning for 
experiments to measure Drell-Yan single transverse-spin asymmetry 
in order to extract 
the Sivers functions and to test the sign change of Sivers functions
when they are measured in Drell-Yan versus SIDIS.  
We then focus on the angular distributions of the lepton 
in the rest frame of the observed lepton pairs in unpolarized hadronic collisions,
to which the Boer-Mulders functions contribute.
We discuss the progress in resolving a long-standing puzzle of the
violation of the Lam-Tung relation observed in pion-induced Drell-Yan experiments,
and the prospect for extracting the Boer-Mulders functions from 
the angular distributions of unpolarized Drell-Yan
process.  In Sec.~4, we briefly 
review the role of generalized Drell-Yan massive lepton-pair production 
via the heavy vector boson, $W/Z$, in measuring the polarized sea distributions 
of proton.  The distinct quark-flavor dependences of the $W, Z, \gamma^*$ 
couplings to quark-antiquark pairs offer a unique possibility for disentangling the flavor
structure of the parton distributions.  
We conclude with the summary and outlook in Sec.~5.

\section{Hadronic production of massive lepton pairs} 
\label{section2}

Throughout the years, hadronic production of massive lepton pairs has
served not only as a channel for discovery of quarkonium states and 
intermediate vector bosons, as well as Higgs boson, but also as a clean, 
precise and controllable probe for short-distance dynamics 
and partonic structure of hadrons.  
In this section, we review briefly key developments for hadronic production 
of massive lepton pairs, as well as puzzles that are still driving us 
to search for a complete understanding of the physics behind the 
massive lepton-pair production, more than forty years after the 
very first measurement of massive muon pairs produced at Brookhaven 
National Laboratory (BNL) \cite{Christenson:1970um}.

\subsection{Drell-Yan mechanism}
\label{subsec:dy}

\begin{figure}[h]
\hskip 0.12\textwidth
\includegraphics[height=1.5in]{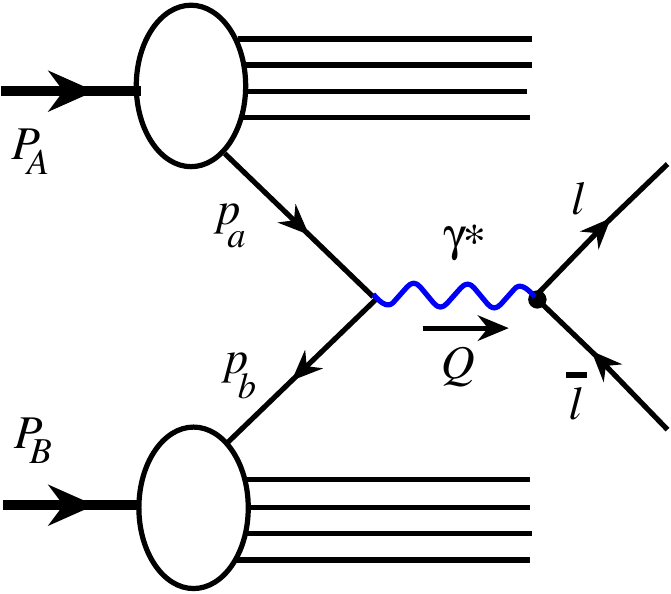}
\hskip 0.14\textwidth
\includegraphics[height=1.4in]{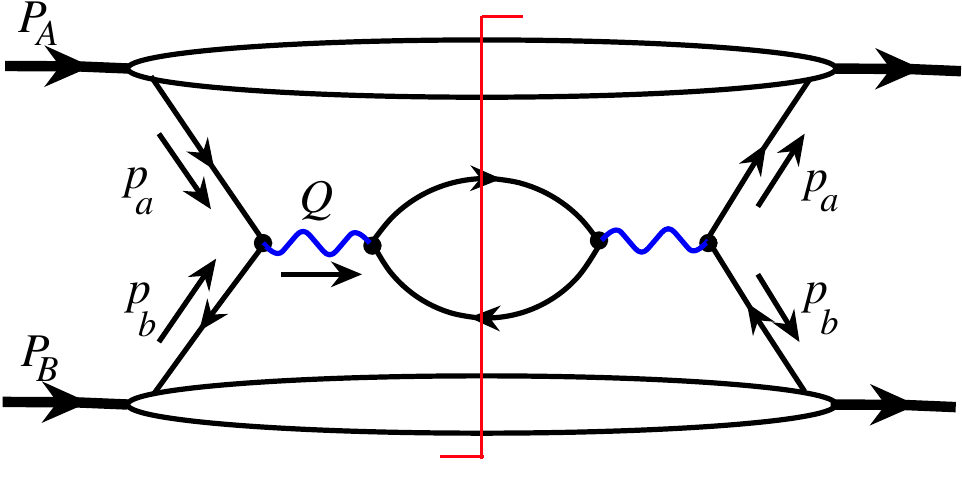}
\caption{Sketch for scattering amplitude of Drell-Yan process (left) and scattering amplitude 
square in the cut diagram notation where the final-state is identified by a vertical line (right). 
}
\label{fig:drell-yan}
\end{figure}

Hadronic production of massive lepton pairs was first studied at BNL by Christenson {\it et al.} 
\cite{Christenson:1970um,Christenson:1973mf} by measuring massive muon pairs 
in collisions of a proton beam on a Uranium target, $p+U \to \mu^+\mu^-(q)+X$, with muon pair mass 
$Q=\sqrt{q^2}=1-6.7$~GeV.  The steep falling of the cross section with $Q$ and an apparent 
slow-down or a shoulder-like structure near $Q\sim 3$~GeV were two very important features 
in the data.  The shoulder-like structure was naturally explained following the discovery of heavy quarkonium J/$\psi$ a few years later \cite{Aubert:1974js,Augustin:1974xw}.  
It was the quick fall-off of the cross section with $Q$ that 
led Sidney Drell and Tung-Mow Yan \cite{Drell:1970wh}
to propose a new production mechanism for massive lepton-pair production in
high energy collisions of hadron $A$ of momentum $P_A$ on hadron $B$ of momentum $P_B$, 
now known as the Drell-Yan mechanism/process, 
\begin{equation}
A(P_A)+B(P_B) \to \gamma^{*}(q)[\to l\overline{l}(q)]+X, 
\label{eq:drell-yan0}
\end{equation}
to explain the data observed at BNL.
In Eq.~(\ref{eq:drell-yan0}), the virtual photon, $\gamma^{*}$, which decays into
the lepton pair, is produced by the annihilation of a parton and an antiparton from the two colliding hadrons, 
as sketched in Fig.~\ref{fig:drell-yan}, where the diagram on the left represents
the production amplitude, while the diagram on the right is in cut diagram notation, 
in which the amplitude and its complex conjugate are combined into 
a forward scattering diagram in which the final state is identified by a vertical line.
The predictive power of the Drell-Yan process is that 
the magnitude and the shape of the cross section are uniquely determined 
by the parton and antiparton distribution functions measured in DIS \cite{Drell:1970wh,Drell:1970yt},
\begin{equation}
\frac{d\sigma_{A+B\to l\bar{l}+X}}{dQ^2 dy}
= \frac{4\pi\alpha_{em}^2}{3Q^4} \sum_{p,\bar{p}}\, 
x_A \phi_{p/A}(x_A) \, x_B \phi_{\bar{p}/B}(x_B) \, ,
\label{eq:drell-yan-pm}
\end{equation}
where $y=\frac{1}{2}\ln(x_A/x_B)$ is the rapidity of the lepton pair, 
and the parton and antiparton momentum fractions are given by
\begin{equation}
x_A = \frac{Q}{\sqrt{S}}\, e^{y} \ \ \mbox{and} \ \
x_B = \frac{Q}{\sqrt{S}}\, e^{-y}
\label{eq:kinematics-pm}
\end{equation}
with total center of mass collision energy squared ${S} = (P_A+P_B)^2$.   
The variables $x_A$ and $x_B$ in Eq.~(\ref{eq:kinematics-pm}) 
are also referred in the literature as $x_1$ and $x_2$, respectively.  
The $\phi_{p/A}(x_A)$ and $\phi_{\bar{p}/B}(x_B)$ are parton and antiparton distributions,
respectively, which are probability distributions to find a parton (antiparton) 
in a fast moving hadron carrying its momentum fraction $x_A$ ($x_B$).

The original formalism of Drell-Yan cross section in Eq.~(\ref{eq:drell-yan-pm}) 
contains no free parameter, allowing 
an absolute prediction once the PDFs are extracted from DIS measurements.  
The formalism was consistent with many features of early data on massive lepton-pair production.
The success of the Drell-Yan mechanism clearly demonstrated 
that, like 
the lepton-hadron DIS cross section observed at SLAC \cite{Bloom:1969kc}, 
the massive lepton-pair production in hadronic collisions 
is also determined 
by the point-like partonic cross sections and the rest of hadrons are effectively 
frozen during the hard collisions.  
The idea to apply the parton model \cite{Feynman:1969ej} outside lepton-hadron 
DIS was revolutionary since one or more partonic interactions could take place 
between two ``partonic beams'' of the two colliding hadrons.

The Drell-Yan process predated the discovery of QCD.  
Recognizing the parton and antiparton as quark and antiquark in QCD, 
the $\sum_{p,\bar{p}}$ in Eq.~(\ref{eq:drell-yan-pm}) is replaced by 
$\sum_{q,\bar{q}}$ over all quark and antiquark flavors, weighted by 
their fractional charge squared, plus an overall color factor $1/N_c$ with
$N_c=3$ for SU($N_c$) color of QCD.  
The Drell-Yan formalism in Eq.~(\ref{eq:drell-yan-pm}),
after taking into account the sum of various quark flavors, 
their fractional charges and the color factor,
can describe many features of  
massive lepton-pair production data. However,
a somewhat large 
$K_{\rm factor}=\sigma_{\rm Exp}/\sigma_{\rm Thy}\sim 2$ is found,
which indicates that the normalization of the predicted cross section is off 
by roughly a factor of 2.

\subsection{QCD improved Drell-Yan mechanism}
\label{sec:dy-inclusive}

The triumph of Drell-Yan mechanism and the formalism in Eq.~(\ref{eq:drell-yan-pm}) 
for the hadronic massive lepton-pair production 
is the {\em factorization} of short-distance production of the massive lepton pair 
from the complicated internal dynamics of colliding hadrons.  
It is this factorization and the process independence of the 
PDFs that give the parameter-free predictive power of Drell-Yan mechanism.  

\begin{figure}[h]
\hskip 0.2\textwidth
\includegraphics[height=1.5in]{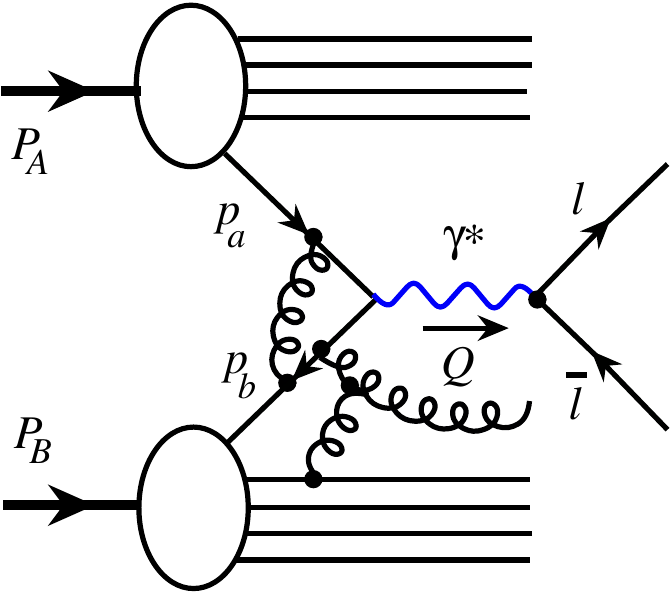}
\hskip 0.14\textwidth
\includegraphics[height=1.5in]{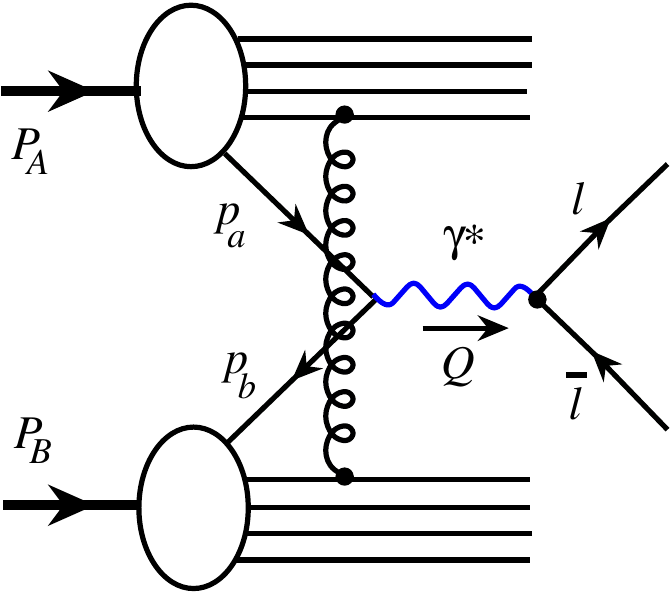}
\caption{Sample QCD modification to the scattering amplitudes of Drell-Yan mechanism: 
gluon radiation and interaction (left) and gluon interaction between spectators (right). 
}
\label{fig:qcd-dy}
\end{figure}

However, the clean and simple factorization of Drell-Yan mechanism could be easily broken 
in QCD by its rich quark-gluon interaction.  In the parton model 
calculation of the Drell-Yan process, 
from which Eq.~(\ref{eq:drell-yan-pm}) was derived, 
the parton and antiparton that annihilate into the observed lepton pair are
assumed to be real and on their mass-shell.  
But, in QCD, the quark and antiquark from colliding hadrons 
that annihilate into a massive lepton pair are always off their mass shell, 
whose virtuality needs to be integrated over.  
In addition to quarks and antiquarks, there are gluons in QCD.
Every quarks and gluons from colliding hadrons could participate in the collisions to 
produce the observed massive lepton pair.  
That is, other than the simplest quark-antiquark annihilation subprocess, 
massive lepton pairs could be produced by the same annihilation process dressed up 
by any number of gluonic interactions, as shown in Fig.~\ref{fig:qcd-dy}.
For example, introducing complicated quark-gluon 
correlations inside the colliding hadron into the production of the lepton pair, 
as sketched in Fig.~\ref{fig:qcd-dy} (left), could go beyond the quark-antiquark 
(or parton-antiparton) annihilation and lead to potential process dependence of the PDFs.
In addition, as sketched in the diagram on the right of Fig.~\ref{fig:qcd-dy}, the
gluon interactions between spectators of two colliding hadrons could alter PDFs of one hadron 
by the presence of another hadron, which would break the universality of PDFs and 
the predictive power of the formalism.
Actually, the inclusive cross section for producing a massive lepton pair 
in a hadronic collision, as defined in Eq.~(\ref{eq:drell-yan0}), 
$d\sigma_{A+B\to l\bar{l}+X}/dQ^2dy$ is {\it not} calculable in QCD perturbation theory. 
Uncanceled infrared (IR) divergences were identified by explicit calculations of
QCD contribution to Drell-Yan cross section \cite{Doria:1980ak,Di'Lieto:1980dt,Brandt:1988xt}.

When the mass of lepton pair, $Q$, the minimum momentum transfer of the collision, 
is much larger than the typical momentum scale of the dynamics inside the colliding hadrons, 
$Q\gg 1/{\rm fm}\sim \Lambda_{\rm QCD}$, the hard collision is localized to a very small 
size $\sim 1/Q \ll$~1 fm.   Pulling an extra quark or gluon from incoming hadron to participate 
in the localized hard collision, like the process in Fig.~\ref{fig:qcd-dy} (left), 
could be suppressed by the power of $\Lambda_{\rm QCD}/Q$.  
Different number of active partons from colliding hadrons to produce the lepton pair 
corresponds to the contributions to the cross section at different power in $1/Q$ \cite{Qiu:1990xy}.  
That is, all QCD contributions to the cross section of massive 
lepton-pair production
could be naturally reorganized in terms of an $1/Q$ power expansion,
\begin{equation}
\frac{d\sigma_{A+B\to l\bar{l} +X}}{dQ^2 dy}
=\sum_{n=0} \frac{d\sigma_{A+B\to l\bar{l} +X}^{(n)}}{dQ^2 dy} 
\left(\frac{\Lambda_{\rm QCD}}{Q}\right)^n.
\label{eq:power-exp}
\end{equation}
Although there is no QCD factorization for full cross section of 
hadronic massive lepton-pair production, it was 
proved \cite{Collins:1989gx}
to all orders in powers of $\alpha_s$ in QCD perturbation theory 
that the leading power (LP) contribution to Drell-Yan cross section,
$d\sigma_{A+B\to l\bar{l} +X}^{\rm (0)}/dQ^2 dy$ in Eq.~(\ref{eq:power-exp}), 
can be systematically factorized into a formalism that is
effectively the same as the original Drell-Yan formalism in Eq.~(\ref{eq:drell-yan-pm}), 
\begin{equation}
\frac{d\sigma^{\rm (LP)}_{A+B\to l\bar{l}+X}}{dQ^2 dy}
=\sum_{ab}\int_0^1 dx_a\int_0^1 dx_b \, \phi_{a/A}(x_a,\mu) \, \phi_{b/B}(x_b,\mu) \,
\frac{d\hat{\sigma}_{a+b\to l\bar{l}}(x_a,x_b,Q,\mu,\alpha_s)}{dQ^2 dy}\, ,
\label{eq:qcd-dy-lp}
\end{equation}
where the superscript ``(LP)'' indicates the leading power contribution 
to the full cross section in its $1/Q$ expansion, 
$\sum_{a,b}$ runs over all parton flavors including quark and antiquark, as well as gluon, 
$\mu\sim Q$ is the factorization scale, 
$\phi$'s are universal PDFs extracted from QCD global analyses
\cite{CTEQ6.6,CT10,MSTW08,NNPDF2.0,NNPDF2.1}.
In Eq.~(\ref{eq:qcd-dy-lp}), $\hat{\sigma}_{a+b\to l\bar{l}}$ is the short-distance part of  
QCD partonic cross section for two incoming partons of flavor $a$ and $b$, respectively, 
to produce a massive lepton pair.  
Since the $\hat{\sigma}_{a+b\to l\bar{l}}$ is insensitive to the long-distance details of 
colliding hadrons, the factorization formalism in Eq.~(\ref{eq:qcd-dy-lp}) should be
also valid for the collision between two asymptotic partons of various flavors.  
In this case, both the partonic scattering cross section on the left and 
the PDFs of colliding partons on the right can be expressed in terms of Feynman
diagrams order-by-order in QCD perturbation theory.  As required by the factorization, 
all collinear divergences of the partonic cross sections on the left should be cancelled 
by the corresponding collinear divergence of the PDFs on the right, 
order-by-order in powers of $\alpha_s$, 
to leave the short-distance hard parts, $\hat{\sigma}_{a+b\to l\bar{l}}$, infrared safe
to all powers of $\alpha_s$.

The predictive power of QCD improved Drell-Yan formalism in Eq.~(\ref{eq:qcd-dy-lp}) 
for the production of massive lepton pairs relies on the approximation to neglect 
all power corrections suppressed by $(\Lambda_{\rm QCD}/Q)^n$, 
the universality of PDFs and our ability to calculate the short-distance 
partonic cross sections, which have been calculated to next-to-next-to-leading order 
(NNLO) in powers of $\alpha_s$ \cite{dy-nnlo}.
With the QCD improved Drell-Yan formalism in Eq.~(\ref{eq:qcd-dy-lp}), and
the universal PDFs extracted from QCD global analysis 
\cite{CTEQ6.6,CT10,MSTW08,NNPDF2.0,NNPDF2.1},  
the $K_{\rm factor} \sim 1$ for all existing data of Drell-Yan type processes 
including massive vector boson $W/Z$ production at collider energies. 

The proof of the QCD factorization formalism in Eq.~(\ref{eq:qcd-dy-lp}) 
for the inclusive production of Drell-Yan massive lepton pairs 
is highly nontrivial involving decades of effort of many people, 
and is best summarized in Refs.~\cite{Collins:1989gx,collins-book}. 
A less technical and intuitive summary of the key steps of the proof 
is given in Appendix~\ref{sec:appendixa}.
The proof is rigorous in the sense that all identified sources of 
leading power contributions are either factorizable or canceled 
in perturbative calculations to all orders in powers of $\alpha_s$.
For example, the contribution involving the quark-gluon correlation of 
one hadron, as shown in Fig.~\ref{fig:qcd-dy} (left) is either suppressed by
power of $1/Q^2$ or included into the definition of PDF to make it gauge invariant, 
while the gluon interaction between spectators 
are effectively canceled \cite{Collins:1989gx,collins-book}.

In addition to the leading power contribution, it was demonstrated 
\cite{Qiu:1990xy,Qiu:1990xxa} that 
the first subleading power corrections, at $1/Q^2$ (or $1/Q$) for unpolarized and 
longitudinally polarized (or single transversely polarized) Drell-Yan 
cross section, can
also be factorized into a sum of convolutions of a perturbatively 
calculable short-distance partonic hard part, a PDF from one colliding hadron, and 
a universal twist-4 four-parton (or twist-3 three-parton) correlation function 
of the other colliding hadron.  

On the other hand, several explicit calculations demonstrated that 
QCD contributions to Drell-Yan cross section beyond the first subleading power corrections 
{\em cannot} be systematically factorized into convolutions
of perturbative hard parts and universal long-distance parton correlation functions 
\cite{Doria:1980ak,Di'Lieto:1980dt,Brandt:1988xt,Basu:1984ba}.
The break of factorization is caused by the non-factorizable power suppressed contribution to
the Drell-Yan cross section from the interactions of physically polarized soft gluons 
between two colliding hadrons at the power of $1/Q^4$ (or $1/Q^2$) 
and higher for unpolarized and longitudinally polarized (or single transversely polarized) 
hadronic collisions \cite{Qiu:1990xy,Qiu:1990xxa}.
 
The QCD factorization formalism in Eq.~(\ref{eq:qcd-dy-lp}) is also valid for 
Drell-Yan differential cross section, $d\sigma/d^4q d\Omega$, when $q_T\sim Q$,
where $\Omega$ is the solid angle of lepton in the pair's rest frame \cite{collins-book}.  
Because of the nature of electromagnetic or weak interaction between 
the observed lepton pair and how the pair was produced, 
inclusive Drell-Yan cross section is a hard and clean probe 
for PDFs and partonic structure of nucleons and nuclei, as well as pions.  
It is especially sensitive to the sea quarks of colliding hadron because 
of its leading quark-antiquark annihilation subprocess, and complementary to 
DIS which is more sensitive to the sum of parton flavors.  
In addition, when $q_T > Q/2$, inclusive Drell-Yan cross section 
is dominated by the quark-gluon ``Compton'' subprocess, and it 
becomes an excellent probe of glue inside a colliding hadron 
\cite{Berger:1998ev,Berger:1999as}.  

Even when $Q\gg q_T$ (or $q_T\gg Q$), so long as both of them $Q, q_T \gg \Lambda_{\rm QCD}$,
the formalism in Eq.~(\ref{eq:qcd-dy-lp}) is still valid.
However, perturbatively calculated short-distance hard part, 
$\hat{\sigma}_{a+b\to l\bar{l}}$, will receive $\alpha_s\ln^2(Q^2/q_T^2)$ (or  
$\alpha_s\ln(q_T^2/Q^2)$) type large logarithms  
for each additional power of $\alpha_s$
\cite{DDTdoublelog,PPbspace,Collins:1984kg,berger2001}. 
Such large logarithmic contributions would ruin the convergence of perturbative expansion
in powers of $\alpha_s$, and need to be resummed.
Technique and formalism to resum these large logarithmic contributions have
been developed and successfully applied, and will be discussed in 
the next section.
In the rest of this section, we review the role of inclusive Drell-Yan process in probing
PDFs and hadron's parton structure. Comprehensive reviews of earlier 
results on the Drell-Yan process can be found in several review 
articles~\cite{kenyon82,schochet86,freudenreich90,stirling93,
pat99,reimer07,dutta13}, 
we focus on the more recent results obtained at the
Fermilab fixed-target experiments and LHC.

\subsection{Up and down sea quark flavor asymmetry}
\label{sec:udasym}

Soon after the discovery of the point-like constituents in the nucleons
in DIS experiments, evidence for the existence of nucleon sea was 
revealed from the observation that the structure functions
continue to rise as $x \to 0$. The important role of the quark-antiquark
pairs in hadronic systems is in sharp contrast to the situation for
atoms, where particle-antiparticle pairs play a relatively minor role.
As a result of the large coupling strength $\alpha_s$, quark-antiquark
pairs are readily produced in strong interactions, and they form
an integral part of nucleon's structure.

The earliest parton models assumed that the proton sea was SU(3) 
flavor symmetric,
even though the valence quark distributions are clearly not
flavor symmetric.
The flavor symmetry assumption was not based on any known physics. 
Indeed, neutrino-induced charm production
experiments~\cite{conrad98} already showed
that the strange-quark content of the nucleon was only about one half
of the up or down sea quarks. This asymmetry was attributed to the
heavier strange-quark mass compared to the up and down quarks. The comparable
masses for the up and down quarks suggest that the nucleon sea should be nearly
up-down flavor symmetric.

\subsubsection{Experimental evidence for $\bar u / \bar d$ flavor asymmetry}

\begin{figure}[h]
\includegraphics[width=0.5\textwidth]{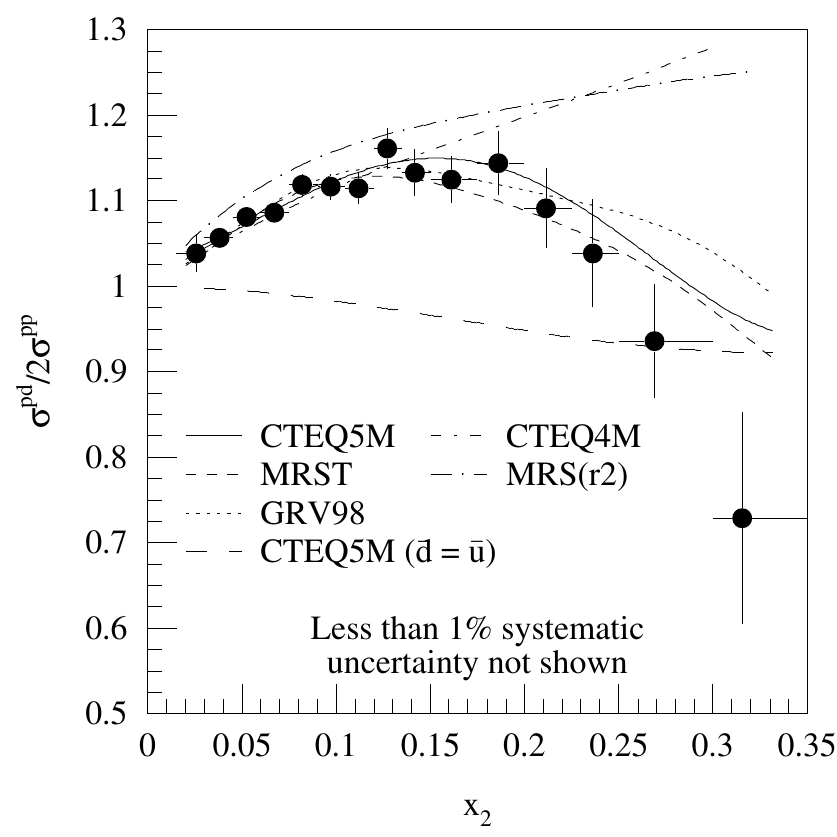}
\includegraphics[width=0.5\textwidth]{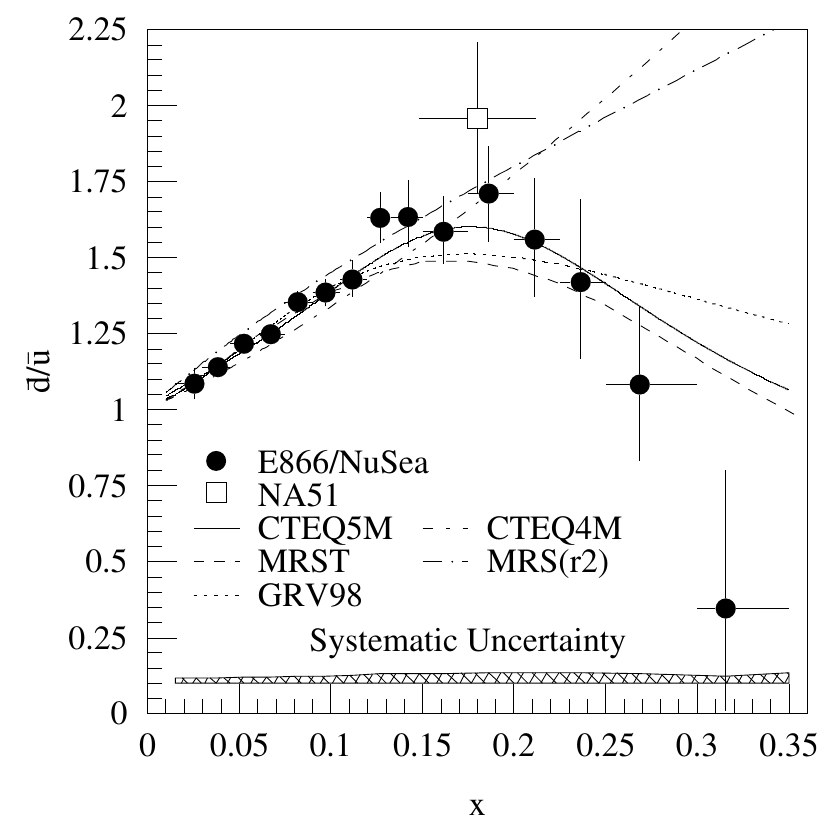}
\caption{Left (a): The Drell-Yan cross section ratios of $p+d$ over $2(p+p)$
versus $x_2$ (momentum fraction
of the target partons) from FNAL E866. The curves are
the calculated next-to-leading-order cross section ratios for the
Drell-Yan using various PDFs including MRS~\cite{MRS}, GRV98~\cite{GRV98},
MRST~\cite{MRST}, CTEQ4M~\cite{CTEQ4M} and CTEQ5M~\cite{CTEQ5M}.
Right (b): $\bar d(x)/\bar u(x)$ versus $x$ extracted from FNAL E866.
Parametrizations from various PDFs and the data point from NA51
are also shown (from ~\cite{e866,e866-1,e866-2}).}
\label{fig3.1}
\end{figure}

The issue of the equality of $\bar u$ and $\bar d$ was first
encountered in measurements of the Gottfried integral~\cite{gott},
given as
\begin{equation}
I_G = \int_0^1 \left[F^p_2 (x) - F^n_2 (x)\right]/x~ dx =
{1\over 3}+{2\over 3}\int_0^1 \left[\bar u_p(x)-\bar d_p(x)\right]dx,
\label{eq3.1}
\end{equation}
where $F^p_2$ and $F^n_2$ are the proton and neutron structure
functions measured in DIS experiments.
Eq.~(\ref{eq3.1}) is derived assuming charge symmetry (CS) at
the partonic level, namely, $u_p(x)=d_n(x),~ \bar
u_p(x) = \bar d_n(x),~d_p(x)= u_n(x),$ and $ \bar d_p(x) = \bar u_n(x)$.
For a $\bar u$, $\bar d$ flavor-symmetric sea in
the proton, the Gottfried Sum Rule (GSR)~\cite{gott}, $I_G
= 1/3$, is obtained. The New Muon Collaboration (NMC)~\cite{nmc91} 
determined the Gottfried integral to be
$ 0.235\pm 0.026$, significantly below 1/3, strongly 
suggesting that $\bar d(x) \ne \bar u(x)$ (the subscript $p$
is dropped for simplicity). This surprising result has
generated much interest. Since the violation of the GSR 
could also be caused by unusual behavior of the parton distributions at 
unmeasured small $x$ region, as well as by violation of the charge
symmetry at the partonic level, an independent experimental test of
the assumption $\bar u(x) = \bar d(x)$ was required.

\begin{figure}[h]
\includegraphics[width=0.42\textwidth]{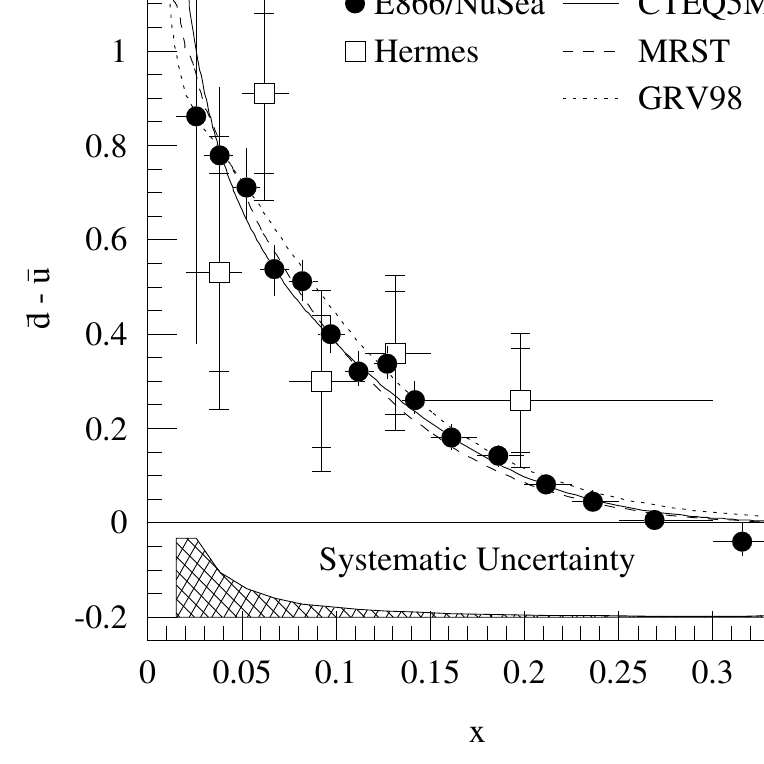}
\hskip 0.5cm
\includegraphics[width=0.46\textwidth]{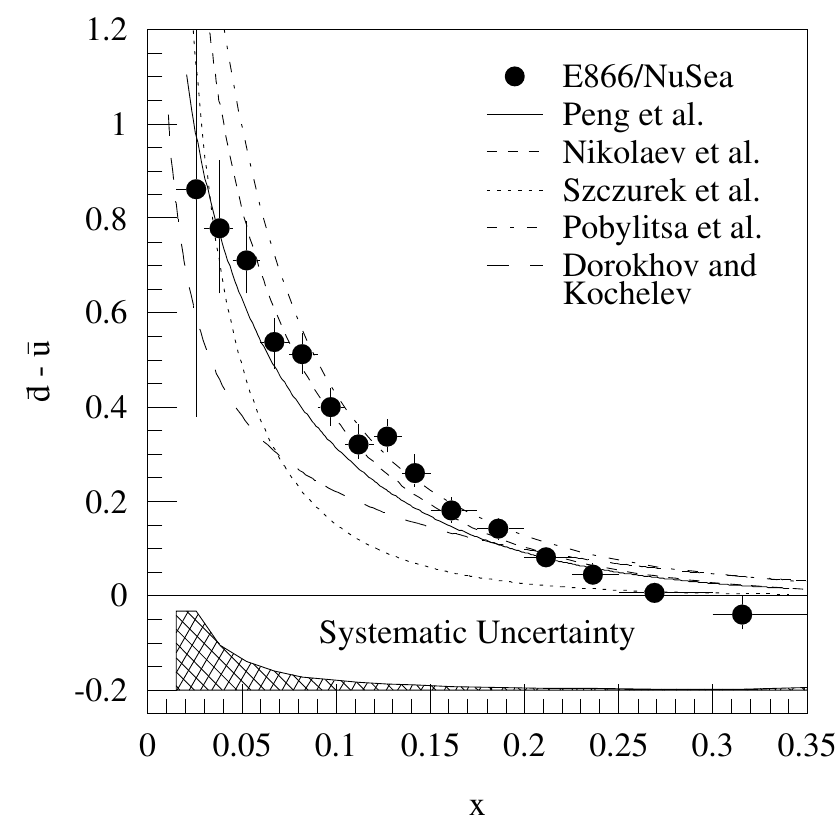}
\caption{
Left (a): $\bar d - \bar u$ as a function of $x$.
The E866 results~\cite{e866-2}, scaled to fixed $Q^2 = 54$~GeV$^2$,
are shown
as the circles. Results from HERMES ($\langle Q^2\rangle = 2.3$~GeV$^2$)
are shown as squares~\cite{hermes98}.
Right (b): Comparison of the measured ${\bar d}(x)-{\bar u}(x)$ at $Q^2$ = 54
\rm{GeV$^2$} to predictions of
several models of the nucleon sea~\cite{e866-2}. The curves correspond
to pion-cloud model by Peng et al.~\cite{e866-1} and
Nikolaev et al.~\cite{nikolaev}, chiral-quark model by
Szczurek et al.~\cite{szczurek96},
chiral-quark soliton model by Pobylitsa et al.~\cite{pobylitsa},
and instanton model by Dorokhov and Kochelev~\cite{dorokhov}.}
\label{fig3.2}
\end{figure}

The $\sigma_{DY}(p+d)/\sigma_{DY}(p+p)$ cross section ratios for 
the proton-induced Drell-Yan 
process provide an
independent means to probe the flavor asymmetry of the nucleon sea~\cite{es}.
At forward rapidity region the Drell-Yan cross section is dominated by the
annihilation of the $u$ quark in the proton beam with the $\bar u$
antiquark in the target nucleon. Assuming CS, one obtains
\begin{equation}
\sigma_{DY}(p+d)/2\sigma_{DY}(p+p) \simeq
\left[1+\bar d(x)/\bar u(x)\right]/2,
\label{eq:3.2}
\end{equation}
\noindent where $x$ refers to the momentum fraction of antiquarks.
Eq.~(\ref{eq:3.2}) shows that an important advantage of the Drell-Yan 
process is that the $x$ dependence of
$\bar d / \bar u$ can be determined. Using a 450 GeV proton beam,
the NA51 collaboration~\cite{na51} obtained
$\bar u(x)/\bar d(x) = 0.51 \pm 0.04 (stat) \pm 0.05 (syst)$
at $x = 0.18$ and $\langle M_{\mu\mu}\rangle = 5.22$ GeV.
The Fermilab E866/NuSea~\cite{e866,e866-1,e866-2} Collaboration
measured the Drell-Yan cross section ratios over a broad range
of $x$ using 800 GeV proton beams. 
As shown in Fig.~\ref{fig3.1}(a), these ratios were found
to be significantly different from those expected for $\bar d = \bar u$ sea
(dashed curve), indicating an asymmetric $\bar d$, $\bar u$
sea over an appreciable range in $x$.
The values of $\bar d(x)/\bar u(x)$ extracted from NLO calculations
of the $\sigma^{p+d}/2\sigma^{p+p}$ Drell-Yan cross section ratios 
are shown in Fig.~\ref{fig3.1}(b). Results from
the NA51 and E866 experiments are consistent, showing that the
$\bar d(x) / \bar u(x)$ ratio increases linearly from unity at $x \to 0$
up to $x \sim 0.15$, reaching a maximal value of $\sim 1.75$, and then
drops off at higher $x$.

The Drell-Yan cross section ratios from E866 were analysed to obtain
$\bar d(x) - \bar u(x)$ over the region $0.02 < x < 0.345$ as shown 
in Fig.~\ref{fig3.2}(a).
The HERMES Collaboration has reported a SIDIS measurement 
of charged pions from hydrogen and deuterium
targets~\cite{hermes98}.
Based on the differences between charged-pion yields from the two targets,
$\bar d(x) - \bar u(x)$ is determined
in the kinematic range, $0.02 < x < 0.3$ and
1 GeV$^2 < Q^2 <$ 10 GeV$^2$. 
It is worth noting that the HERMES results
are consistent with
the E866 results obtained at much higher $Q^2$.
In Table~\ref{tab:1} we list the
values of the integral $\int_0^1 [\bar d(x) - \bar u(x)] dx$ determined
from the NMC, HERMES, and FNAL E866 experiments. The agreement among these
results, obtained using different techniques including DIS, semi-inclusive
DIS, and Drell-Yan, is quite good.

\begin{table}
\caption{Values of the integral
$\int_0^1 [\bar d(x) - \bar u(x)] dx$ determined from the DIS,
semi-inclusive DIS, and Drell-Yan (DY) experiments.}
\label{tab:1}       
\begin{center}
\begin{tabular}{lll}
\hline\noalign{\smallskip}
Experiment & $\langle Q^2 \rangle$ (GeV$^2$/c$^2$) &
$\int_0^1 [\bar d(x) - \bar u(x)] dx$  \\
\noalign{\smallskip}\hline\noalign{\smallskip}
NMC/DIS & 4.0 & $0.147 \pm 0.039$ \\
HERMES/SIDIS & 2.3 & $0.16 \pm 0.03$ \\
FNAL E866/DY & 54.0 & $0.118 \pm 0.012$ \\
\noalign{\smallskip}\hline
\end{tabular}
\end{center}
\end{table}

\subsubsection{Theoretical interpetation of the $\bar d / \bar u$ flavor 
asymmetry}

The Drell-Yan data on $(p+d)/(p+p)$ have been included in all recent global
fits to determine nucleon parton distribution functions.
Most of these global fits are performed at next-to-leading order (NLO),
while several higher order (NNLO) global fits are also becoming available.
Some examples of the recent NLO PDFs include the fits performed by
the CTEQ/CT group (CTEQ6.6~\cite{CTEQ6.6} and CT10~\cite{CT10}),
the MSTW group (MSTW08~\cite{MSTW08}), and the NNPDF collaboration
(NNPDF2.0~\cite{NNPDF2.0} and NNPDF2.1~\cite{NNPDF2.1}).
At the initial $Q_0^2$ scale, the $\bar d(x) - \bar u(x)$ flavor asymmetry
together with the valence quark, light quark sea and gluon distributions,
are usually parametrized with some functional forms (with the
exception of the NNPDF, which adopts neural network methodology). The fact that 
$\bar d(x) - \bar u(x)$ flavor asymmetry is already present at the 
initial $Q_0^2$ 
scale reflects the non-perturbative nature of this asymmetry. It is
conceivable that other non-perturbative features of the nucleon
sea, such as the valence-like intrinsic sea and
the $s(x)-\bar s(x)$ asymmetry (to be discussed later), will also be
included for future PDF parametrizations. 

\begin{figure}[t]
\includegraphics[width=0.5\textwidth]{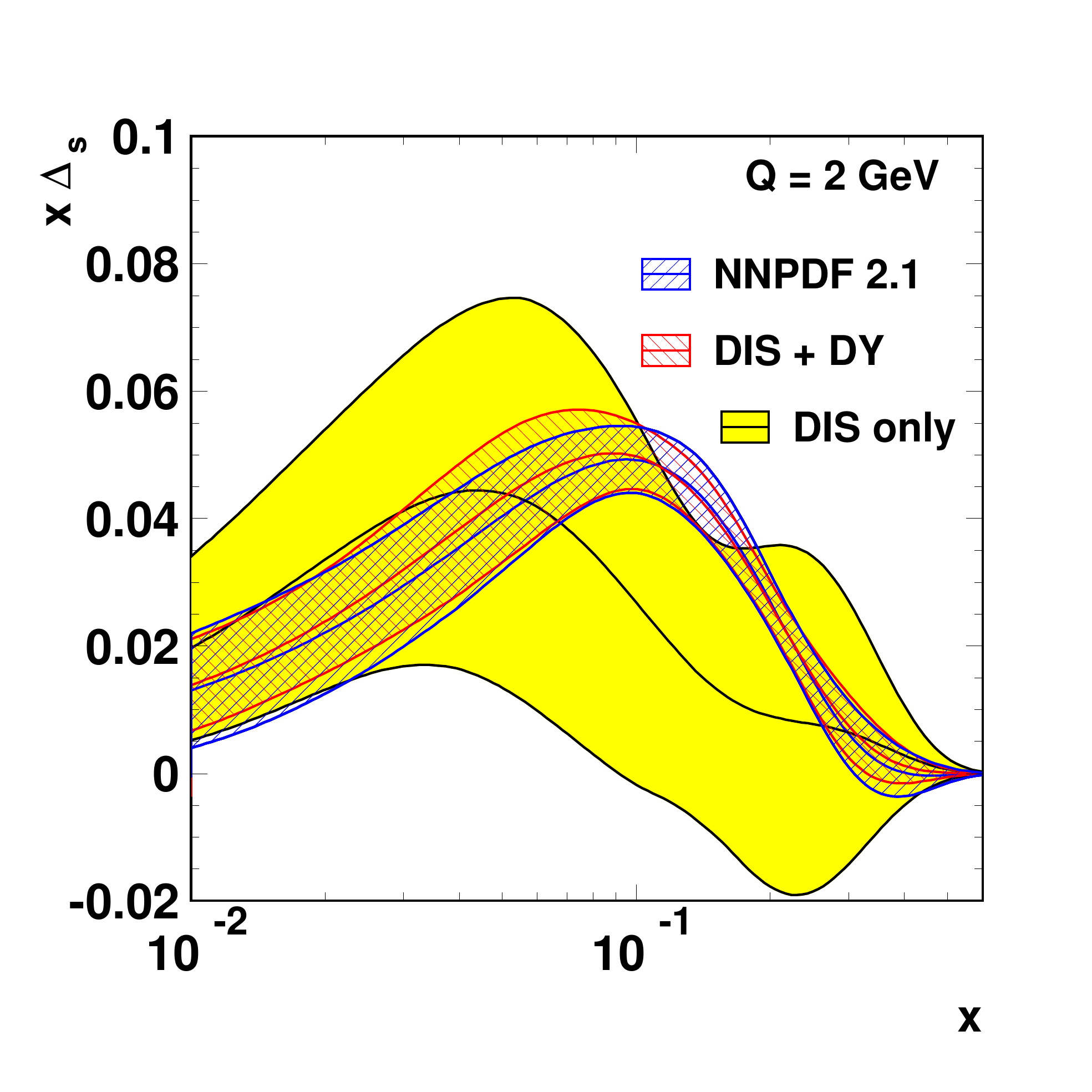}
\includegraphics[width=0.5\textwidth]{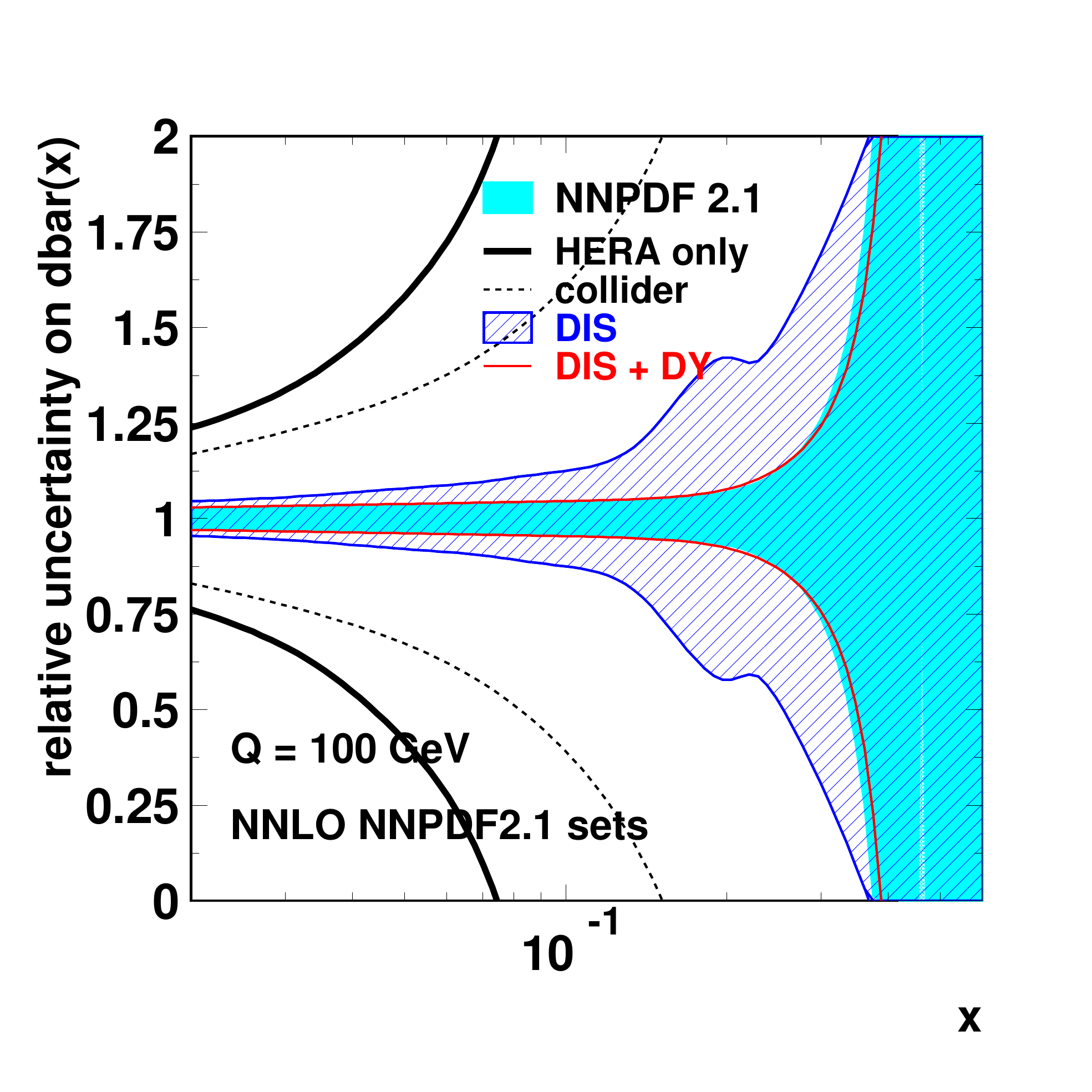}
\caption{Left (a): The asymmetry of the light sea 
$x \Delta_s(x) = x(\bar{d}(x) - \bar{u}(x))$ and its one standard deviation 
uncertainty at $Q = 2$~GeV as obtained from the NNPDF analysis, when only 
DIS data are included in the fit (yellow band), when Drell-Yan data are 
included in addition (red band), and from the reference NNPDF2.1 
fit (blue band).
Right (b): One standard deviation uncertainties on
$\bar{d}(x)$ at $Q = 100$~GeV,
as obtained from the NNPDF2.1 global fit (filled area), and from the same fit
but applied to a subset of experimental data. From~\cite{perez12}.}
\label{fig3.3}
\end{figure}

To illustrate the unique role of proton-induced
Drell-Yan for constraining the sea quark PDFs, Fig.~\ref{fig3.3} (a) shows 
the analysis of the NNPDF collaboration on how various experimental data
constrain the $\bar d(x) - \bar u(x)$ at $Q=2$ GeV~\cite{perez12}. 
The yellow band
corresponds to result obtained from fits to DIS data. The inclusion of 
the Drell-Yan fixed-target $p+p$ and $p+d$ 
data dramatically improves the accuracy (red band). Further inclusion of 
the Tevatron data only reduce the uncertainty slightly, as shown by 
the blue band in Fig.~\ref{fig3.3} (a). Figure~\ref{fig3.3} (b) illustrates
how the fixed-target DY data significantly reduce the uncertainty of
$\bar d(x)$ in the region of $0.05 < x < 0.3$. Future experiments are
anticipated to further constrain $\bar d(x) - \bar u(x)$ at larger $x$,
as discussed later.

Many theoretical models, including meson-cloud model~\cite{thomas83}, 
chiral-quark model~\cite{eichten}, Pauli-blocking model~\cite{field}, 
instanton model~\cite{dorokhov}, chiral-quark soliton
model~\cite{wakamatsu}, and statistical model~\cite{bourrely}, 
have been proposed to explain the $\bar d/ \bar u$ flavor asymmetry. Details of these models can be found in
several review articles~\cite{kumano98,speth98,vogt00,garvey02}. Most 
of these models
emphasize the important contribution of meson cloud to nucleon's 
sea quark content. This was first pointed out by Sullivan~\cite{sullivan} 
in the context of DIS, and later considered by Thomas~\cite{thomas83}
to predict the $\bar d(x) - \bar u(x)$ flavor asymmetry. It is remarkable
that meson cloud, usually considered to be important only at
low $Q^2$, can lead to such striking effect even at very large $Q^2$
scale probed by the DY experiments ($\langle Q^2 \rangle \sim 
54$ GeV$^2$ for FNAL E866). Fig.~\ref{fig3.2} (b) shows that  
$\bar d(x) - \bar u(x)$ can be reasonably well described by many different
theoretical models. To shed new lights on the origin of the 
$\bar d(x) - \bar u(x)$ asymmetry and to further test these models,
it is important to measure other sea-quark observables, such as
the polarized $\Delta \bar u(x)$ and $\Delta \bar d(x)$,
as well as $s(x) - \bar s(x)$~\cite{garvey02}. These will be discussed in later sections.

The $\bar d(x) - \bar u(x)$ data have been
utilized recently to extract the intrinsic light-quark sea content of the 
nucleons~\cite{chang11,chang11-1,chang11-2}.
The possible existence of the so-called
``intrinsic charm" in the nucleon via the $u u d c \bar c$ five-quark
Fock state was proposed some time ago by Brodsky,
Hoyer, Peterson, and Sakai (BHPS)~\cite{brodsky80,brodsky80-1}. The ``intrinsic 
charm" is to be distinguished from the conventional ``extrinsic
charm" produced in the splitting of gluons into $c \bar c$ pairs. 
While the extrinsic charm is expected to be localized at the small $x$ region, 
the intrinsic charm is predicted to be valence-like with a distribution
peaking at larger $x$ and could account for certain forward charm production
data~\cite{vogt95}. The CTEQ collaboration~\cite{pumplin06} 
performed a global fit to existing data allowing an intrinsic charm component,
and concluded that the data neither confirm nor rule out 
the existence of intrinsic charm. 
The probability for the $|u u d Q \bar Q\rangle$ Fock state is 
expected to be approximately proportional to $1/m_Q^2$,
where $m_Q$ is the mass of the quark $Q$~\cite{brodsky80,brodsky80-1}. Therefore,
the light five-quark states $|u u d u \bar u\rangle$, $|u u d d \bar
d\rangle$ and $|u u d s \bar s\rangle$ are likely to have
significantly larger probabilities than the $|u u d c \bar c\rangle$
state, and could be more readily observed experimentally. 
The challenge is to separate the intrinsic from the
extrinsic seas. It is essential
to consider experimental observables which have little or no
contributions from the extrinsic sea. The
$\bar d(x) - \bar u(x)$ is an example of such observables,
since the perturbative $g \to Q \bar Q$ processes will generate 
$u \bar u$ and $d \bar d$ extrinsic sea with equal probabilities.
A comparison between the calculations using the BHPS model for the
intrinsic sea with the $\bar d(x) - \bar u(x)$
data show good agreement. By including also the $s(x)+\bar s(x)$
and $\bar u(x) + \bar d(x) - s(x) - \bar s(x)$ data  
the probabilities for the $|uudu\bar{u}\rangle$, $|uudd\bar{d}\rangle$,
and $|uuds\bar{s}\rangle$ five-quark Fock states in the proton
are extracted~\cite{chang11,chang11-1,chang11-2}. Results from
this analysis support the concept of intrinsic sea in the nucleons.

\subsubsection{$\bar d(x) / \bar u(x)$ at large $x$}

\begin{wrapfigure}{R}{0.4\textwidth}
\begin{center}
\includegraphics[width=0.4\textwidth]{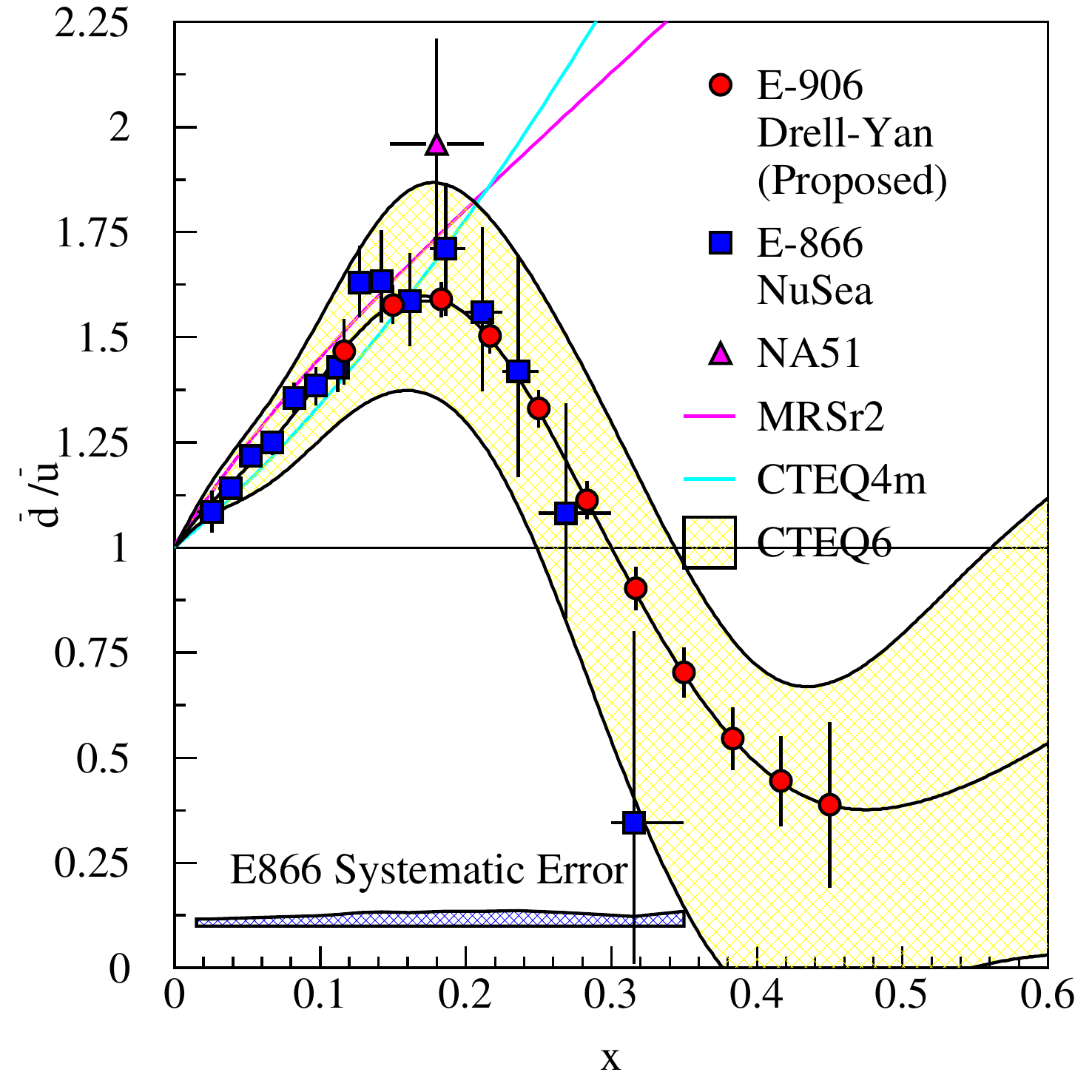}
\end{center}
\caption{Expected sensitivity for the E906/SeaQuest 
experiment~\cite{reimer11}.}
\label{fig_e906}
\end{wrapfigure}

While various theoretical models can describe the general trend of
the $\bar d(x) / \bar u(x)$
asymmetry, they all have difficulties explaining the $\bar d(x) / \bar u(x)$
E866 Drell-Yan data at large $x$ ($x>0.2$), where this ratio appears 
to fall below
unity, i.e. $\bar d (x)< \bar u$(x).  If confirmed by more 
precise measurements,
this intriguing $x$ dependence of the $\bar d(x) / \bar u(x)$ asymmetry could
shed important new  light on the nature of the nucleon sea.
Thus, it would be
very important to have new measurements sensitive to the $\bar d(x) /
\bar u(x)$ ratios at large $x$ ($x>0.2$).
For given values of $x_1$ and $x_2$ the Drell-Yan cross section
is proportional to $1/S$, where $S$ is the center-of-mass energy squared.
Hence the Drell-Yan measurements at lower beam energies could more effectively
probe the large $x$ region. The Fermilab E906/SeaQuest~\cite{e906}
experiment will
utilize 120 GeV proton beam, and the expected statistical
accuracy for $\sigma (p+d)/ 2 \sigma (p+p)$ is shown in Fig.~\ref{fig_e906}.
A definitive measurement of
the $\bar d(x)/ \bar u(x)$ over the region $0.25 < x < 0.5$ 
is expected in the near future.

The new 50 GeV
proton accelerator, J-PARC, presents another opportunity for
extending the $\bar d(x)/ \bar u(x)$ measurement to even larger 
$x$ ($0.25 < x < 0.7$)~\cite{peng00}. Since only 30 GeV proton beam is 
available at the initial
phase of J-PARC, the first measurements would focus on $J/\Psi$ production
at 30 GeV. An important feature of $J/\Psi$ production using 30 GeV proton
beam is the dominance of the quark-antiquark annihilation subprocess.
This is in striking contrast to $J/\Psi$ production at 800 GeV, where 
the dominant process is the gluon-gluon fusion. This suggests 
that $J/\Psi$
production at J-PARC can be used as an alternative method to probe antiquark
distribution, including $\bar d(x)/\bar u(x)$ at 
large $x$~\cite{peng00}.

\subsection{Flavor dependence of the EMC effect}
\label{sec:flavor-emc}

The first definitive evidence for the modification of parton distributions
in nuclei was observed in muon DIS
experiment~\cite{emc1}. This surprising finding, called the EMC effect, was
confirmed later by other DIS experiments with electron, muon,
and neutrino
beams~\cite{slac,muons,neutrinos}. Many 
theoretical models~\cite{geesaman_review,norton_review} have been proposed
to explain the EMC effect. While these
models are capable of describing some features of the EMC effect, they
span a broad range of physics ideas. The true physics origin of the EMC effect
remains to be better understood.

An effective test of the various theoretical models can be
obtained from
the quark flavor dependences of the EMC effect. This was nicely
demonstrated by the measurement~\cite{drell_yan} of 
the nuclear dependence of the 
proton-induced Drell-Yan process, which is primarily sensitive to
the EMC effect of $\bar u$ distributions. The absence of the nuclear
enhancement of $\bar u$ distributions in these measurements has 
provided strong constraints on many EMC models~\cite{drell_yan}.

To further understand the origin of the EMC effect, it is
important to examine other quark flavor dependences.
The possibility for a flavor-dependent modification of quark distributions 
was suggested recently  
by Clo\"{e}t, Bentz, and
Thomas (CBT)~\cite{ianemc,ian}, who pointed out that the isoscalar and
isovector mean fields in a nucleus will modify the quark distributions in the
bound nucleons according to the Nambu-Jona-Lasinio model.
An interesting consequence of the presence of the
isovector vector meson ($\rho^0$) mean field in an $N \ne Z$ nucleus is that
the $u$ and $d$ quarks in the bound nucleons are modified
differently, leading
to flavor dependences of nuclear quark distributions.

\begin{figure}[tbp]
\includegraphics[width=0.5\textwidth]{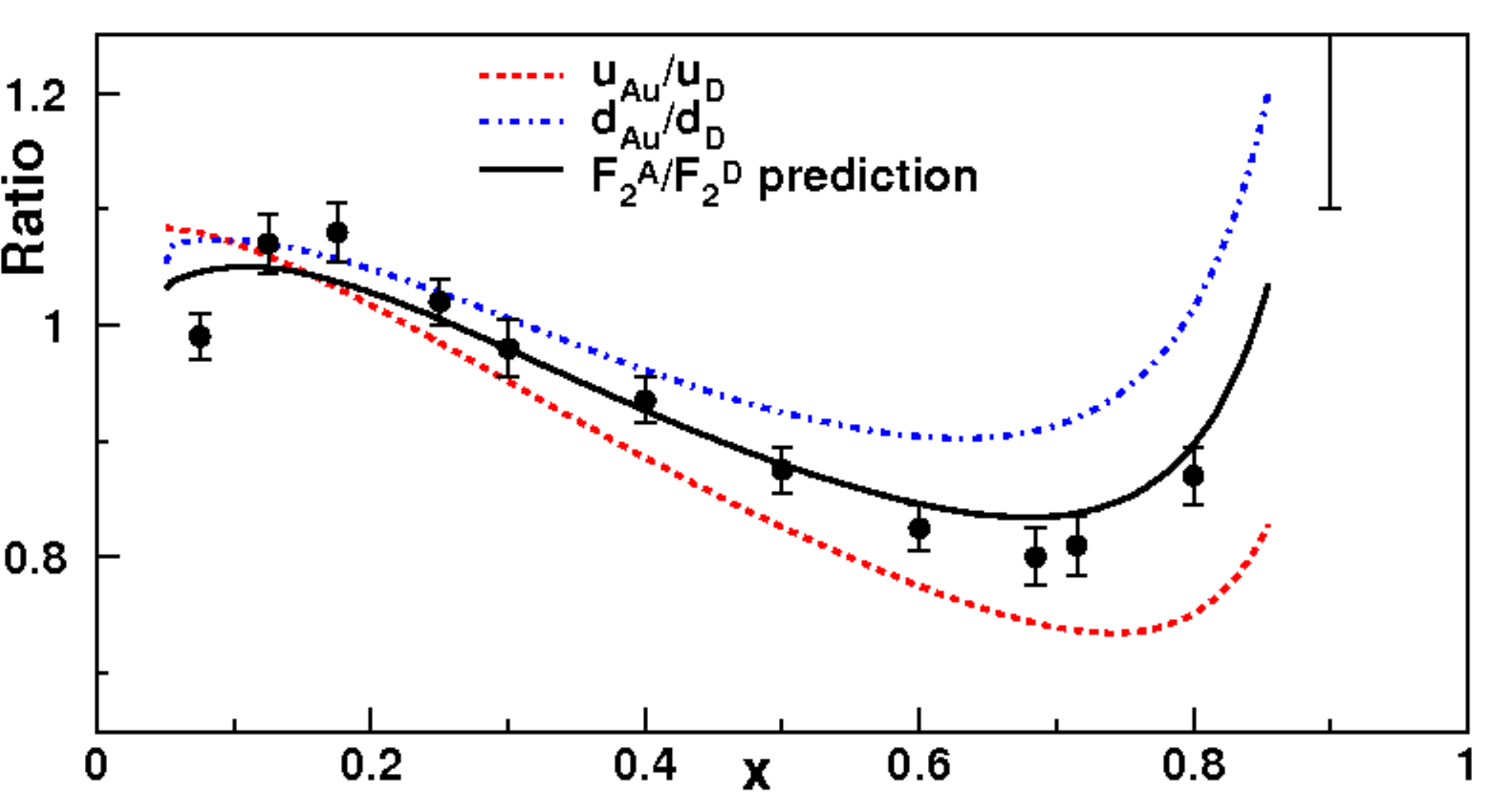}
\includegraphics[width=0.5\textwidth]{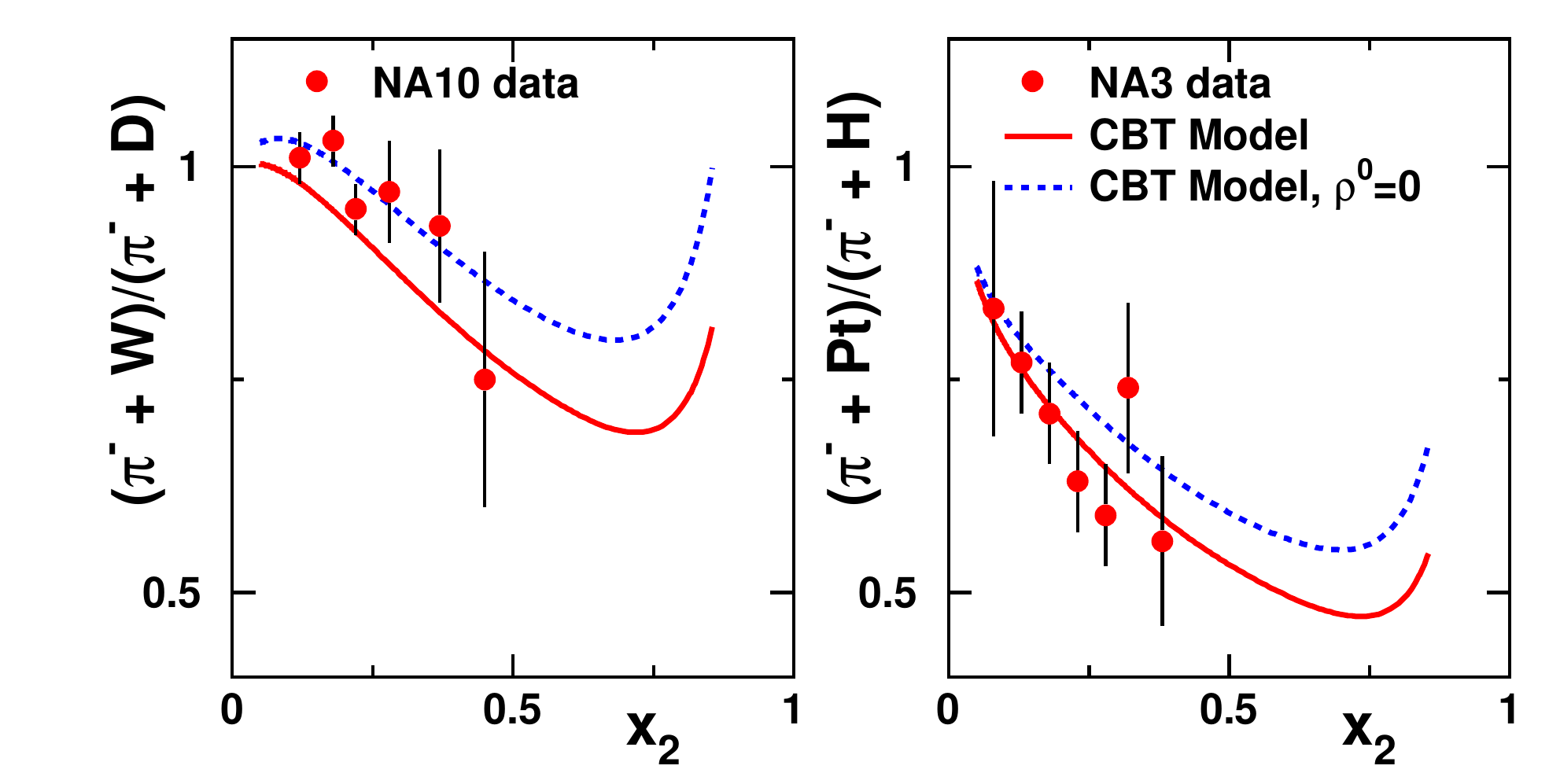}
\caption{Left (a): Ratios of quark distributions and structure functions in
nuclear matter versus the deuteron plotted as a function of Bjorken-$x$,
at $Q^2 = 10\,$GeV$^2$.
Filled circles are data for $N=Z$ nuclear matter from
Ref.~\cite{day-sick}. Curves are results from the CBT model~\cite{ianemc,ian}
where the solid line is the calculation of $F_2^A/F_2^D$ for $N=Z$
nuclear matter, and dashed and dot-dashed
curves are the ratios of $u$ and $d$ quark distributions in a gold nucleus
over those in a deuteron. Right (b): Comparison between NA3 and NA10 data 
with calculations from the CBT model~\cite{dutta2011}. The solid line 
is the full flavor-dependent result and the dashed line is obtained by 
setting the $\rho^0$ mean field to zero.}
\label{dutta}
\end{figure}

Figure~\ref{dutta} (a) shows some CBT results for nuclear PDFs.
The solid curve is for the usual EMC effect in symmetric
nuclear matter, where $F^A_2$ and $F_2^D$
are the per-nucleon structure functions of the nucleus and the deuteron,
respectively. Inclusive DIS experiments measure the weighted sum of 
$u$ and $d$ quark distributions, making it difficult to extract
the quark-flavor dependence of the EMC effect.
Several new measurements sensitive to the flavor-dependent EMC effects
have been considered. They include the parity-violating DIS
asymmetry proposed at the JLab 12 GeV facility~\cite{pr12007,ian12},
the semi-inclusive DIS on nuclear target first considered by Lu and
Ma~\cite{ma2006} and recently proposed at JLab~\cite{pr12004}.

Pion-induced Drell-Yan processes provide another experimental tool that
is sensitive to flavor-dependent effects in the nuclear quark 
distributions~\cite{dutta2011}.
Keeping only the dominant terms in Drell-Yan cross section,
one obtains 
\begin{equation}
R^{-}_{A/D} = \frac{\sigma^{DY}(\pi^- + A)}{\sigma^{DY}(\pi^- + D)}
\approx \frac{u_{A}(x)}{u_{D}(x)};~~~~~~
R^{-}_{A/H} = \frac{\sigma^{DY}(\pi^- + A)}{\sigma^{DY}(\pi^- + H)}
\approx \frac{u_{A}(x)}{u_p(x)},
\end{equation}
The target PDFs have a subscript $A$, and the up quark distribution in
the deuteron and the proton are labeled $u_D$ and $u_p$, respectively.

In Fig.~\ref{dutta} (b) the calculation of the pion-induced Drell-Yan
cross section ratios are compared with the existing NA3~\cite{badier81}
and NA10~\cite{bordalo87} data.
The solid curves in Fig.~\ref{dutta} (b) are calculations using the
flavor-dependent $u_A(x)$ and $d_A(x)$ from the CBT model.
The dashed curves in Fig.~\ref{dutta} (b) are obtained using
PDFs from the CBT model with no flavor-dependent nuclear effects,
that is, with the $\rho^0$ mean field set to 0. The NA10 data
exhibit a preference for the flavor-independent nuclear
PDFs. In contrast, the NA3 data favor the calculation using
flavor-dependent CBT nuclear PDFs.

The existing Drell-Yan data are not sufficiently accurate to draw
any firm conclusions. Precise future pion-induced Drell-Yan experiments,
such as the proposed measurement by the COMPASS collaboration with
160-GeV pion beams, offer an opportunity for a precise
test~\cite{dutta2011} of flavor dependence of the EMC 
effect. A recent paper also
explored the feasibility of using $W$ production in proton-nucleus
collision at LHC to probe the flavor-dependent EMC effect~\cite{chang_w}.

\subsection{Gluon distribution in proton versus neutron}

As discussed earlier, the measurement of the $(p+d)/(p+p)$ Drell-Yan
cross section ratios has led to the conclusion that $\bar u(x)$ and
$\bar d(x)$ distributions in the proton are different. Strictly speaking,
this measurement is probing directly the ratio of the $\bar u(x)$
distribution of the neutron over that of the proton. The Drell-Yan process,
dominated by the $u - \bar u$ annihilation subprocess, leads to the
relation $\sigma (p + d)_{DY}/2 \sigma (p+p)_{DY} \approx \frac {1}{2}
(1 + \bar u_n(x_2)/\bar u_p (x_2))$, where $\bar u_{p,n}$ refers to
the $\bar u$ distribution in the proton and neutron, respectively.
An interesting question is whether the gluon distribuions are identical
for proton and neutron.
While it is generally assumed that the gluon
distributions in the proton and neutron are identical, this assumption
should be tested experimentally.
Charge symmetry violation at the parton level could lead to a
difference for the gluon distributions in the proton and neutron~\cite{piller}.
Since quarkonium production at high energies is dominated by the
gluon-gluon fusion subprocess~\cite{peng95,vogt99}, it can probe 
the gluon distributions in
the colliding hadrons. In particular, the $\sigma(p+d \to J/\Psi (\Upsilon)X)
/2 \sigma(p+p \to J/\Psi (\Upsilon))$ ratio for $J/\Psi$ or $\Upsilon$ 
production can probe the gluon content in the neutron relative to that 
in the proton~\cite{piller}.

\begin{figure}[tb]
\begin{center}
\includegraphics[width=0.6\textwidth]{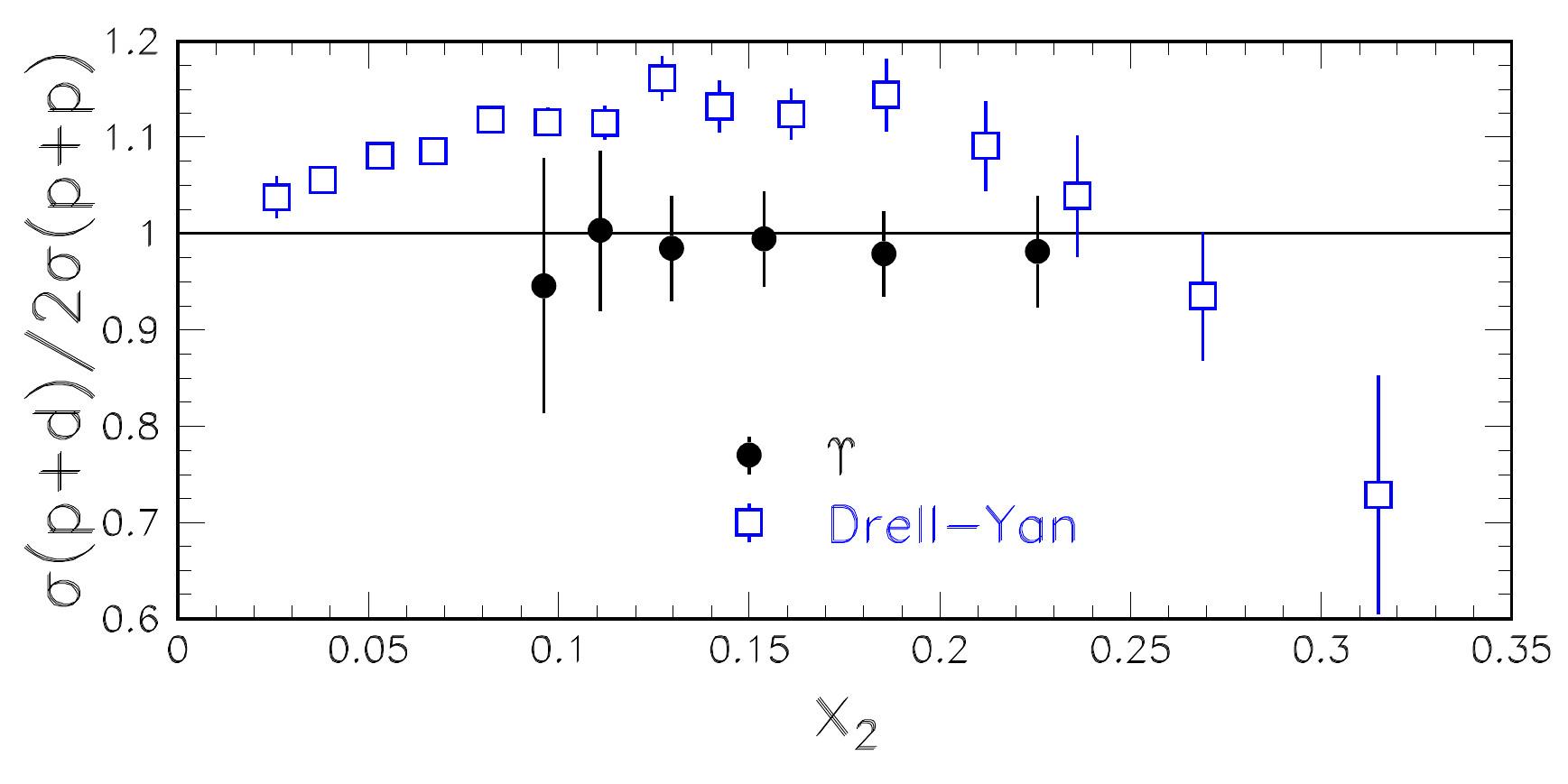}
\end{center}
\caption{The E866 $\sigma(p + d) /2 \sigma(p + p)$ cross section
ratios for $\Upsilon$ resonances as a function of $x_2$~\cite{zhu_ups}. 
The ratios for the E866 Drell-Yan cross 
sections~\cite{e866,e866-1,e866-2}
are also shown.}
\label{crosfig5}
\end{figure}

The Fermilab E866 measured the $\sigma (p+d)/2\sigma(p+p)$ 
ratios for $\Upsilon (1S+2S+3S)$ production~\cite{zhu_ups}.
As shown in Fig.~\ref{crosfig5}, these ratios are consistent with
1 over the range of $0.09 < x_2 < 0.25$.
The dominance of the gluon-gluon fusion subprocess implies that 
$\sigma (p+d \to \Upsilon)/ 2 \sigma(p+p \to \Upsilon)\approx \frac
{1}{2} (1+g_n(x_2)/g_p(x_2))$. 
Figure~\ref{crosfig5} shows that
the gluon distributions in the proton ($g_p$) and neutron ($g_n$)
are very similar, consistent with no charge symmetry breaking 
for nucleon's gluon distributions. Similar results were 
also obtained by the NA51
collaboration for the measurement of $J/\Psi$ and $\Psi^\prime$
on hydrogen and deuterium targets using 450 GeV 
proton beam~\cite{abreu98}.

The upcoming Fermilab E906 experiment offers the opportunity to
extend the measurement to $J/\Psi$ production. A high statistics measurement
of $\sigma (p+d \to J/\Psi)/ 2 \sigma(p+p \to J/\Psi)$ at 120 GeV
proton energy could be obtained in E906. 
A fixed-target experiment at LHC using 7 TeV proton beam extracted by a bent
crystal has also been considered~\cite{brodsky13}. A large number
of $J/\Psi (\sim 10^9)$ and $\Upsilon (\sim 10^6)$ would be
detected in $p+p$ and $p+d$ collisions. Both the E906 and 
the LHC fixed-target experiments could lead to
further sensitive test for charge symmetry breaking in nucleon's
gluon distributions.

\subsection{Strange quark distributions in the proton}

The question regarding the strange quark contents of the nucleon continues 
to attract much attention. 
Motivated by earlier indication that nucleon's mass has a significant
contribution from the $s \bar s$ component, several parity-violating
electron-nucleon elastic scattering experiments have been performed
in order to determine the strangeness electric and magnetic form factors.
As discussed in a recent review article~\cite{armstrong}, the 
strange quark form factors
were found to be small, in good agreement with lattice QCD and some
other model calculations.

In the large $Q^2$ domain, the strange quark distributions in the
nucleon can be probed by various hard processes.
Unlike the up and down quarks, there is no 
net strangeness, or equivalently, no net valence strange quarks,
in the nucleon. Therefore, a comparison between the strange quark 
sea and the 
up and down quark seas could shed some light on how  
the flavor structure of the nucleon sea depends on the
flavors of the valence quarks. It is also interesting
to investigate how the SU(2) flavor asymmetry of the $\bar d / \bar u$
sea is extended to SU(3) flavor asymmetry with the inclusion
of the $s$ and $\bar s$ sea.

In this section, we discuss two aspects of the
strange quark distributions in the nucleon. First, we examine 
the $x$-integrated as well as the $x$-dependent 
strange quark distributions relative
to those of the up and down sea quarks. Second, we compare the
$s(x)$ versus the $\bar s(x)$ distributions in the proton.

\subsubsection{Asymmetry between strange and up/down seas}

Assuming SU(3) flavor symmetry and neglecting effects of the valence
$u$ and $d$ quarks on the sea quark flavor structure, a symmetric
$u, d, s$ nucleon sea would be expected. However, this expectation is
not realistic, as the SU(3) flavor symmetry is broken by the larger 
mass of the strange quark. Moreover, the large $\bar d(x)/\bar u(x)$
asymmetry suggests that the valence quarks can affect the flavor
structure of the nucleon sea. 

\begin{figure}[tbp]
\includegraphics[width=0.5\textwidth]{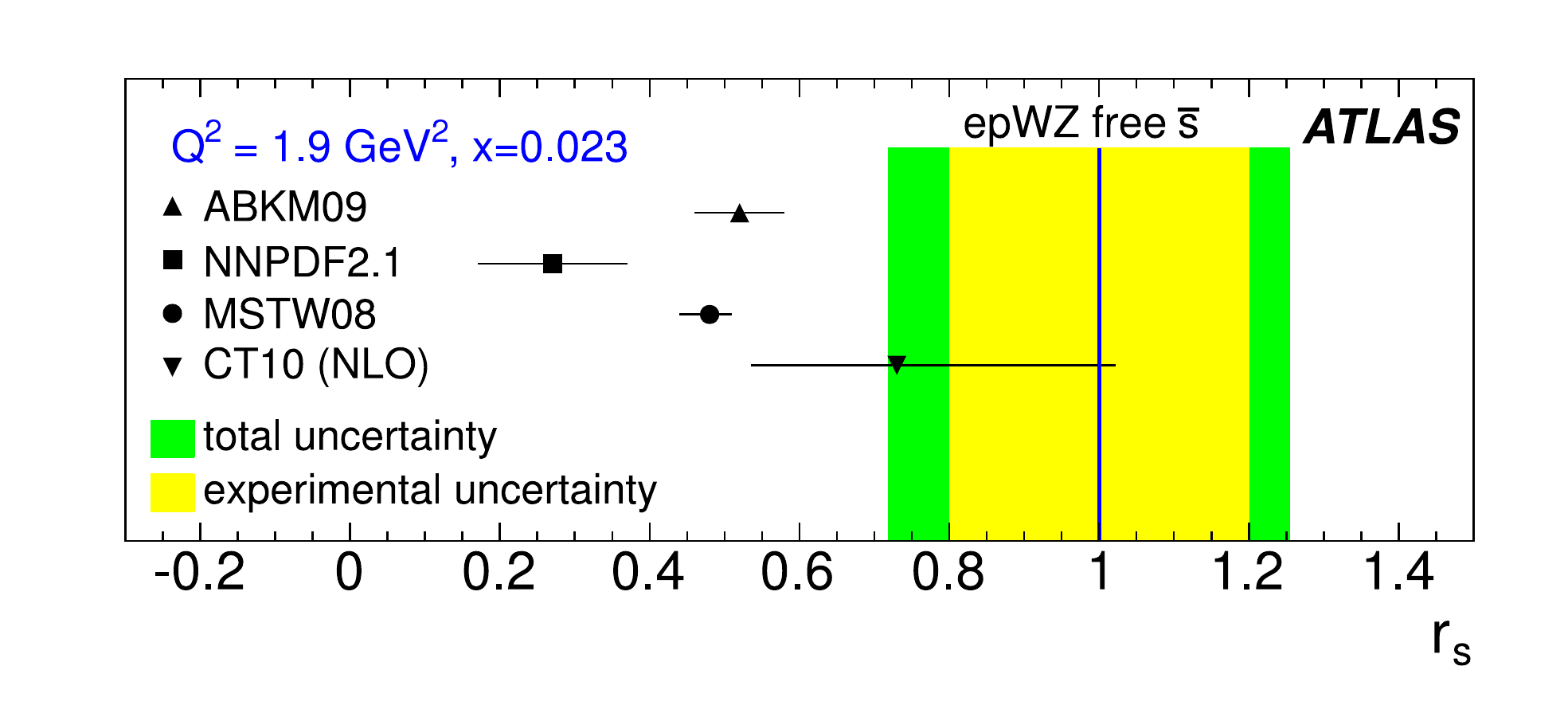}
\includegraphics[width=0.5\textwidth]{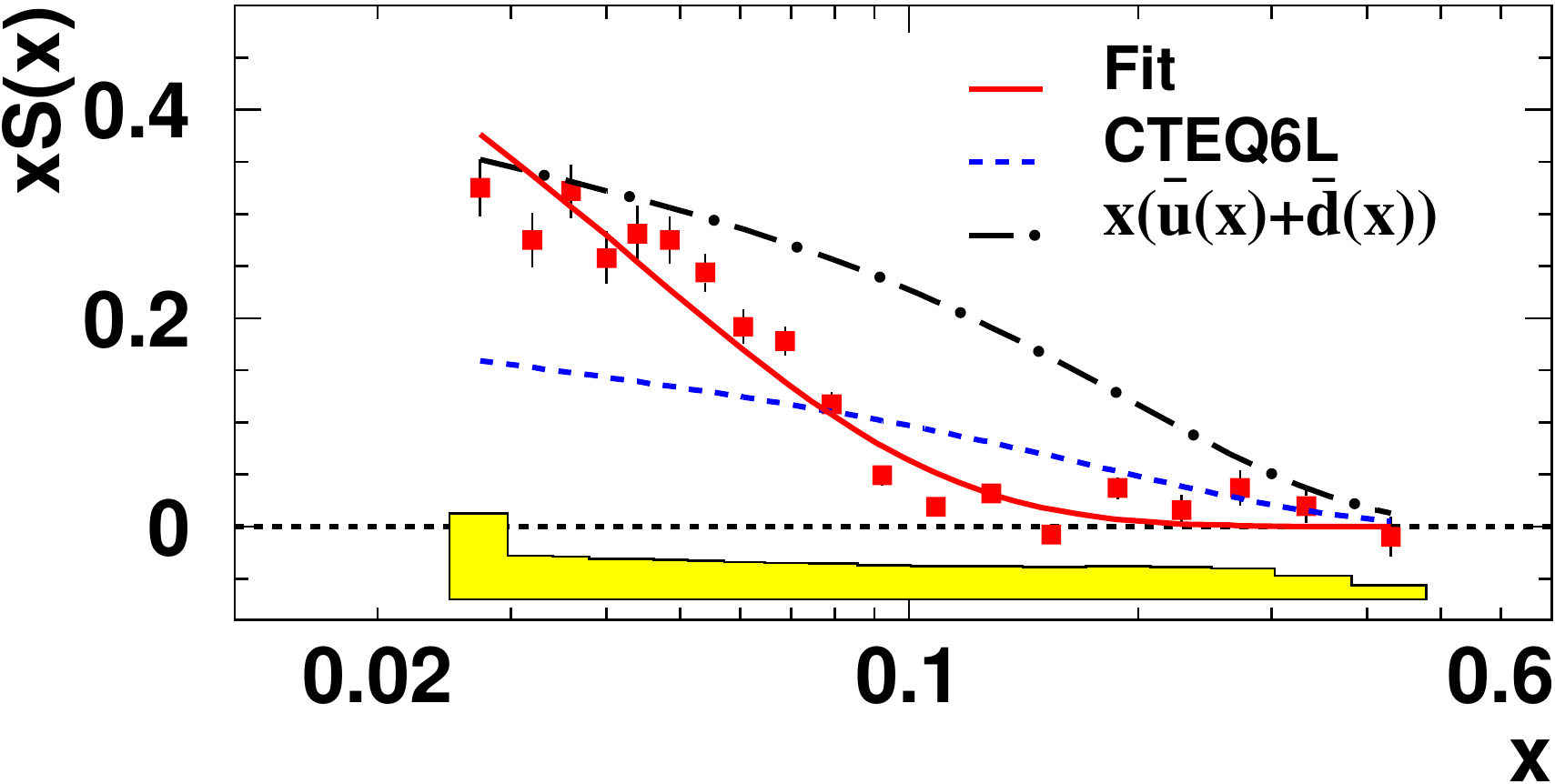}
\caption{Left (a): Ratio $r_s = (s+\bar s)/2\bar d$ at 
$Q^2 = 1.9$ GeV$^2$ and $x = 0.023$ from ATLAS (bands) and 
various PDFs~\cite{atlas_s}. 
Right (b): The strange parton distribution $x(s(x)+\bar s(x)$ from HERMES
semi-inclusive kaon production measurement~\cite{hermes08}. 
The dot-dash curve is 
$x (\bar u(x) + \bar d(x))$ from CTEQ6L~\cite{CTEQ6.6}.}
\label{atlas_hermes}
\end{figure}

Information on the strange quark contents was primarily extracted from
neutrino DIS experiments. These experiments allowed the determination of
the ratio, $\kappa$, of nucleon momentum carried by strange versus 
lighter non-strange sea quarks:
\begin{equation}
\kappa = \left[\int^1_0 x(s(x)+\bar s(x))~dx\right] /\left[\int^1_0 x(\bar u(x)+\bar d(x))~dx\right].
\end{equation}
\noindent A value of $\sim 0.5$ was found for $\kappa$, indicating that
strange quark sea is suppressed by roughly a factor of two relative 
to the lighter quark sea, presumably due to its heavier mass. The 
extraction of $\kappa$ typically assumes identical shape ($x$-dependence) 
for $u, d$ and$s$ quark seas. However, the neutrino DIS data only cover
a relatively narrow range in $x$, and can not test this assumption. 
As discussed earlier, the ratio $\bar d(x)/ \bar u(x)$ is strongly
dependent on $x$, and it is conceivable that similar situation also
occurs for the $(s(x) + \bar s(x))/(\bar u(x) + \bar d(x))$ ratio.

The assumption of identical shapes for up, down, and strange quark seas is
not in agreement with some recent experiments. From $W$ and $Z$ boson 
production
in $pp$ collisions at 7 TeV, the ATLAS collaboration at LHC 
determined~\cite{atlas_s} the 
$r_s=(s + \bar s)/2\bar d$ ratio to be 
$1.0 +0.25 -0.28$ at $x=0.023$ and $Q^2 = 1.9$ GeV$^2$,
suggesting a symmetric light quark sea at small $x$, see 
Fig.~\ref{atlas_hermes} (a).
Since the neutrino DIS data showed a suppression of strange relative to
up and down quark seas  by a factor of $\sim 2$ at larger $x$, the 
$(s + \bar s) / (\bar u + \bar d)$ ratio clearly has a 
strong $x$ dependence. It should be cautioned that the ATLAS result
on $r_s$ has a large uncertainty and is higher than the values
of recent PDFs, as shown in Fig.~\ref{atlas_hermes} (a). Much improved
statistics for LHC W,Z data is anticipated. Moreover, W production
events in coincidence with charm jet, which is expected to be a sensitive
probe for $s$ and $\bar s$ distributions, are becoming 
available~\cite{cms_charm,stirling13}.

Another important recent development is that the 
HERMES collaboration has extracted the $x (s(x) + \bar s(x))$
distribution at $Q^2 = 2.4$ GeV$^2$ from semi-inclusive DIS kaon
production on deuteron, shown in Fig.~\ref{atlas_hermes} 
(b)~\cite{hermes08}. The 
dot-dash curve in Fig.~\ref{atlas_hermes} (b) is the 
$x (\bar u(x) + \bar d(x))$ 
distribution from the CTEQ6L~\cite{CTEQ6.6} global fit. 
While $\bar u(x) + \bar d(x)$
is significantly greater than $\bar s(x)$ at large $x$ region ($x > 0.1$),
Fig.~\ref{atlas_hermes} (b) shows that the nucleon sea becomes largely
SU(3) flavor symmetric at $x<0.05$, consistent with the recent ATLAS result.

\begin{figure}[tbp]
\includegraphics[width=0.23\textwidth]{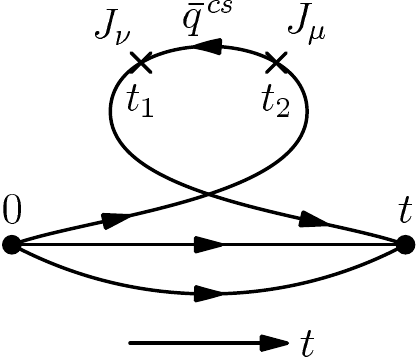}
\includegraphics[width=0.23\textwidth]{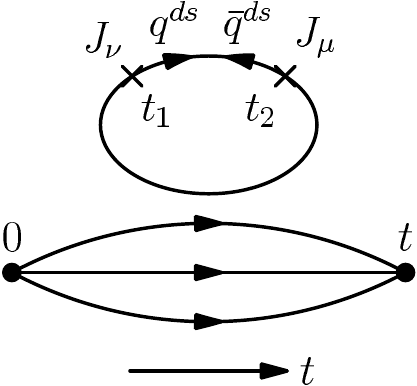}
\includegraphics[width=0.40\textwidth]{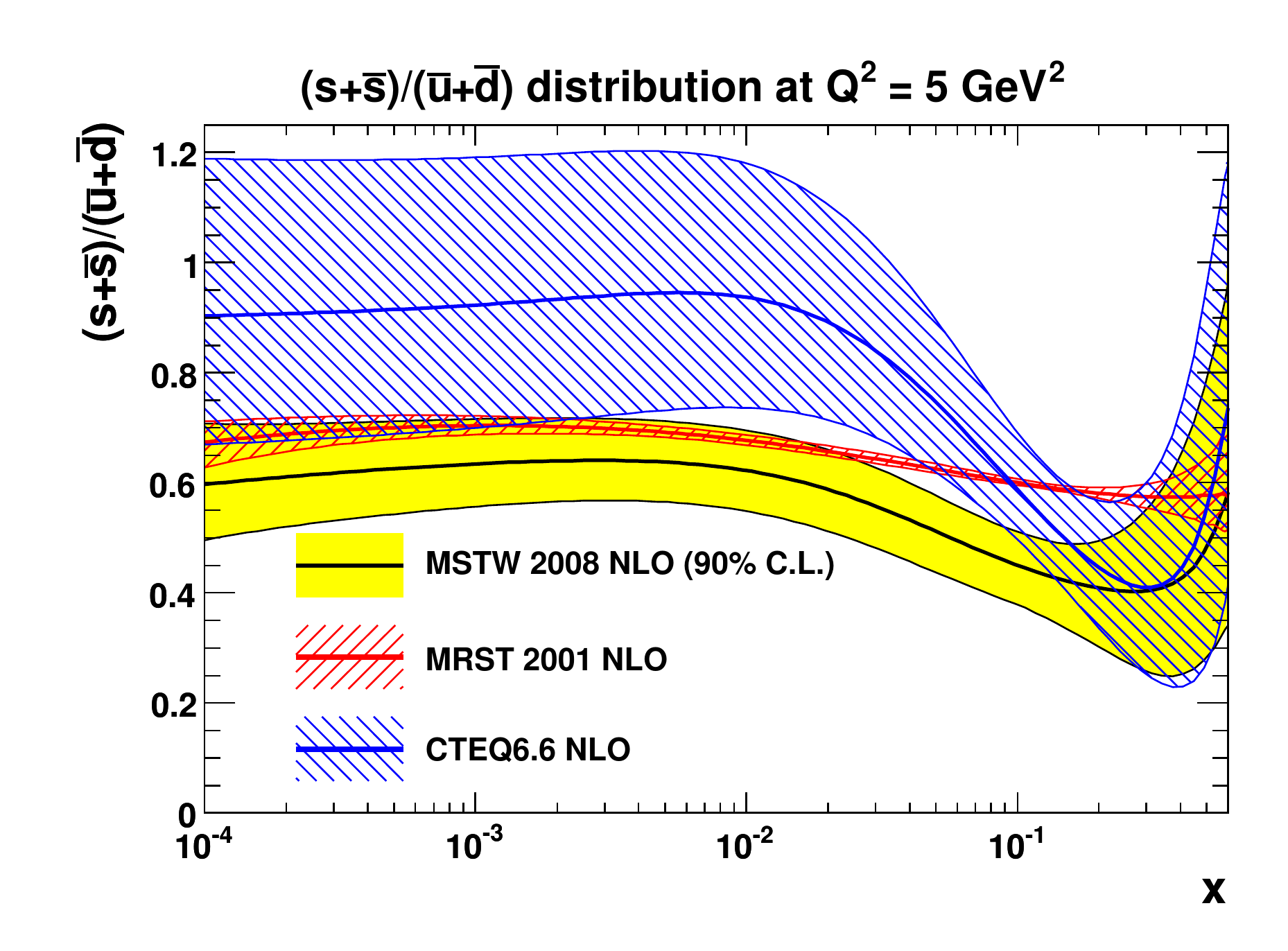}
\caption{Left (a): Two gauge invariant and topologically distinct diagrams
for connected sea and disconnected sea~\cite{kfliu12}.
Right (b): Ratio of $s + \bar s$ over $\bar u + \bar d$ versus
$x$ at $Q^2 = 5$ GeV$^2$ from various recent
PDFs~\cite{MSTW08}.}
\label{fig_liu_pdf}
\end{figure}

Some insight on the $x$-dependence of the 
$(s(x) +\bar s(x)) / (\bar u(x) + \bar d(x))$ ratio 
was recently presented by Liu et al.~\cite{kfliu12}. According to
path-integral formalism, there are two distinct gauge invariant
diagrams for the nucleon sea, shown in Fig.~\ref{fig_liu_pdf} (a).
Antiquarks in the nucleons originate from the connected sea (CS) and the 
disconnected sea (DS), shown as the first and second diagram, respectively.
These two difference sources of sea quarks have distinct quark-flavor
and $x$ dependences~\cite{kfliu12}. 
While $\bar u$ and $\bar d$ have both CS and DS
contributions, $\bar s$ (and $\bar c$) can only come from DS.
The absence of the CS component for the strange quarks implies
that any difference between the $s(x) + \bar s(x)$ and 
$\bar u(x) + \bar d(x)$ distributions
must originate primarily from the CS diagram. The CS and DS are also
expected to have different $x$ distributions. At small $x$, the CS
has a distribution of $\bar q(x)^{cs} \propto x^{-1/2}$ due to
reggeon exchange, while the DS has
$\bar q(x)^{ds} \propto x^{-1}$ as a result of pomeron exchange.
Therefore, the DS component is expected to dominate at small
$x$. Recent lattice QCD calculation~\cite{doi08} shows that DS is nearly flavor
independent, and the ratio $R$ for the momentum fraction carried
by $\bar s$ versus $\bar u$ (or $\bar d)$ is $0.857 \pm 0.040$, consistent
with the results from ATLAS and HERMES. At larger $x$, the $\bar u $
and $\bar d$ sea has the additional contribution from CS, resulting
in smaller $(s(x) + \bar s(x))/(\bar u(x) + \bar d(x))$ ratios. 
This expectation is in 
qualitative agreement with the recent HERMES result.

Figure~\ref{fig_liu_pdf} (b) shows the ratio 
$(s(x) + \bar s(x))/(\bar u(x) +\bar d(x))$ at $Q^2 = 5$ GeV$^2$
from a selection of recent PDF sets. The uncertainty band for MRST2001
is rather narrow, presumably due to the assumption of 
identical $x$-dependence for $s + \bar s$
and $\bar u + \bar d$, namely, 
$s(x) + \bar s(x) = \kappa_s (\bar u(x) + \bar d(x))$ at the initial
$Q_0$ scale. This constraint was relaxed for both CTEQ6.6 and
MSTW2008, although CTEQ6.6 introduces more freedom in the
functional form for $s(x) + \bar s(x)$ than MSTW2008.
While the error bands for CTEQ6.6 and MSTW2008 are quite
broad, they exhibit the same trend that the ratio 
$(s(x) + \bar s(x))/(\bar u(x) +\bar d(x))$ is strongly $x$-dependent,
being nearly constant at $x<0.01$ and falls with increasing $x$ in the region
$0.01<x< 0.3$. Both PDFs have not included the ATLAS $W, Z$ production
and the HERMES data. It is anticipated that these data as well
as new $W,Z$ production data from LHC will significantly improve our knowledge
on the flavor structure of the $u, d, s$ quark sea.

\subsubsection{Asymmetry between $s$ and $\bar s$ seas}

The absence of net valence strange quarks in the nucleon implies
$\int_0^1(s(x)-\bar s(x)) dx = 0$. However, no known physics requires
$s(x) = \bar s(x)$. Indeed, both perturbative and
nonperturbative processes can lead to asymmetric $s(x)$, $\bar s(x)$
distributions in the nucleon. For the nonperturbative contribution,
the role of meson cloud for generating $s(x) \ne \bar s(x)$ was first
considered by Signal and Thomas~\cite{signal87}. In particular,
the $K\Lambda$ and $K\Sigma$ kaon-hyperon components naturally lead to an
asymmetric $s(x)$ and $\bar s(x)$, since $\bar s$ residing in a kaon
is expected to have a different momentum distribution compared with
$s$ contained in a hyperon. This asymmetry is a consequence of the
kaon cloud very much analogous to the $\bar d / \bar u$ asymmetry
caused by the pion cloud discussed earlier. Burkardt 
and Warr~\cite{burkardt92} later showed a sizeable $s(x)$, $\bar s(x)$
asymmetry resulting from calculation using the chiral Gross-Neveu 
model. Brodsky and Ma~\cite{brodsky96} then described the kaon-hyperon
fluctuation utilizing the light-cone two-body wave functions. The magnitude
and shape of $s(x) - \bar s(x)$ depend sensitively on the choice of the
kaon-hyperon splitting function as well as the $s$ and $\bar s$ distribution
functions in the hyperons and kaon. In general, the meson-cloud model
predicts $s(x) < \bar s(x)$ at the $x>0.2$ 
region~\cite{holtmann96,wally97,christiansen98,cao99}, while the light-cone
model predicts a softer $\bar s(x)$ distribution with $s(x) > \bar s(x)$
for $x>0.2$. The inclusion of the $K^* \Lambda$ channel was shown
to affect the $s(x) - \bar s(x)$ distribution significantly~\cite{cao03}.
The $s(x) - \bar s(x)$ asymmetry is also predicted in the chiral quark
model in which constituent quarks can couple to kaons~\cite{szczurek96,ding05},
as well as in the chiral-quark soliton model~\cite{wakamatsu05}. Both the
chiral-quark and the soliton models predict $s(x) > \bar s(x)$ at
$x>0.2$.

It was pointed out that perturbative QCD can also lead to an $s(x)/\bar s(x)$
asymmetry~\cite{catani04}. Up to next-to-leading order (NLO) in $\alpha_s$,
QCD evolution generates symmetric $s(x)$, $\bar s(x)$ sea. However, at
the level of the NNLO evolution, the probability 
for $u \to s (d \to s)$ splitting
is different from the $u \to \bar s (d \to \bar s)$ splittings. Therefore,
an $s(x) / \bar s(x)$ asymmetry can result from NNLO evolution even with
$s(x) = \bar s(x)$ at the initial scale $Q_0$~\cite{catani04}. 
This is analogous to the familiar case of $\bar d(x) / \bar u(x)$ 
flavor asymmetry.
While LO evolution preserves a flavor 
symmetric $\bar d(x) / \bar u(x)$ sea, the NLO
evolution generates $\bar d(x) / \bar u(x)$ flavor 
asymmetry due to the difference 
in the $u \to \bar u$ and $d \to \bar d$ splitting~\cite{ross79,broadhurst04}.  
Using the 3-loop splitting functions derived by Moch et al.~\cite{moch04},
non-zero value of $s(x) - \bar s(x)$  
has been obtained~\cite{catani04,feng12}. While the 
nonperturbative contributions to $s(x) - \bar s(x)$ dominates in the
larger $x$ region, the perturbative contributions are more significant
in the small $x$ ($0.02 < x < 0.03$) region~\cite{feng12}.

The possible role of $s(x) / \bar s(x)$ asymmetry in explaining, at
least partially, the so-called ``NuTeV anomaly" has attracted much interest.
From a comparison of the neutral and charged current $\nu$ and $\bar \nu$
cross sections on an iron target, the NuTeV collaboration reported
a $3\sigma$ deviation from the Standard Model value 
of $\sin^2 \theta_W$~\cite{zeller02}. It was later pointed out that this
discrepancy could be reduced if $s$ quarks carry more momentum than the
$\bar s$ quarks~\cite{davidson02}, namely, 
\begin{equation}
[S^-] = \int^1_0 x [s(x)-\bar s(x)]dx > 0.
\end{equation}
\noindent In particular, Kretzer et al.~\cite{kretzer04} estimated that
a value of $[S^-] = 0.004$ at $Q^2=20$ GeV$^2$ would shift 
$\sin^2\theta_W$ by $\sim -0.005$ and
completely remove the NuTeV anomaly. The exact value would clearly depend
on the $x$-dependence of $s(x) - \bar s(x)$. It is worth noting
that the $\bar d(x) / \bar u(x)$ flavor asymmetry corresponds to 
$\int^1_0 x[\bar d(x) - \bar u(x)]dx \approx 0.007$ at 
$Q^2 = 20$ GeV$^2$~\cite{garvey02}. Therefore, a value of $[S^-] = 0.004$ 
would represent a sizable asymmetry between $s(x)$ and $\bar s(x)$.
Since most of the theoretical
models predict $s(x) > \bar s(x)$ at $x>0.2$, resulting in $[S^-] > 0$, 
this could partially account for the NuTeV anomaly. 
Clearly, the $s(x) - \bar s(x)$ distribution and the value of $[S^-]$
must be determined from experiments.

The first indication of an $s(x) - \bar s(x)$ asymmetry came from a global
analysis~\cite{barone00} of charged lepton and $\nu(\bar \nu)$ DIS, 
and the Drell-Yan
cross section data, suggesting a harder $s(x)$ 
distribution than $\bar s(x)$ with $[S^-] = 0.0020 \pm 0.0005$ at 
$Q^2 = 20$ GeV$^2$. However, when the CCFR-NuTeV 
$\nu (\bar \nu)$ DIS data~\cite{yang01} 
were added to the global 
fit, $[S^-] = (1.8 \pm 3.8) \times 10^{-4}$ was obtained~\cite{barone06},
which is consistent with no $s(x) / \bar s(x)$ asymmetry. 

The $s(x)$ and $\bar s(x)$ seas can be directly accessed through the
measurement of dimuon events in neutrino and antineutrino DIS. For the
$\nu_\mu N \to \mu^- \mu^+ X$ reaction, a significant contribution
comes from the charged-current subprocess
$\nu_\mu + s \to \mu^- + c$ followed by the semi-leptonic decay of the
charmed hadron. These dimuon events provide a sensitive probe for the
$s$-quark sea in the nucleon. Similarly, the
$\bar \nu_\mu N \to \mu^+ \mu^- X$ events would measure the 
$\bar s$-quark sea via the $\bar \nu_\mu + \bar s \to \mu^+ + \bar c$
subprocess. The availability of high statistics NuTeV dimuon
data~\cite{mason07} prompted several global fits~\cite{MSTW08,mason07,nnpdf1.2,
cteq6.5,akp08,pedro10} at NLO allowing
for a non-zero $s(x) - \bar s(x)$ at the initial $Q_0$ scale.
In Table~\ref{tab:2} are listed the values of $[S^-]$ 
obtained from these PDFs. While all PDFs give a positive
central value for $[S^-]$, the rather large uncertainties
prevent a definitive conclusion on the existence of
$s(x) / \bar s(x)$ asymmetry from these global fits. As a result,
some recent PDFs (CTEQ6.6 and CT10, for example) continue to
assume $s(x) = \bar s(x)$ at the initial scale. 
In an assessment of recent status, Bentz et al.~\cite{bentz10}
also obtained a vlaue of $[S^-] = 0.0 \pm 0.002$, consistent with
no strangeness asymmetry.
\begin{table}
\caption{Values of the integral
$[S^-]=\int_0^1 x[s(x) - \bar s(x)] dx$ determined from various
NLO global fits. The values of $Q^2$ are also listed.}
\label{tab:2}       
\begin{center}
\begin{tabular}{lll}
\hline\noalign{\smallskip}
PDF & $Q^2$ (GeV$^2$) & $[S^-]$ \\
\noalign{\smallskip}\hline\noalign{\smallskip}
NuTeV~\cite{mason07} & 16.0 & $0.00196^{+0.0016}_{-0.0013}$ \\
CTEQ6.5~\cite{cteq6.5} & 1.69 & $0.0018^{+0.0016}_{-0.0014}$ \\
AKP08~\cite{akp08} & 20.0 & $0.0013\pm 0.0009$ \\
GJR08~\cite{pedro10} & 20.0 & $0.0008 \pm 0.0005$ \\
MSTW08~\cite{MSTW08} & 10.0 & $0.0018^{+0.0018}_{-0.0012}$ \\
NNPDF1.2~\cite{nnpdf1.2} & 20.0 & $0.0005 \pm 0.0086$ \\
\noalign{\smallskip}\hline
\end{tabular}
\end{center}
\end{table}

\begin{figure}[h]
\includegraphics[width=0.5\textwidth]{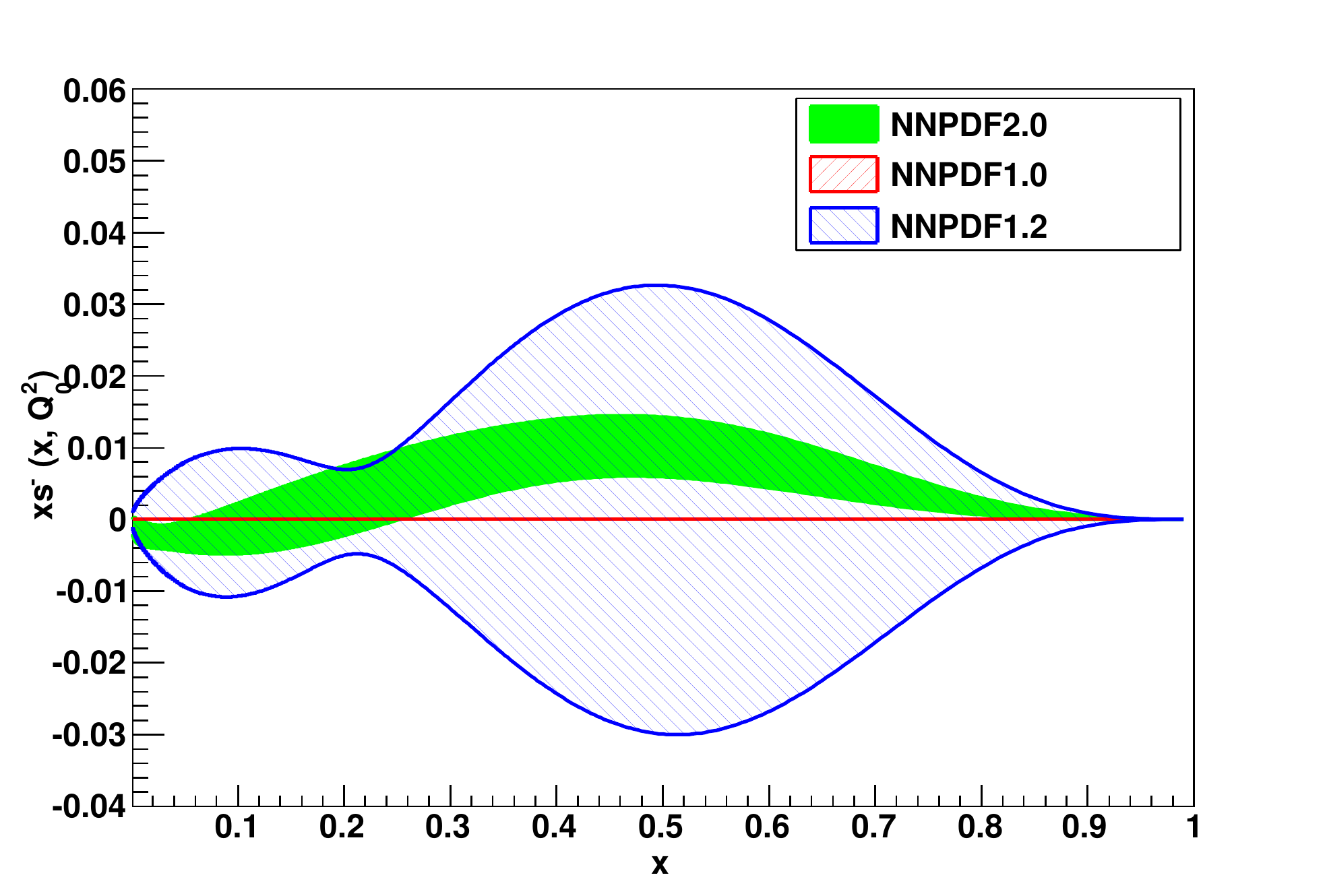}
\includegraphics[width=0.5\textwidth]{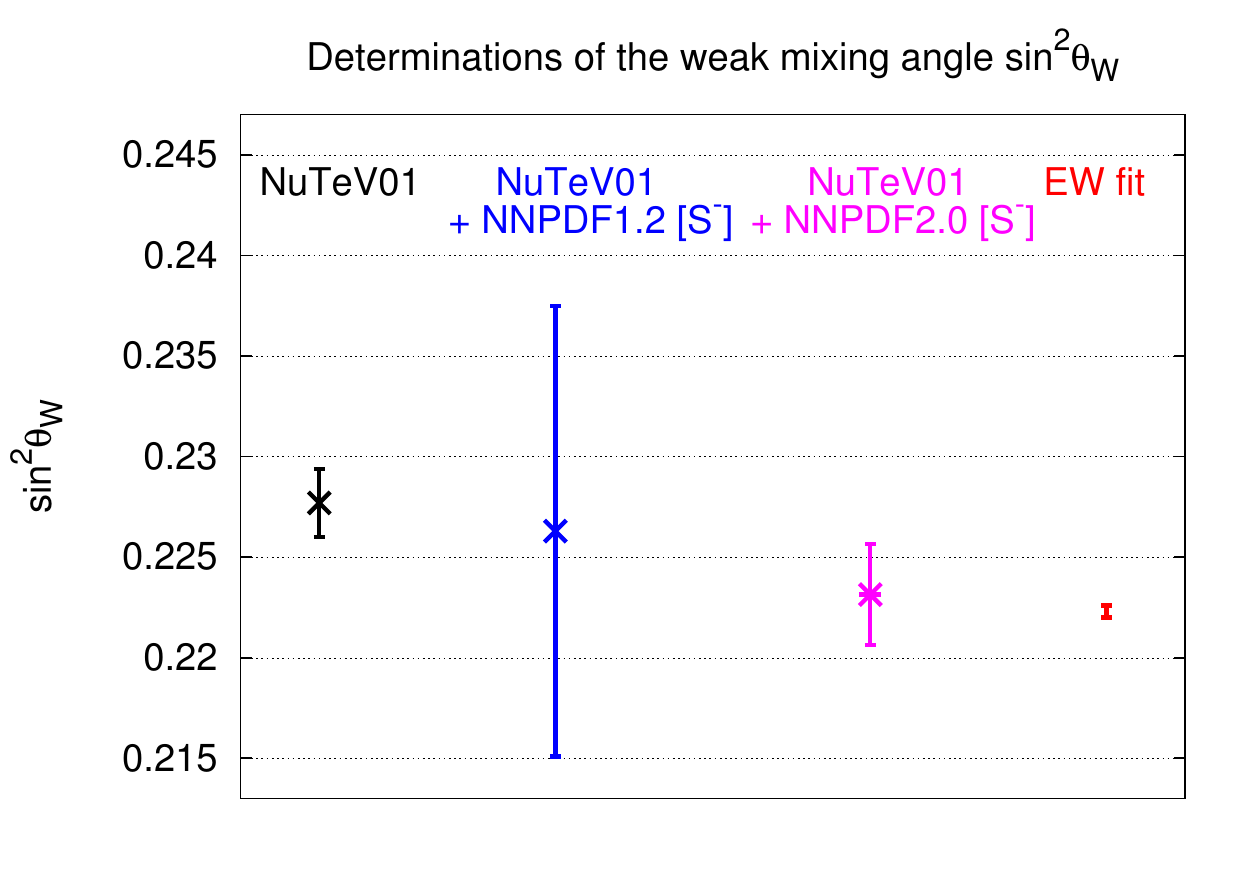}
\caption{Left (a): $x [s(x)-\bar s(x)]$ from three NNPDF sets~\cite{NNPDF2.0}.
NNPDF1.0 assumes $s(x)=\bar s(x)$, while NNPDF1.2 and NNPDF2.0
allow $s(x) \ne \bar s(x)$. The NNPDF2.0 analysis includes fixed-target
Drell-Yan and Tevatron $W$ and $Z$ production data.
Right (b): Weak mixing angle determined from the NuTeV data~\cite{zeller02},
plus correction using $[S^-]$ determined from 
NNPDF1.2~\cite{nnpdf1.2}, or from NNPDF2.0~\cite{NNPDF2.0}. 
Only statistical uncertainty is included in the NNPDF2.0 result.
The Weinberg
angle from global electroweak fit is also shown. Figures are 
from~~\cite{NNPDF2.0}.}
\label{fig_nnpdf_ssbar}
\end{figure}

The NNPDF collaboration recently reported a more
precise determination of $s(x) - \bar s(x)$ and the
value of $[S^-]$ in the NNPDF2.1 PDF set~\cite{NNPDF2.0}.
After including the fixed-target Drell-Yan and the
Tevatron $W$ and $Z$ production data in their global fits, they
found that the uncertainty of $s(x) - \bar s(x)$ is much
reduced, as shown in Fig.~\ref{fig_nnpdf_ssbar} (a).
It is remarkable that the Drell-Yan data, which are 
not sensitive to $s$ and $\bar s$ distributions per se, can 
lead to tight constraints on the $s(x) - \bar s(x)$ 
distributions. This is attributed to the fact 
that $\bar d(x)$, $\bar u(x)$ are precisely determined
by the Drell-Yan data, allowing a more accurate flavor
decomposition of the nucleon sea. 
However, it is
curious why similar results were not obtained by other
PDFs listed in Table~\ref{tab:2}, which also include
Drell-Yan data in their global fits.
Figure ~\ref{fig_nnpdf_ssbar} (b)
shows that the NuTeV anomaly practically disappears when the
$[S^-]$ determined from NNPDF2.0 is adopted. 

Future experiments such as semi-inclusive DIS with $K^+$ and
$K^-$ production at the JLab 12 GeV upgrade and at the EIC,
as well as kaon-induced Drell-Yan at COMPASS and J-PARC, could
provide new information on the intriguing and yet unsettled
issue of possible flavor
asymmetry between the $s$ and $\bar s$ seas.

\section{Drell-Yan process with $Q^2\gg q_T^2$}
\label{sec:two-scales}

The Drell-Yan massive lepton-pair production for $Q\gg q_T\sim \Lambda_{\rm QCD}$ 
has attracted tremendous attention from both experimental and theoretical sides in recent years.
It might be one of the best processes to measure transverse motion of quarks inside a hadron
because of its sensitivity to the quark's confined motion at the scale of $\Lambda_{\rm QCD} \sim 1/$fm.
It is also the much needed process to test the predicted sign change of two most interested
TMDs, known as Sivers function and Boer-Mulders function, when they are measured 
in SIDIS in comparison with that measured in the Drell-Yan process.

\subsection{QCD TMD factorization for Drell-Yan process}
\label{sec:tmd-fac}

With the transverse momentum of the massive lepton pair, $q_T\ll Q$, Drell-Yan cross section
has effectively two observed, but, very different momentum scales.  The perturbatively calculated
$\hat{\sigma}$ in Eq.~(\ref{eq:qcd-dy-lp}) will have $\alpha_s \ln^2(Q^2/q_T^2)$-type 
huge logarithmic contributions from every power of $\alpha_s$ in its perturbative expansion,
which could clearly ruin the convergence of the perturbative expansion in powers of $\alpha_s$.
Furthermore, when $q_T\sim \Lambda_{\rm QCD}$, the approximation to neglect the transverse
momentum of active partons in QCD collinear factorization approach, the approximation in
Eq.~(\ref{eq:collinear}), is clearly not valid, and the transverse momentum of active partons
could directly influence the $q_T$ of produced lepton pair.  
The QCD improved Drell-Yan formalism in Eq.~(\ref{eq:qcd-dy-lp}) should not be valid for 
this kinematic regime of Drell-Yan process.

QCD TMD factorization generalizes the short-distance production of massive lepton pairs by
the collision of two on-shell collinear partons in Eq.~(\ref{eq:qcd-dy-lp}) to the collision of 
two on-shell partons with both collinear as well as transverse momentum components, 
and the PDFs to transverse momentum dependent PDFs or TMDs
\cite{collins-book,collins-qiu-fac,collins-metz-fac},
\begin{eqnarray}
\frac{d\sigma_{A+B\to l\bar{l}+X}(\vec{S}_a,\vec{S}_b)}
       {dQ^2dy\, d^2{\bf q_T}}
&=&
\sum_{a,b} \int dx_a\, dx_b\, d^2{\bf p}_{a\perp}\, d^2{\bf p}_{b\perp}\,
\delta^2({\bf q}_T-{\bf p}_{a\perp}-{\bf p}_{b\perp})
\nonumber\\
&\ & \times
f^{\rm DY}_{a/A}(x_a,{\bf p}_{a\perp},\vec{S}_a,\mu)\,
f^{\rm DY}_{b/B}(x_b,{\bf p}_{b\perp},\vec{S}_b,\mu)\, 
\frac{d\hat{\sigma}_{a+b\to l\bar{l}}(x_a,x_b,Q,y,\mu)}{dQ^2dy}
\nonumber\\
&\ & + \ Y(q_T,Q,y) + {\cal O}\left((\Lambda_{\rm QCD}/Q)^\alpha\right),
\label{eq:dy-tmd} 
\end{eqnarray}
where ${\vec S}_a$ (and ${\vec S}_b$) is the spin vector of colliding hadron $A$ (and $B$), 
$y$ is the rapidity of the observed lepton pair, 
and $Y(q_T,Q,y)$ is a perturbative contribution extending the TMD-factorization 
formalism to the region where $q_T\sim Q$ \cite{Collins:1984kg}.  
The partonic hard part $\hat{\sigma}_{a+b\to l\bar{l}}(x_a,x_b,Q,y,\mu)$ 
depends on the type of intermediate vector boson ($\gamma$, or $W$, or $Z$) 
\cite{Collins:1984kg}.
In Eq.~(\ref{eq:dy-tmd}), $f^{\rm DY}_{a/A}(x_a,{\bf p}_{a\perp},\vec{S}_a,\mu)$
and $f^{\rm DY}_{b/B}(x_b,{\bf p}_{b\perp},\vec{S}_b,\mu)$ 
are TMDs 
with factorization scale $\mu$ defined in Ref.~\cite{collins-book}, 
also in Appendix~\ref{sec:appendixb}.
The TMD factorization formalism in Eq.~(\ref{eq:dy-tmd}) was proved to the same rigor as 
the collinear factorization formalism in Eq.~(\ref{eq:qcd-dy-lp}) with corrections down by powers of 
$1/Q$ \cite{collins-book,Ji:2004xq}, and the technical details of the proof 
is well documented 
in Ref.~\cite{collins-book}.  Some simple and intuitive arguments for the validity of Eq.~(\ref{eq:dy-tmd}) 
are given in Appendix~\ref{sec:appendixb}.  The same TMD factorization formalism was also
derived by using Soft-Collinear Effective Theory of QCD \cite{GarciaEchevarria:2011rb}.
Although it seemingly used a different approach to what was introduced 
by Collins in Ref.~\cite{collins-book}, recently, Collins and Rogers showed \cite{Collins:2012uy}
that these two approaches are effectively the same, apart from an apparent difference in 
their choices of normalization schemes.

With the dependence on parton transverse momentum, there are more TMDs than corresponding
collinear PDFs \cite{Accardi:2012hwp,tmd-duke}.  For example, there are a total of eight 
leading power quark TMDs while there are only three leading power collinear quark PDFs.  
TMDs provide direct information not only on parton's confined transverse motion inside a fast 
moving hadron, but also on quantum correlation between the preference of parton's transverse motion 
and the spin direction of a transversely polarized hadron.   TMDs extracted from experimental data
could provide the picture of 3D motion of quarks and gluons inside a hadron.  
The validity of TMD factorization in Eq.~(\ref{eq:dy-tmd}) is necessary to bridge 
the confined partonic motion inside a hadron to 
the measurements of production cross section of massive lepton pairs and 
the pairs' momentum distribution.

However, unlike PDFs, TMDs are not necessarily universal and could be process dependent.  
Such process dependence could ruin the predictive power of the TMD factorization formalism
in Eq.~(\ref{eq:dy-tmd}).  Fortunately, due to the parity and time-reversal invariance of QCD, 
the process dependence of TMDs is only up to a sign change when they are measured 
in different physical processes.  For example, as shown in Appendix~\ref{sec:appendixb},
the parity and time-reversal invariance of QCD relates the TMD for finding an unpolarized 
quark inside a transversely polarized hadron measured in SIDIS and Drell-Yan process as
follows \cite{collins02,Kang:2009bp},
\begin{equation}
f_{q/h^\uparrow}^{\rm SIDIS}
(x,\mathbf{p}_\perp,\vec{S})
=
f_{q/h^\uparrow}^{\rm DY}
(x,\mathbf{p}_\perp,-\vec{S})\, .
\label{eq:pt-inv}
\end{equation}
The Sivers function is defined to be proportional to the difference, 
$[f_{q/h^\uparrow}(x,\mathbf{p}_\perp,\vec{S})-f_{q/h^\uparrow}(x,\mathbf{p}_\perp,-\vec{S})]/2$.
Eq.~(\ref{eq:pt-inv}) effectively requires the Sivers function measured in SIDIS to 
have an opposite sign from that measured in Drell-Yan process 
\cite{collins02,collins-metz-fac,Kang:2009bp}.
This sign change is the prediction of TMD factorization approach, and 
is one of the most important tests of QCD dynamics and 
factorization approaches to hadronic cross sections.  

Experiments to test this sign change have been proposed to several hadron facilities 
around the world, including COMPASS \cite{COMPASSdy} at CERN, 
RHIC \cite{RHICspin} at BNL, PAX \cite{PAX} and PANDA \cite{PANDA} at GSI, 
NICA \cite{NICA-JINR} at JINR in Dubna, 
SPASCHARM \cite{SPASCHARM} at IHEP in Protvino (Russia), 
J-PARC \cite{J-PARC}, and E1027 \cite{FNALspin} at Fermilab.

The Boer-Mulders function, $h_{1q}^\perp(x)$, is defined to be the sum,
$[f_{h_{1q}/h^\uparrow}(x,\mathbf{p}_\perp,\vec{S})
+f_{h_{1q}/h^\uparrow}(x,\mathbf{p}_\perp,-\vec{S})]/2$, where
$f_{h_{1q}/h^\uparrow}(x,\mathbf{p}_\perp,\vec{S})$ is the TMD with 
a tensor spin projection, defined by replacing the vector spin projection 
$\gamma^+$ of the TMD, $f_{q/h^\uparrow}(x,\mathbf{p}_\perp,\vec{S})$ 
in Eq.~(\ref{eq:pt-inv}) by a spin projection proportional to $\sigma^{+\perp}$.
That is, the Boer-Mulders function, $h_{1q}^\perp(x)$, is proportional to the 
spin-averaged part of the TMD, $f_{h_{1q}/h^\uparrow}(x,\mathbf{p}_\perp,\vec{S})$.
Similar to Eq.~(\ref{eq:pt-inv}), the parity and time-reversal invariance of 
QCD requires
\begin{equation}
f_{h_{1q}/h^\uparrow}^{\rm SIDIS}
(x,\mathbf{p}_\perp,\vec{S})
=
-f_{h_{1q}/h^\uparrow}^{\rm DY}
(x,\mathbf{p}_\perp,-\vec{S})\, ,
\label{eq:pt-inv_tensor}
\end{equation}
which effectively requires that the Boer-Mulders function measured in Drell-Yan process 
to have a sign opposite to that measured in SIDIS.

Like PDFs, TMDs also depend on the scale where they are probed.  
The scale dependence (or often referred to as scaling violation in QCD) 
is a prediction of QCD dynamics.  
The scale dependence of PDFs is determined by the DGLAP evolution equations
\cite{Dokshitzer:1977sg,Gribov:1972ri,Altarelli:1977zs}, which effectively resums 
leading logarithmic scale dependence and has been successfully tested 
\cite{CTEQ6.6,CT10,MSTW08,NNPDF2.0}.  
The scale dependence of TMDs is systematically determined in QCD factorization 
framework by the often-called Collins-Soper equation \cite{collins-soper-b2bjets},
which effectively resum the double logarithmic contributions from gluon shower
of active colliding partons.  When applied to unpolarized Drell-Yan process, 
the resummation of large $\alpha_s\ln^2(Q^2/q_T^2)$-type double logarithms
are taken care of by the so-called CSS formalism, due to the work of 
Collins, Soper and Sterman 
(CSS)~\cite{Collins:1984kg,collins-soper-b2bjets,collins-soper-pdfs}.
Because of small parton transverse momentum, which is often a non-perturbative 
scale, the evolution of TMDs is best described in the impact parameter $b$-space,
a Fourier transform of parton's transverse momentum 
\cite{PPbspace,Collins:1984kg,APPdy}.  
The TMD factorization formalism in Eq.~(\ref{eq:dy-tmd}) can be written 
in $b_T$-space as \cite{Collins:1984kg},
\begin{equation}
\frac{d\sigma_{A+B\to l\bar{l}+X}}{dQ^2\, dy\, dq_T^2} =
\frac{1}{(2\pi)^2}\int d^2b\, e^{i\vec{q}_T\cdot \vec{b}}\,
\widetilde{W}(b,Q,x_A,x_B) + Y(q_T,Q,x_A,x_B)\, ,
\label{eq:css-gen}
\end{equation}
where $x_A$ and $x_B$ are defined in Eq.~(\ref{eq:kinematics-pm}), and 
$\widetilde{W}$ is proportional to the product of two TMDs in $b$-space when 
$b$ is small and $1/b$ is a perturbative scale \cite{Collins:1984kg}, 
\begin{equation}
\widetilde{W}^{\rm pert}_{ij}(b,Q,x_A,x_B) = 
{\rm e}^{-S(b,Q)}\, 
f_{i/A}(x_A,\mu=\frac{c}{b})\, f_{j/B}(x_B,\mu=\frac{c}{b})\, ,
\label{eq:css-W-pert}
\end{equation}
where $S(b,Q)$ is the Sudakov form factor resumming all leading and next-to-leading power of 
logarithmic contributions to the Drell-Yan cross 
section \cite{Collins:1984kg,collins-soper-b2bjets}, 
and the functions $f_{i/A}$ and $f_{j/B}$ in Eq.~(\ref{eq:css-W-pert})
are the modified PDFs \cite{Collins:1984kg}, 
\begin{equation}
f_{i/A}(x_A,\mu) = \sum_a 
  \int_{x_A}^1\frac{d\xi}{\xi}\, 
  C_{i/a}(\frac{x_A}{\xi},\mu))\, \phi_{a/A}(\xi,\mu)
\label{eq:mod-pdf}
\end{equation}
where $\sum_a$ runs over all parton flavors.  In Eq.~(\ref{eq:mod-pdf}),
$\phi_{a/A}(\xi,\mu)$ is the normal PDFs of flavor $a$
of hadron $A$, and $C_{i/a}=\sum_{n=0} C_{i/a}^{(n)} (\alpha_s/\pi)^n$
are perturbatively calculable coefficient functions for finding a
parton of flavor $i$ from a parton of flavor $a$ \cite{Collins:1984kg}.

The predictive power of this $b$-space TMD-factorization formalism in Eq.~(\ref{eq:css-gen})
is very sensitive to how we extrapolate the perturbatively calculated 
$\widetilde{W}^{\rm pert}_{ij}(b,Q,x_A,x_B)$ in Eq.~(\ref{eq:css-W-pert}),
or in general, the $b$-space TMDs, which are valid at small $b$, into
the large $b$ region so that we can perform the Fourier transform from $b$ to $q_T$
in Eq.~(\ref{eq:css-gen}) \cite{catani2000,Qiu:2000hf}.  
With a proper prescription to extrapolate into the 
large $b$ non-perturbative region, the CSS formalism for the evolution of TMDs 
has been extremely successful in interpreting and predicting the unpolarized Drell-Yan and W/Z 
production \cite{Qiu:2000hf,Landry:1999an,Qiu:2000ga,Landry:2002ix,Kang:2012am}, 
as well as $\Upsilon$ \cite{Berger:2004cc} production, at collider energies. 

Recently, the CSS factorization formalism for TMDs was extended to processes 
that involve the Sivers functions \cite{Aybat:2011ge,boer2001,Idilbi:2004vb}.  
Since QCD evolution is insensitive to 
the details of long-distance hadron states, Collins-Soper evolution equations should be
valid for TMDs of hadrons with and without polarization.  Sivers function is defined 
to be the spin-dependent part of the TMD to find a unpolarized quark inside 
a transversely polarized hadron, so that it is the transverse momentum derivative of 
Sivers function that satisfies the same Collins-Soper evolution equations 
\cite{Ji:2004xq,Aybat:2011ge,boer2001,Idilbi:2004vb}.

\subsection{Single transverse-spin asymmetry of the Drell-Yan process}
\label{eq:ssa-dy}

Single transverse-spin asymmetry (SSA) in hadronic collisions is defined to 
be the ratio of the difference and the sum of the cross sections when the 
transverse spin vector $\vec{S}_\perp$ of one of the hadrons is flipped,
\begin{equation}
A_N \equiv \frac{\sigma(\vec{S}_\perp)-\sigma(-\vec{S}_\perp)}
{\sigma(\vec{S}_\perp)+\sigma(-\vec{S}_\perp)}\, .
\label{eq:an}
\end{equation} 
Large SSAs, as large as 30-40 percent, have been consistently observed 
in various experiments involving one polarized hadron 
at different collision energies \cite{ssa_review}, and 
presented a challenge to the leading power collinear QCD factorization formalism 
\cite{Kane:1978nd}.

Tremendous effort has been devoted to understanding the physics behind the 
measured large SSAs in both theory and experiment.  
Two widely discussed theoretical approaches 
have been proposed to evaluate the observed SSAs in QCD.
For observables with a single large momentum scale, such as 
hadronic pion production at large $p_T$,
one attributes the SSA to the quantum interference of scattering amplitudes 
with different numbers of active partons, which is a result of QCD 
collinear factorization extended to the first subleading power 
\cite{Efremov,Qiu:1991pp,Qiu:1991wg}. 
The size of the asymmetry is determined by new kind of 
three-parton correlation functions \cite{Kang:2008ey}. 
For observables with two very different momentum scales,
such as Drell-Yan lepton-pair production when $q_T\ll Q$, 
one factorizes $\sigma(\vec{S}_\perp)$ in terms of the TMDs
as in Eq.~(\ref{eq:dy-tmd}), and attributes the SSAs to the nonvanishing Sivers function
\cite{sivers90}.  
There is one crucial difference between these two approaches 
besides the difference in kinematic regimes where they apply. 
The Sivers function in TMD factorization approach can be process dependent, 
while in the collinear factorization approach, all distribution functions are universal
and the process dependence is taken care of by perturbatively 
calculated hard parts. 
As demonstrated in Sec.~\ref{sec:tmd-fac}, the Sivers function extracted from 
SSAs of Drell-Yan process should have an opposite sign from that extracted from
SIDIS \cite{collins02}.

\begin{figure}[h]
\begin{center}
\includegraphics[width=0.4\textwidth,angle=-90]{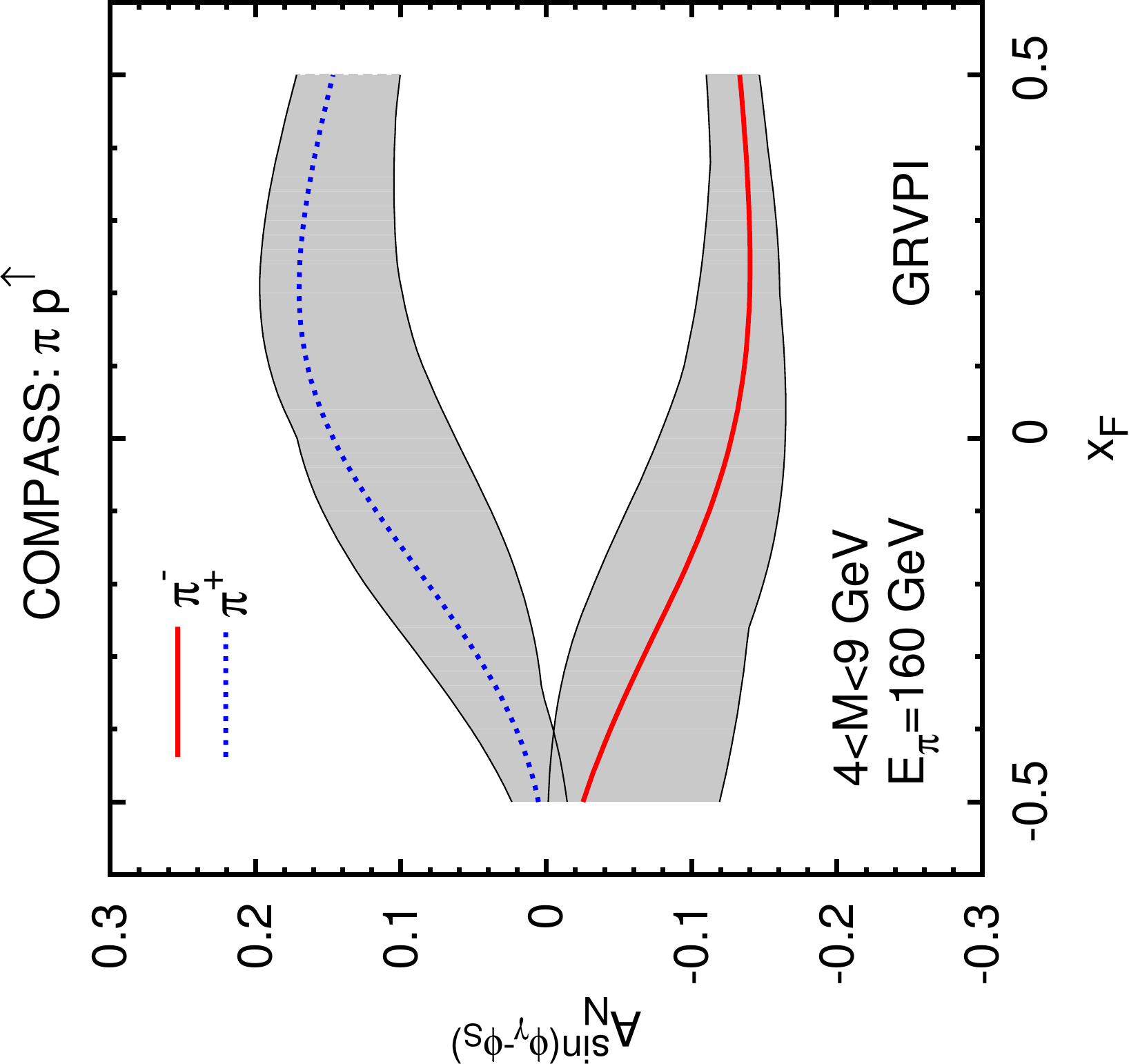}
\hskip 0.1\textwidth
\includegraphics[width=0.4\textwidth,angle=-90]{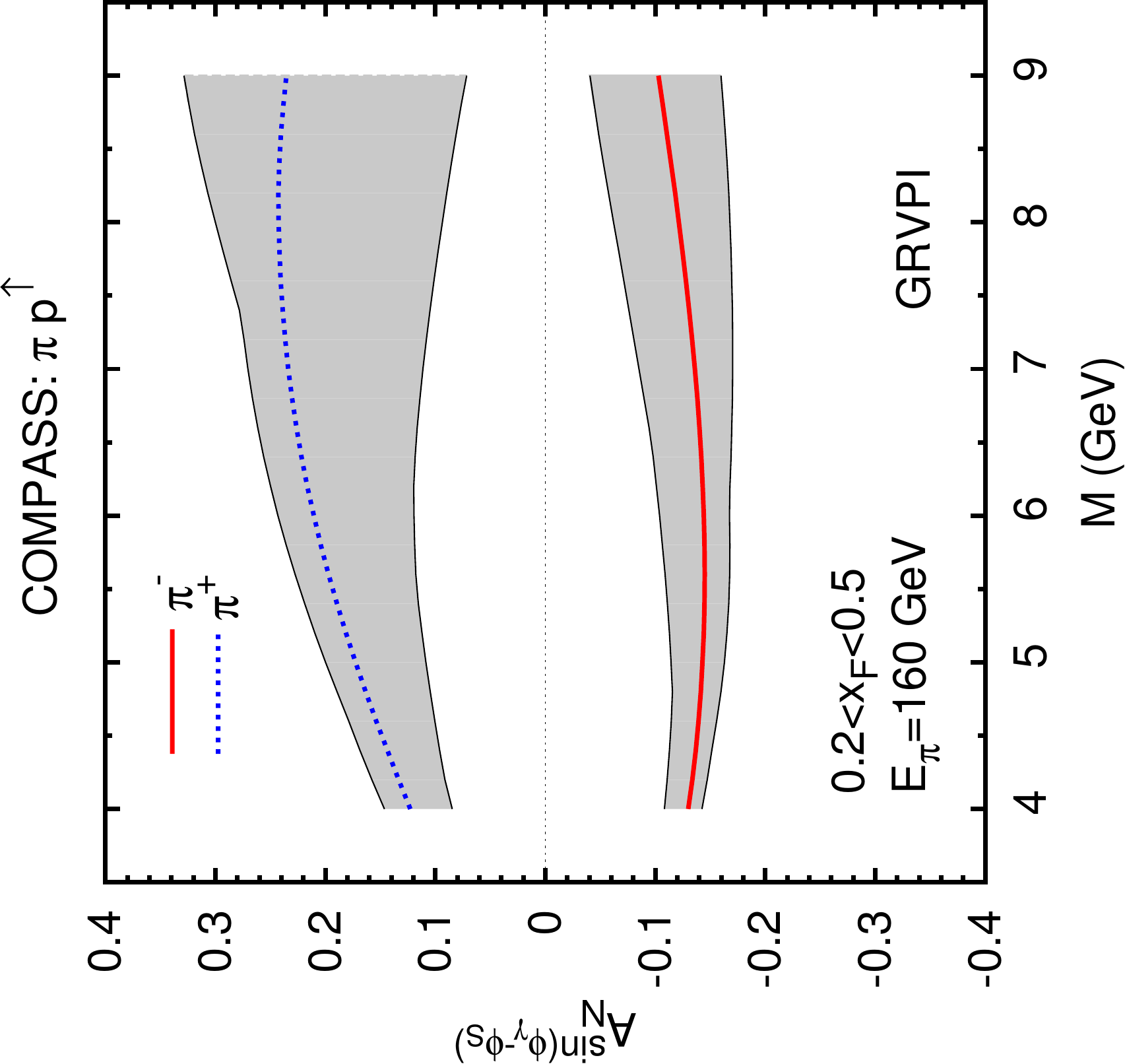}
\caption{\label{fig:an_dy_compass}
Predictions for SSAs of Drell-Yan process: $\pi^\pm+(p\uparrow) \to \mu^+\mu^-(M)+X$ 
at COMPASS as a function of $x_F=x_1-x_2$ (left) and invariant mass $M$ (right) of the lepton pair
with a pion beam energy of 160 GeV ($\sim \sqrt{S}=17.4$~GeV), with phase space
integration:  $0\leq q_T\leq 1$~GeV, $4< M < 9$~GeV, and $0.2 < x_F < 0.5$ 
\cite{Anselmino:2009st}.
}
\end{center}
\end{figure}

\begin{figure}[h]
\begin{center}
\includegraphics[width=0.4\textwidth,angle=-90]{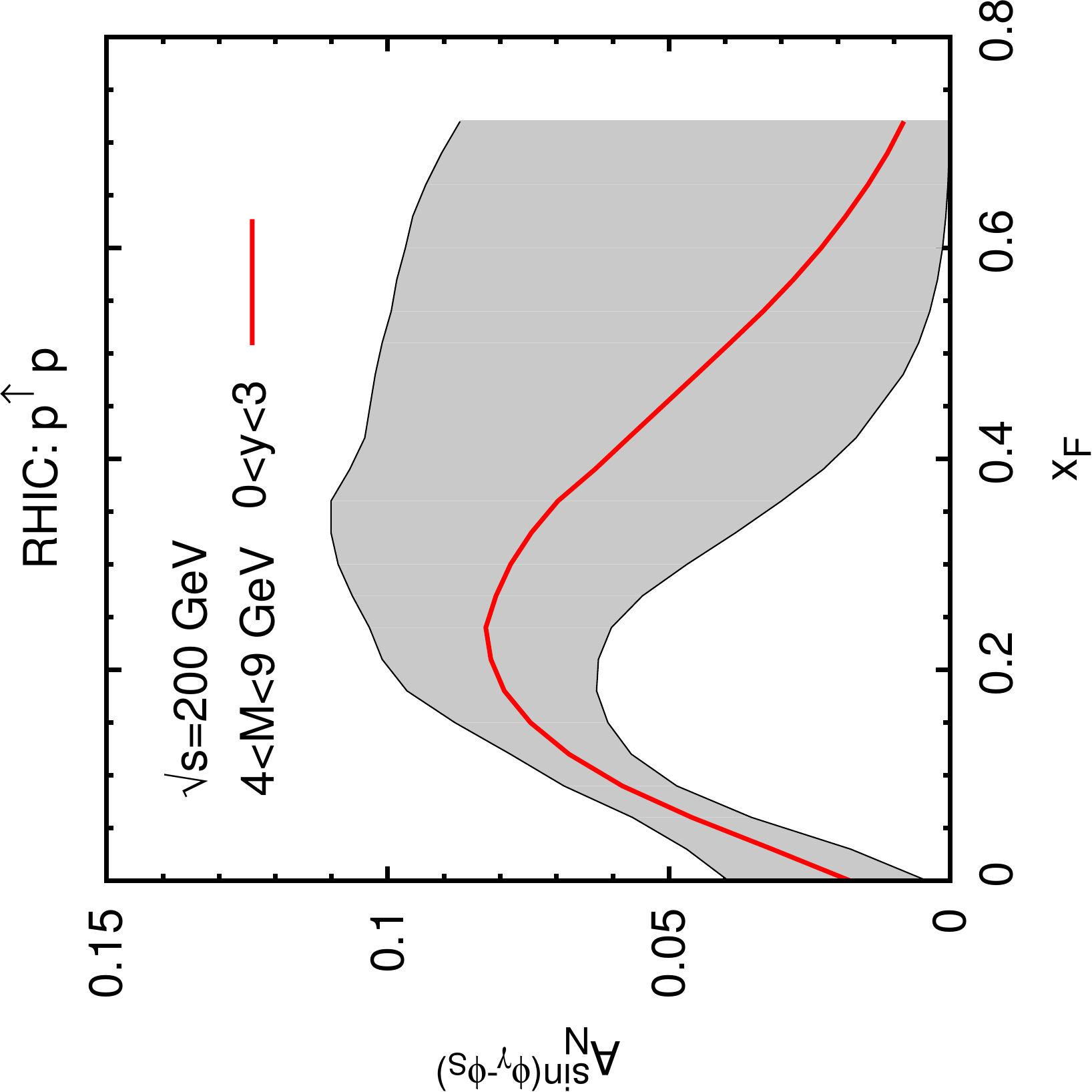}
\hskip 0.1\textwidth
\includegraphics[width=0.4\textwidth,angle=-90]{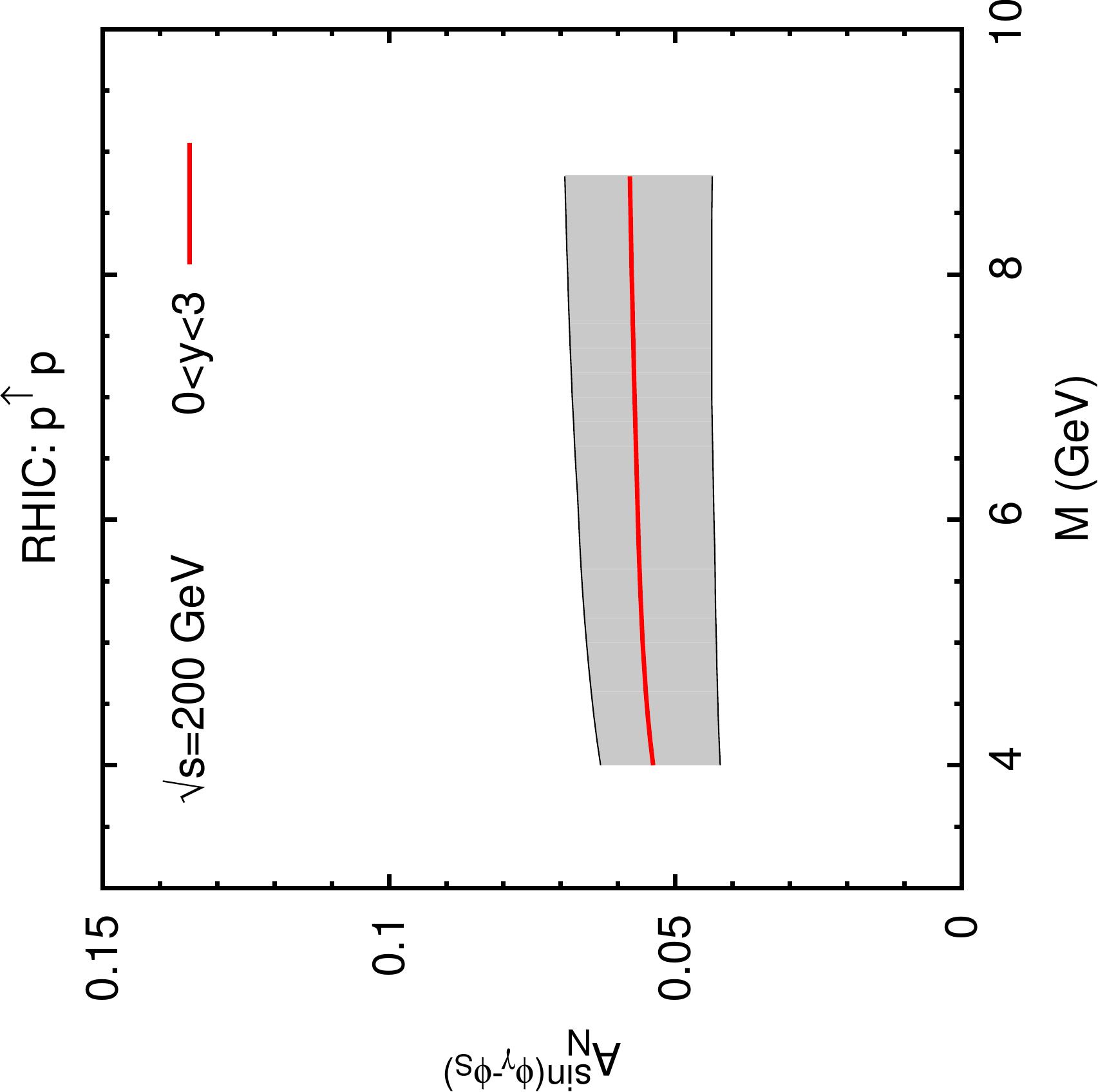}
\caption{\label{fig:an_dy_rhic}
Predictions for SSAs of Drell-Yan process: $p(\uparrow)+p\to \mu^+\mu^-(M)+X$ 
at RHIC as a function of $x_F$ (left) and invariant mass $M$ (right) of the lepton pair at 
$\sqrt{S}=200$~GeV, with phase space integration: $0\leq q_T\leq 1$~GeV,  
$4< M < 9$~GeV, and $0 < y < 3$ \cite{Anselmino:2009st}.
}
\end{center}
\end{figure}

Collins {\it et al.} \cite{Collins:2005rq} made predictions for SSAs of Drell-Yan process 
at RHIC, based on their fit to the Sivers function from SIDIS.  More recently, with
better data on SSAs in SIDIS from HERMES \cite{hermes05,hermes07,hermes05-1} and 
COMPASS \cite{compass05,compass08}, a much improved set of Sivers functions 
was extracted by Anselmino {\it et al.} \cite{Anselmino:2008sga}.  Using this new set of 
Sivers functions, Anselmino {\it et al.} updated the predictions by Collins {\it et al.}, 
and extend their predictions to SSAs of various Drell-Yan experiments proposed  around the world \cite{Anselmino:2009st}.  In Figs.~\ref{fig:an_dy_compass} and
\ref{fig:an_dy_rhic}, we present the predictions of Drell-Yan SSAs 
by Anselmino {\it et al.} for COMPASS and RHIC kinematics, respectively.
The asymmetry was predicted at a few percents.  
Measurements at COMPASS and RHIC should be 
able to tell the difference in the sign of the asymmetry, as well as the sign of the Sivers function.

The sign change of the Sivers function is an important test of the 
TMD factorization in QCD.  If the test fails for a sufficiently large range of $Q$ and small $q_T$, 
it should force us to question the whole TMD factorization approach, which is 
very important for providing the theoretical framework to measure the confined motion of quarks and 
gluons inside a colliding hadron.  With the ability to vary $q_T$ of Drell-Yan process
experimentally, we can move from TMD factorization to the region relevant to collinear factorization,
in which the sign changes are dynamical from perturbatively calculated hard parts, while all 
twist-3 correlation functions are process independent.  Drell-Yan process is in a very unique
position to critically test the dynamics of QCD beyond the very successful leading power 
collinear factorization formalism.

\subsection{Lepton angular distribution of the Drell-Yan process}

By measuring the angular distribution of the observed lepton in 
the lepton-pair's rest frame, 
Drell-Yan process can directly measure quantum interference between scatterings 
with the virtual photon (or in general, an intermediate vector boson) 
in different polarization states~\cite{tangerman95,arnold09}.
Despite the success of perturbative QCD in describing the
Drell-Yan cross sections, it remains a challenge to understand the angular
distributions of the Drell-Yan process.

With the lepton momenta in the pair's rest frame defined as 
\begin{eqnarray}
l^\mu &=&
\frac{Q}{2} \left(
1, \sin\theta \cos\phi, \sin\theta \sin\phi, \cos\theta \right) 
\nonumber\\
\bar{l}^\mu &=&
\frac{Q}{2} \left(
1, -\sin\theta \cos\phi, -\sin\theta \sin\phi, -\cos\theta 
\right)\, , 
\label{leptonM} 
\end{eqnarray}
we can write the Drell-Yan cross section for a massive lepton pair production
in terms of four independent ``helicity'' structure functions, 
assuming dominance of the single-photon process \cite{lam78},
\begin{eqnarray}
\frac{d\sigma}{d^4q d\Omega}
&=&
\frac{\alpha_{\rm em}^2}{2(2\pi)^4 S^2 Q^2}
\left[W_T (1+\cos^2\theta) + W_L (1-\cos^2\theta) \right.
\nonumber\\
&& 
\left.
+ W_{\Delta} (\sin2\theta \cos\phi) + W_{\Delta\Delta} (\sin^2\theta\cos 2\phi) 
\right]\, .
\label{x-sec-angular}
\end{eqnarray}
These helicity structure functions $W_T$, $W_L$,  
$W_{\Delta}$, and $W_{\Delta \Delta}$ depend on $Q$, $q_\perp$, rapidity 
$y$, and on the center-of-mass energy $\sqrt{S}$ of the 
production process.  
The angular-integrated cross section is then expressed in terms of 
two structure functions, $W_T$ and $W_L$, as
\begin{eqnarray}
\frac{d\sigma}{d^4q}
= \frac{\alpha_{\rm em}^2}{12 \pi^3 S^2 Q^2}
\left[2 W_T  + W_L \right] 
= \frac{\alpha_{\rm em}^2}{12 \pi^3 S^2 Q^2}
\left[-g^{\mu\nu}\, W_{\mu\nu}\right] 
\ ,
\label{x-sec-integ}
\end{eqnarray}
given in terms of the hadronic tensor of the Drell-Yan process,
\begin{eqnarray}
W_{\mu\nu} 
&=&
S \int d^4z\, {\rm e}^{i q\cdot z}\,
\langle P_A P_B | J_{\mu}^{\dagger}(0)\, J_{\nu}(z) |P_A P_B\rangle ,
\nonumber\\
&=&
- \left(g_{\mu\nu}-T_\mu T_\nu\right) 
  \left(W_T + W_{\Delta\Delta}\right)
- 2 X_\mu X_\nu W_{\Delta\Delta} 
\nonumber\\
&\ &
+ Z_\mu Z_\nu \left( W_L - W_T - W_{\Delta\Delta} \right)
- \left( X_\mu Z_\nu + X_\nu Z_\mu \right) W_{\Delta}\, ,
\label{W-def}
\end{eqnarray}
\noindent where $J_\mu$ is the electromagnetic current, $T_\mu = q_\mu/Q$, and 
$X$, $Y$, and $Z$ are orthogonal unit vectors of 
the virtual photon's rest frame, defined to have $X$ and $Z$ in the plan 
spanned by $P_A$, $P_B$, and $q$ with 
$X^2=Y^2=Z^2=-1$ and $q_\mu X^\mu = q_\mu Y^\mu = q_\mu Z^\mu = 0$.
The angular-integrated Drell-Yan cross section 
corresponds to a sum of contributions from production of a transversely
polarized virtual photon plus that of a longitudinally polarized virtual photon.

Let $\epsilon_\lambda^\mu(q)$ be the virtual photon's 
polarization vector in the photon's rest frame, and
$\epsilon_\pm^\mu = (\mp X^\mu - i Y^\mu)/\sqrt{2}$, 
$\epsilon_0^\mu = Z^\nu$ for three polarization states, 
$\lambda=\pm 1, 0$ \cite{lam78}.
In terms of these polarization vectors, 
the four helicity structure functions 
are given by
\begin{eqnarray}
W_T 
&=& 
W_{\mu\nu}\, \epsilon_1^{\mu*}\epsilon_1^\nu\, ,
\nonumber\\
W_L 
&=& 
W_{\mu\nu}\, \epsilon_0^{\mu*}\epsilon_0^\nu\, ,
\nonumber\\
W_{\Delta} 
&=& 
W_{\mu\nu} 
\left( \epsilon_1^{\mu*}\epsilon_0^\nu
     + \epsilon_0^{\mu*}\epsilon_1^\nu \right)/\sqrt{2}\, ,
\nonumber\\
W_{\Delta\Delta} 
&=& 
W_{\mu\nu}\, \epsilon_1^{\mu*}\epsilon_{-1}^\nu\, ,
\label{helicity-f-def}
\end{eqnarray}
These structure functions, $W_T$, $W_L$, $W_{\Delta}$, and $W_{\Delta \Delta}$
correspond, respectively, to the transverse spin, 
longitudinal spin, single spin-flip, and double spin-flip contributions 
to the Drell-Yan cross section.  

From Eqs.~(\ref{x-sec-angular}) and (\ref{x-sec-integ}), 
we can define the normalized Drell-Yan angular distribution as 
\begin{eqnarray}
\frac{dN}{d\Omega}
&\equiv &
\left( \frac{d\sigma}{d^4q} \right)^{-1}
\frac{d\sigma}{d^4q d\Omega}
\nonumber \\
&=&
\frac{3}{4\pi}\left(\frac{1}{\lambda +3}\right)
\bigg[ 1 + \lambda \cos^2\theta
+ \mu \sin 2\theta \cos\phi
+\frac{\nu}{2}\sin^2\theta\cos 2\phi \bigg] 
\label{eq:eq1}
\end{eqnarray}
with the coefficients of the angular dependence given by
helicity structure functions,
\begin{eqnarray}
\lambda 
&=& \frac{W_T-W_L}{W_T+W_L}\, ,
\nonumber \\
\mu 
&=& \frac{W_{\Delta}}{W_T+W_L}\, ,
\nonumber \\
\nu 
&=& \frac{2W_{\Delta\Delta}}{W_T+W_L}.
\label{angularpar}
\end{eqnarray}
In the ``naive"
Drell-Yan model, where the transverse momentum of the quark is ignored
and no gluon emission is considered, $\lambda =1$ and $\mu = \nu =0$ are
obtained. QCD effects~\cite{chiappetta86} and
non-zero intrinsic transverse momentum of the quarks~\cite{cleymans81}
can both lead to $\lambda \ne 1$ and $\mu, \nu \ne 0$. However,
$\lambda$ and $\nu$ should still
satisfy the relation
$1-\lambda = 2 \nu$~\cite{lam78}. This so-called Lam-Tung
relation, obtained as
a consequence of the spin-1/2 nature of the quarks, is analogous
to the Callan-Gross relation~\cite{callan69}
in DIS. While QCD effects can significantly
modify the Callan-Gross relation, the Lam-Tung relation
is predicted to be largely unaffected by QCD corrections~\cite{lam80}.

\begin{figure}[h]
\includegraphics[width=0.55\textwidth]{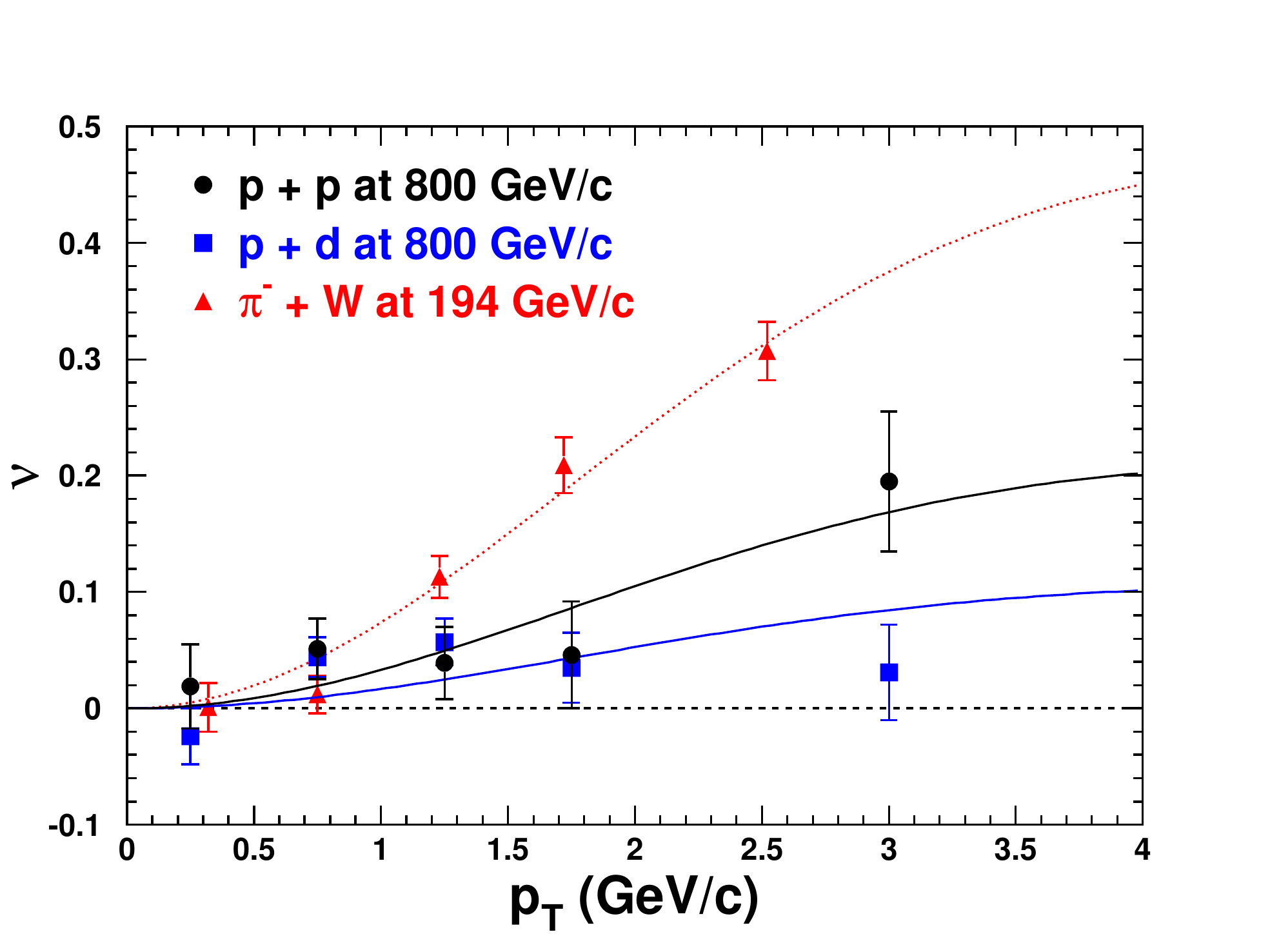}
\includegraphics[width=0.4\textwidth]{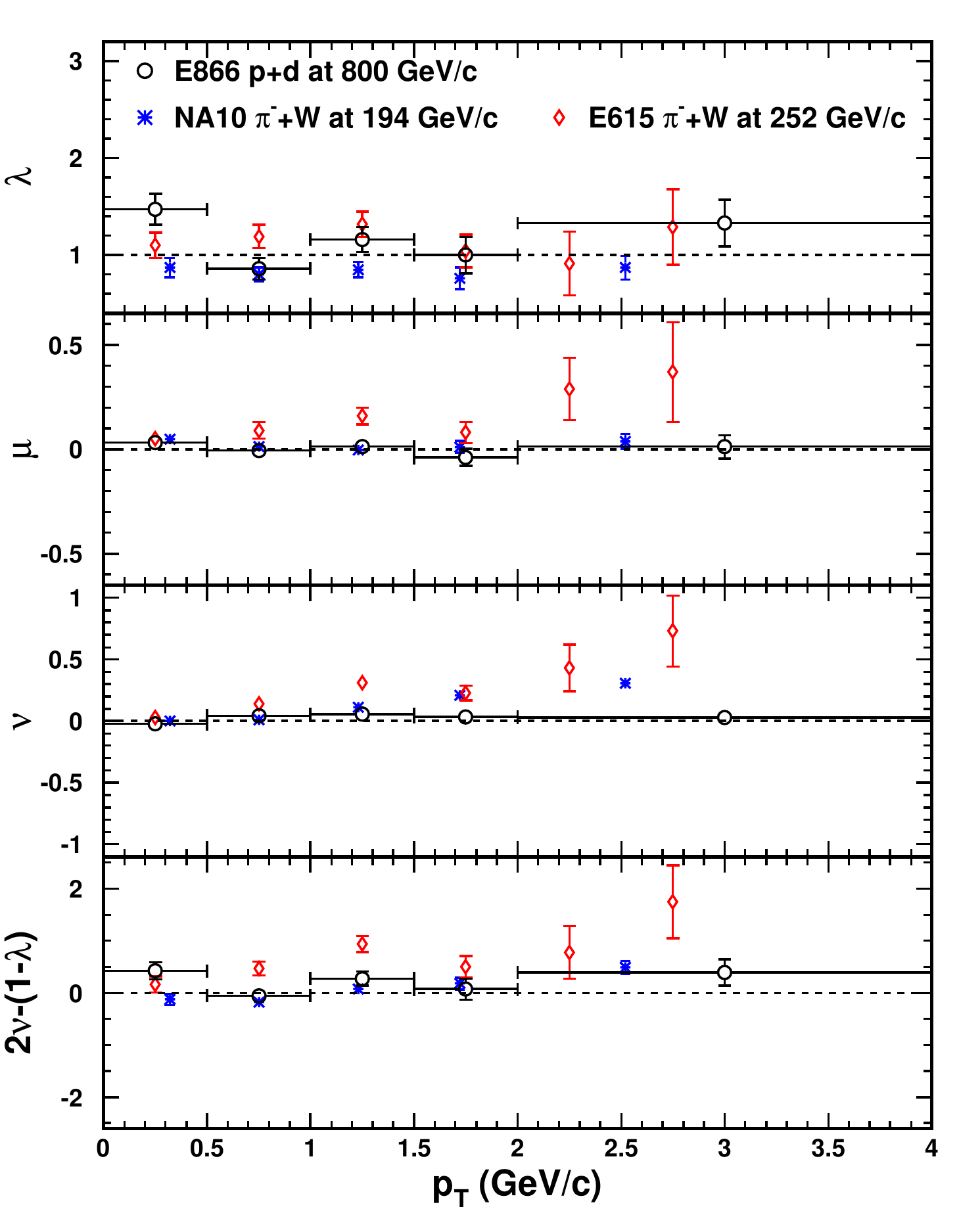}
\caption{Left (a): Parameter $\nu$ vs.\ $p_T$ in the Collins-Soper
frame for several Drell-Yan
measurements~\cite{falciano86,zhu07,zhu09}. Fits to the data using 
an empirical formula~\cite{boer99} are
also shown.
Right (b): Parameters $\lambda, \mu, \nu$ and $2\nu - (1-\lambda)$
vs.\ $p_T$ in the Collins-Soper frame. Solid circles are for
E866 $p+d$ at 800 GeV/c~\cite{zhu07}, crosses are for NA10 $\pi^- + W$ 
at 194 GeV/c~\cite{falciano86}, and
diamonds are E615 $\pi^- + W$ at 252 GeV/c~\cite{conway89}.}
\label{fig3.5}
\end{figure}

The Drell-Yan angular distribution was
measured by the NA10 Collaboration for $\pi^- + W$ at 140, 194, and
286 GeV/c~\cite{falciano86}.
The $\cos 2 \phi$ angular modulation shows sizable values of $\nu$,
increasing with dimuon transverse momentum 
($p_T$, which is the same as $q_T$) 
and reaching a value
of $\approx 0.3$ at $p_T = 2.5$
GeV/c (see Fig.~\ref{fig3.5} (a)). This could not
be explained by perturbative QCD, which predicts much smaller
values of $\nu$~\cite{chiappetta86}. The Fermilab E615 Collaboration
subsequently measured $\pi^- + W$ Drell-Yan angular distribution
at 252 GeV/c with a broad coverage in the
decay angle $\theta$~\cite{conway89}.
The E615 results showed that
$\lambda$ deviates from 1 at large values of $x_1$ with both $\mu$ and
$\nu$ having large non-zero values. The Lam-Tung relation, 
$2\nu = 1 - \lambda$, is clearly violated for the E615 data (see 
Fig.~\ref{fig3.5} (b)).

The results on the Drell-Yan angular distributions strongly
suggest new effects beyond perturbative QCD.
Brandenburg, Nachtmann and Mirke considered a factorization-breaking
QCD vacuum, which can lead to a
correlation between the transverse spin of the quark in the nucleon and
that of the antiquark in the pion~\cite{brandenburg93}.
This would generate a non-zero
$\cos 2\phi$ angular dependence. Boer {\em et al.} noted that 
a possible
source for a factorization-breaking
QCD vacuum is helicity flip in the instanton model~\cite{boer05}.
Higher-twist effects from quark-antiquark binding
in pions~\cite{brandenburg94,eskola94}, motivated by earlier work of
Berger and Brodsky~\cite{berger79}, have also been considered. 
While they 
predict behavior of $\mu$ and $\nu$ in qualitative agreement with the
data, they are applicable
only in the $x_1 \to 1$ region.

\begin{figure}[tbp]
\begin{center}
\includegraphics*[width=0.5\linewidth]{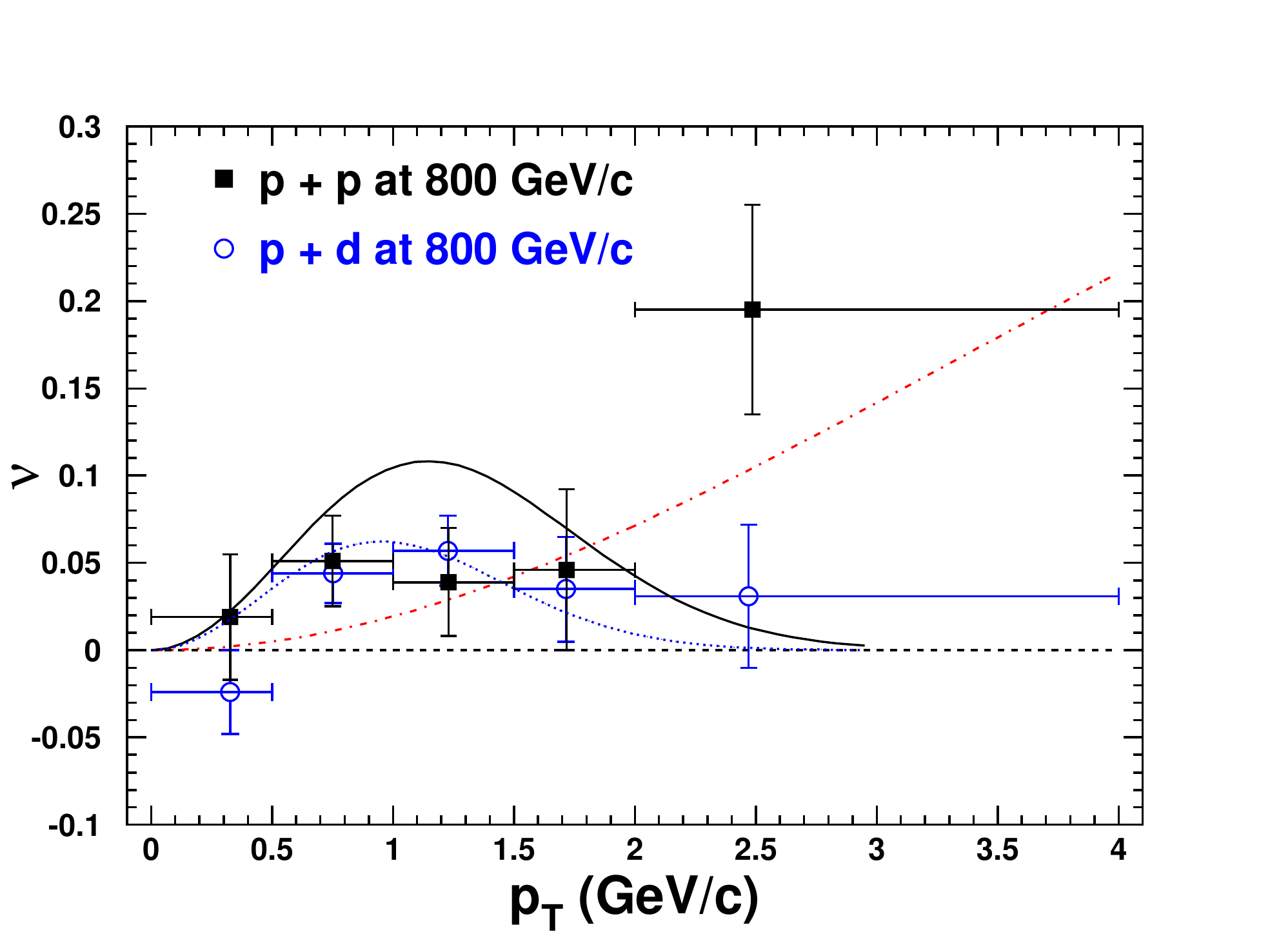}
\end{center}
\caption{Parameter $\nu$ versus $p_T$ in the Collins-Soper
frame for the $p+p$ and $p+d$ E866 Drell-Yan data~\cite{zhu07,zhu09}.
The solid and dotted curves are
calculations~\cite{zhang08} for $p+p$ and $p+d$, respectively,
using parametrizations
based on a fit to the $p+d$ data. The dot-dashed curve is the
contribution from the QCD process~\cite{boer06,berger07}.}
\label{ppfig2}
\end{figure}

Boer pointed out~\cite{boer99}
that the $\cos 2 \phi$ angular dependences could be due to the 
TMD Boer-Mulders function, which 
characterizes the correlation between a quark's transverse 
momentum, $k_T$, and its transverse spin in an unpolarized hadron.
This correlation, when combined with the transverse spin correlation 
in quark-antiquark annihilation, would lead to a preferred transverse
momentum direction resulting in a $\cos 2 \phi$ dependence. The Boer-Mulders
function is a time-reversal odd object and is the 
analog of the Collins fragmentation
function~\cite{collins93}. Model 
calculations~\cite{gamberg03,boer03,bacchetta04,lu05}
for the nucleon (pion) Boer-Mulders functions 
in the framework of quark-diquark (quark-spectator-antiquark)
model can
describe the $\cos 2 \phi$ behavior observed in NA10~\cite{lu05}.

The first measurement of the azimuthal angular 
dependence of the proton-induced Drell-Yan process was reported for
$p+p$ and $p+d$ interactions
at 800 GeV/c~\cite{zhu07,zhu09}. In contrast to pion-induced Drell-Yan,
significantly smaller cos$2\phi$ azimuthal angular modulation
was observed (see Fig.~\ref{fig3.5} (a)). 
While pion-induced Drell-Yan process
is dominated by annihilation between a valence antiquark in the pion
and a valence quark in the nucleon, the
proton-induced Drell-Yan process involves a valence quark 
annihilating with a sea antiquark in the nucleon. Therefore, the
$p+p$ and $p+d$ results show~\cite{zhang08,zhu07,zhu09} that
the Boer-Mulders functions for
sea quarks are significantly smaller than those for valence quarks.
While the Lam-Tung relation is clearly violated in the pion-induced
Drell-Yan data, Fig.~\ref{fig3.5} (b)
shows that the proton Drell-Yan data are consistent with the Lam-Tung relation.
This lends further support to the interpretation that violation
of the Lam-Tung relation in pion-induced Drell-Yan is due to the sizable
valence quark Boer-Mulders functions in pions and nucleons.

Figure~\ref{ppfig2} shows $\nu$ versus $p_T$ for the $p+p$ and
$p+d$ Drell-Yan data~\cite{zhu07,zhu09}. 
The solid and dotted curves are
calculations~\cite{zhang08} for $p+p$ and $p+d$ using
parametrizations of the Boer-Mulders functions deduced from a fit
to the $p+d$ Drell-Yan data.
The prediction for larger values of $\nu$ for $p+p$ versus
$p+d$ in the region of $p_T \sim 1.5$ GeV/c are not
observed. Furthermore, the shape of
the predicted $p_T$ dependence differs
from that of the data.
This suggests that there could
be other mechanisms contributing to the $\cos 2\phi$ angular
dependence at large $p_T$. 

Recently, two groups looked into the Drell-Yan angular 
distributions from the point of view of QCD resummation, which 
is important for the region $q_T \ll Q$   
\cite{boer06,berger07,berger07-1}.  
In Ref.~\cite{boer06}, Boer and Vogelsang carefully
investigate the logarithmic behavior of the order $\alpha_s$ 
perturbative contributions to the helicity structure functions.
At order $\alpha_s$, they find that, like $W_T$, 
both $W_L$ and $W_{\Delta\Delta}$ have a $\ln(Q^2/q_T^2)$ 
logarithmic divergence, but not the $1/q_T^2$ power divergence 
seen in $W_T$, and that $W_\Delta$ has no logarithmic divergence at 
this order in the Collins-Soper frame.  
They notice that the logarithmic contribution to
$W_L$ and $W_{\Delta\Delta}$ from quark-gluon (or gluon-quark)
subprocess is different from that for $W_T$ and does not fit 
the pattern expected for the perturbative expansion of 
the CSS resummation formalism to order $\alpha_s$ \cite{Collins:1984kg}.  

Since only $W_T$ shows the leading $\alpha_s\ln^2(Q^2/q_T^2)/q_T^2$-type 
divergence as $q_T\to 0$, 
previous resummation calculations were carried out only 
for $W_T$ in the same way as for the angular-integrated
cross section.   
As shown in Refs.~\cite{chiappetta86,bqy-wz,Ellis:1997sc},
resummation removes the perturbative power divergence in 
$W_T$ and creates a large change in the relative size of $W_T$ 
versus the rest of the helicity structure functions for 
which no resummation is performed.  
This result is not quite consistent with general expectations 
about the relative size of helicity structure functions in the Collins-Soper frame,
as $q_T^2/Q^2 \to 0$.  For example, one expects $W_{\Delta\Delta}/W_T
 \rightarrow q_T^2$ as
$q_T^2\to 0$ \cite{cs-frame}.

Different from all previous resummation calculations, 
Berger {\it et al.}~\cite{berger07,berger07-1} observe that 
the four helicity structure functions cannot 
be independent as $q_T\to 0$ since there are only two
independent structure functions at $q_T=0$.
Guided by electromagnetic current conservation of the hadronic tensor, 
they find that the leading logarithmic behavior of the 
different helicity structure functions, $W_T$, $W_L$, and 
$W_{\Delta\Delta}$, has a common origin. 
That is, current conservation uniquely ties the perturbative divergences as 
$q_T/Q\to 0$ of the otherwise independent helicity structure functions 
$W_T, W_L,$ and $W_{\Delta\Delta}$ to the divergence 
of the angular-integrated cross section \cite{berger07,berger07-1},
\begin{eqnarray}
W_T^{\rm Resum}
&=& 
\left(1-\frac{1}{2} \frac{q_T^2/Q^2}{1+q_T^2/Q^2} \right)
\frac{W^{\rm Resum}}{2} \, ,
\nonumber \\
W_L^{\rm Resum}
&=& 
\frac{q_T^2/Q^2}{1+q_T^2/Q^2}\,
\frac{W^{\rm Resum}}{2}\, ,
\nonumber \\
W_{\Delta\Delta}^{\rm Resum}
&=& 
\frac{1}{2}\,\frac{q_T^2/Q^2}{1+q_T^2/Q^2}\,
\frac{W^{\rm Resum}}{2} \, 
\label{st-resum-cs}
\end{eqnarray}
where $W^{\rm Resum}$ is given by the resummed 
angular-integrated Drell-Yan cross section~\cite{Collins:1984kg},
\begin{equation}
\frac{\alpha_{\rm em}^2}{12\pi^3 S^2 Q^2}\,W^{\rm Resum} 
=
\frac{1}{(2\pi)^2}\int d^2b\, e^{i\vec{Q}_\perp \cdot \vec{b}}\,
\widetilde{W}(b,Q,x_A,x_B) ,
\label{w-reum-b}
\end{equation}
which is a result of comparing Eq.~(\ref{eq:css-gen}) with 
Eq.~(\ref{x-sec-integ}).  From Eq.~(\ref{angularpar}), one 
obtains asymptotically when $q_T^2\ll Q^2$,
\begin{eqnarray}
\lambda 
&=& \frac{W_T-W_L}{W_T+W_L}
\approx \frac{W_T^{\rm Resum}-W_L^{\rm Resum}}
             {W_T^{\rm Resum}+W_L^{\rm Resum}}
= \frac{1-\frac{1}{2} q_T^2/Q^2}
         {1+\frac{3}{2} q_T^2/Q^2}\, ,
\nonumber \\
\nu 
&=& \frac{2W_{\Delta\Delta}}{W_T+W_L}
\approx \frac{2W_{\Delta\Delta}^{\rm Resum}}
             {W_T^{\rm Resum}+W_L^{\rm Resum}}
= \frac{q_T^2/Q^2}
         {1+\frac{3}{2}q_T^2/Q^2}\, , 
\label{angularpar-resum}
\end{eqnarray}
which satisfy the Lam-Tung relation: $1-\lambda - 2\nu=0$.
That is, the approximate Lam-Tung relation is an all-orders consequence 
of current conservation for the leading 
perturbatively divergent terms. 
This conclusion should not be too surprising.  Lam-Tung relation
is a result of helicity conservation.  The leading resummed contribution from 
the collinear and infrared region should not violate the helicity conservation.

From Eq.~(\ref{angularpar-resum}), the
predicted QCD contribution, the same for $p+p$ and $p+d$ due to the
identical kinematic coverage for the two reactions,
is shown as the dot-dashed curve in Fig.~\ref{ppfig2}. From 
Fig.~\ref{ppfig2} it is
evident that the QCD contribution is expected to become
more important at high $q_T (p_T)$ while the
Boer-Mulders functions contribute primarily at lower $q_T$.
An analysis combining both effects is
required in order to extract reliably the
Boer-Mulders functions from the $p+p$ and $p+d$ data.
It is remarkable that a non-perturbative correlation between the 
quark spin and its transverse motion, described by 
the Boer-Mulders function, could be probed in unpolarized
SIDIS and Drell-Yan reactions. 

Unlike the Sivers functions which have only been extracted from
the SIDIS data so far, some information on the Boer-Mulders
functions has already been obtained from both the SIDIS and the 
Drell-Yan experiments. It is natural to raise the question whether
these data could already test the prediction that
the Boer-Mulders function changes sign from SIDIS to Drell-Yan
(see Eq.~(\ref{eq:pt-inv_tensor})). Several theoretical models predict
that the valence quark Boer-Mulders functions should be
negative~\cite{yuan03,pasquini05,burkardt08,gamberg08}. This
expectation is consistent with the analysis of the existing
SIDIS data on the $\cos 2\phi$ angular dependence
by Barone et al.~\cite{barone10}, showing that both $h^\perp_{1u}$
and $h^\perp_{1d}$ in SIDIS are negative. However, SIDIS data
do not allow the extraction of sea-quark Boer-Mulders functions,
the effects of which are overshadowed by those of the valence
quarks. On the other hand, the $\cos 2\phi$ dependence in $p+p$ and
$p+d$ Drell-Yan is proportional to the product of the valence
and the sea Boer-Mulders functions, namely,
$\nu \sim h^\perp_{1q} (x_1) h^\perp_{1\bar q} (x_2)$,
allowing the extraction of sea-quark Boer-Mulders function
in the Drell-Yan experiment. Assuming $u$-quark dominance,
the positive values of $\nu$ shown in Fig.~\ref{ppfig2} 
already suggest that the Boer-Mulders functions for $u$ and $\bar u$
have the same sign in Drell-Yan. The
Boer-Mulders function for $u$ quark, determined in SIDIS
to have negative sign, is expected to become positive 
in Drell-Yan according to
Eq.~(\ref{eq:pt-inv_tensor}). It follows that both the $u$ and $\bar u$ 
Boer-Mulders functions in Drell-Yan
have positive signs. This is in agreement with the
results obtained by Lu and Schmidt~\cite{lu10} and
by Barone et al.~\cite{barone10a}. In order to
test the sign-change prediction of Eq.~(\ref{eq:pt-inv_tensor}),
an important yet unknown input is the sign of the sea-quark Boer-Mulders
functions in SIDIS. Future SIDIS experiments at the
12 GeV JLab and EIC could provide such information. At this moment,
one can only conclude that all existing data are not in disagreement
with the predicted sing-change of the Boer-Mulders functions predicted
in Eq.~(\ref{eq:pt-inv_tensor}).

\begin{figure}[tbp]
\begin{center}
\includegraphics*[width=0.45\linewidth]{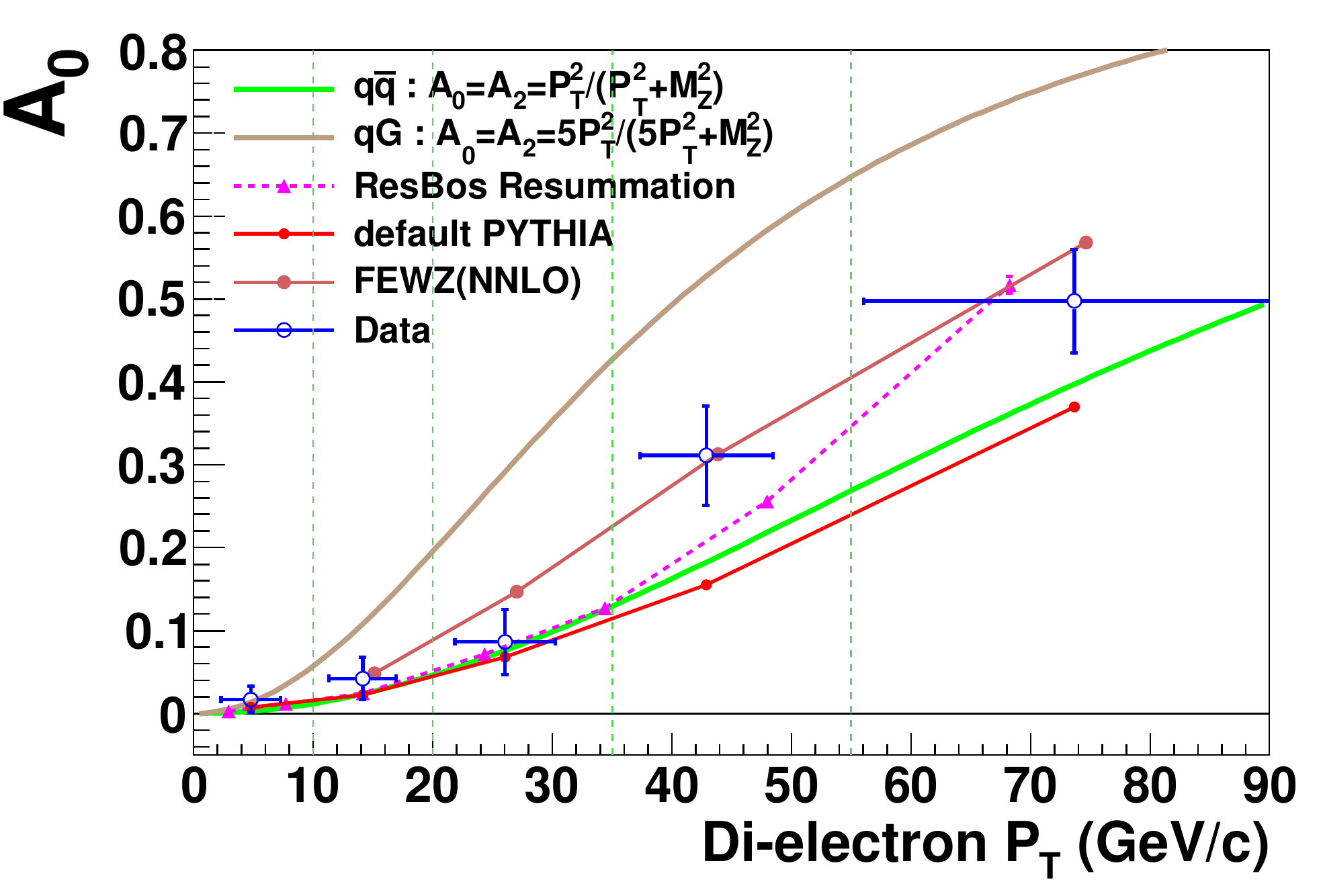}
\includegraphics*[width=0.45\linewidth]{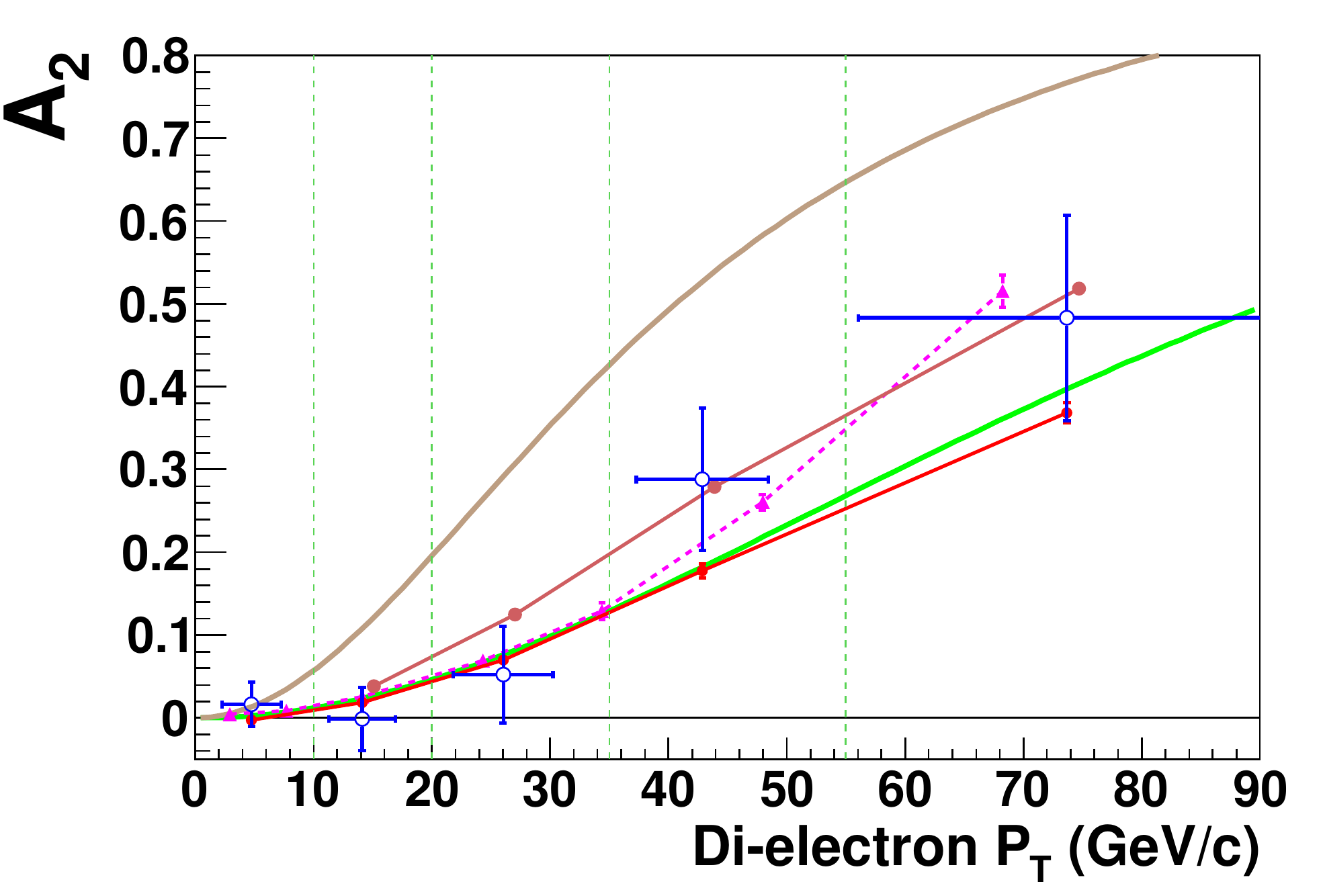}
\end{center}
\caption{Coefficients $A_0$ and $A_2$ for $e^+ e^-$
pairs ($66 < M_{ee} < 116$ GeV) from CDF~\cite{han11,cdf11}.
Predictions from various calculations are also shown.}
\label{fig:cdf}
\end{figure}

The CDF collaboration recently reported the measurement of the
decay angular distribution of Drell-Yan $e^+e^-$ pairs in the
$Z$ mass region from $p \bar p$ collisions at $\sqrt s = 1.96$
TeV~\cite{han11,cdf11}. The following expressions~\cite{mirkes94} 
for the polar and azimuthal angular distributions were
adopted:
\begin{eqnarray}
\frac{d\sigma}{d\cos \theta} &\propto& (1+\cos^2\theta)
+ \frac{1}{2} A_0 (1-3\cos^2\theta) + A_4 \cos\theta;
\nonumber\\
\frac{d\sigma}{d\phi} &\propto& 1 + \beta_3 \cos \phi
+ \beta_2 \cos 2\phi + \beta_7 \sin \phi + 
\beta_5 \sin 2\phi,
\label{cdf_dy}
\end{eqnarray}
\noindent where the $\cos 2\phi$ coefficient is 
$\beta_2 \equiv A_2/4$. Comparing with Eq.~(\ref{eq:eq1}),
some additional terms in Eq.~(\ref{cdf_dy}) are due
to the parity-violating contributions from $Z$ boson
in the $p \bar p \to (\gamma^*/Z) X \to e^+ e^- X$.
It is straightforward to obtain
\begin{eqnarray}
\lambda = \frac{2-3A_0}{2+A_0};~~~~\nu = \frac{2A_2}{2+A_0}.
\label{cdf_lam_tung}
\end{eqnarray}
\noindent The Lam-Tung relation, $1 - \lambda = 2 \nu$,
becomes $A_0 = A_2$. Figure~\ref{fig:cdf} shows that $A_0$
and $A_2$ for $66<M_{ee}<116$ GeV are nonzero and increase
with the transverse momentum ($p_T$) of the $e^+e^-$ pair. 
However, the Lam-Tung relation, $A_0 - A_2 = 0$, is
well satisfied with $\langle A_0 - A_2 \rangle = 0.02 \pm 0.02$~\cite{cdf11}.
Since the Boer-Mulders mechanism is important only at
low $p_T$, the Lam-Tung relation is expected to be valid
at the very large $p_T$ region covered by the CDF data,
which is consistent with the conclusion of pQCD resummation approach
as described right below Eq.~(\ref{angularpar-resum}).
The ability to extract the azimuthal angular distribution information 
at collider energies, as demonstrated by the CDF result,
bodes well for further measurements at LHC~\cite{lu11}.

\section{Generalized Drell-Yan process}
\label{sec:wz}

In this section, we briefly review the role of generalized Drell-Yan 
massive lepton-pair production 
via the heavy vector boson, $W/Z$, in measuring the PDFs, and parton helicity distributions of proton.  
The distinct quark-flavor dependences of the $W, Z, \gamma^*$ 
couplings to quark-antiquark pairs offer a unique possibility for disentangling the flavor
structure of the parton distributions.  

\subsection{$\bar d / \bar u$ from $W$ production}
  
At the $p+p$ colliders at RHIC and LHC, the $\bar d/ \bar u$ asymmetry 
can be measured via the ratio of $W^+$ 
over $W^-$ production~\cite{peng95,doncheski94,bourrely94}.
This method has some important
advantages. First, it does not
require the assumption of the validity of charge symmetry.  The
existing measurements of $\bar d/ \bar u$ asymmetry all depend
on the comparison of DIS or Drell-Yan cross sections off
hydrogen versus deuterium targets. Charge symmetry was
assumed in order to relate the parton distributions in neutron to those
in proton. The possible violation of charge
symmetry at the parton level has been discussed by
several authors~\cite{ma1,ma2,sather,rodionov,benesh1,londergan2},
and was recently reviewed in~\cite{londergan11}.
Ma and collaborators~\cite{ma1,ma2} pointed out that Drell-Yan
experiments, such as NA51 and E866, are subject to both flavor
asymmetry and charge symmetry violation effects. In fact, a 
larger flavor asymmetry is required to compensate for the
charge symmetry violation effect~\cite{steffens96}. An analysis
combining $W$ production in $p+p$ collision with the NA51 and
E866 Drell-Yan experiments could separate the flavor asymmetry from
the charge symmetry violation effects.

\begin{figure}[t]
\includegraphics[width=0.5\textwidth]{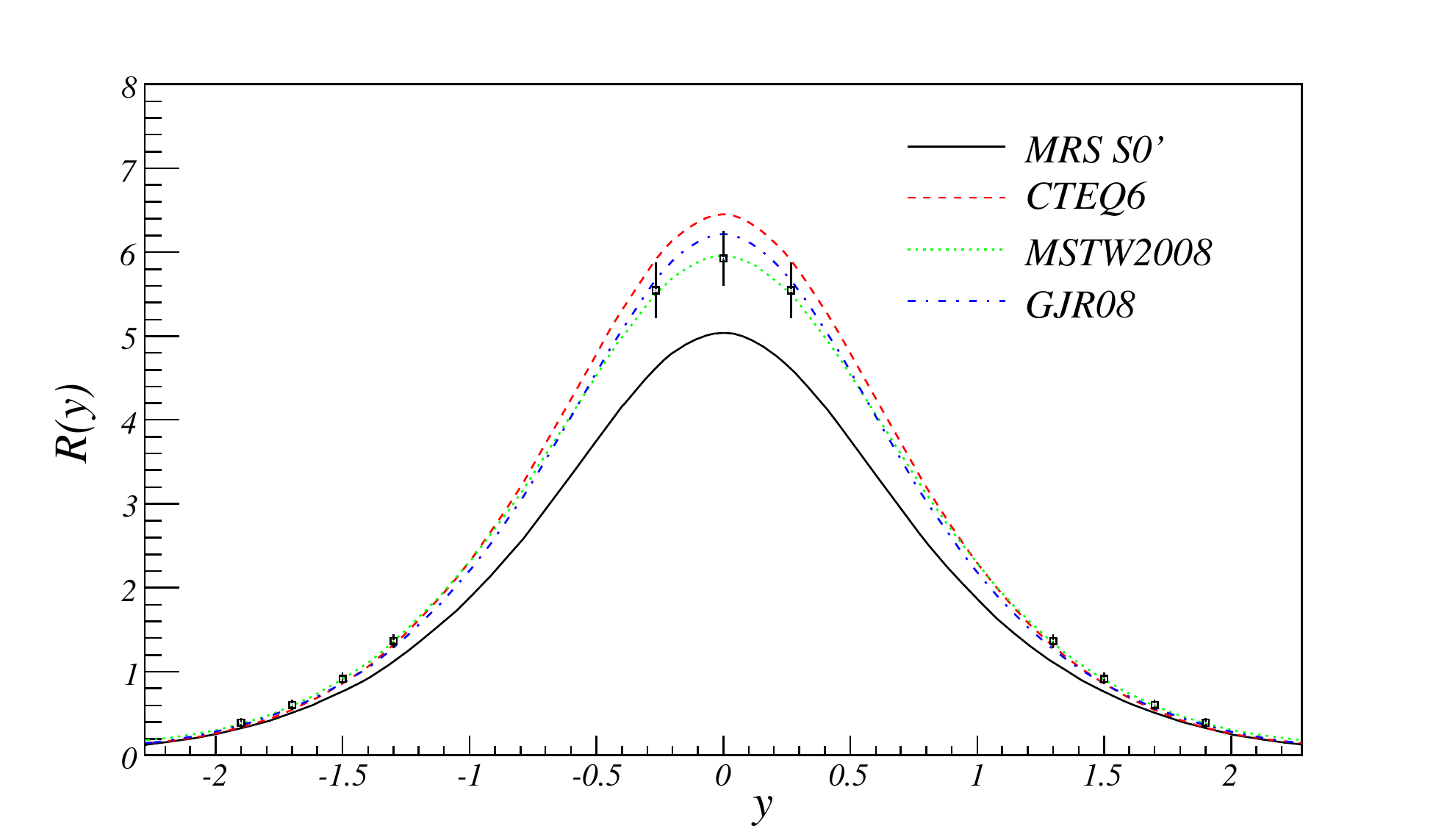}
\includegraphics[width=0.5\textwidth]{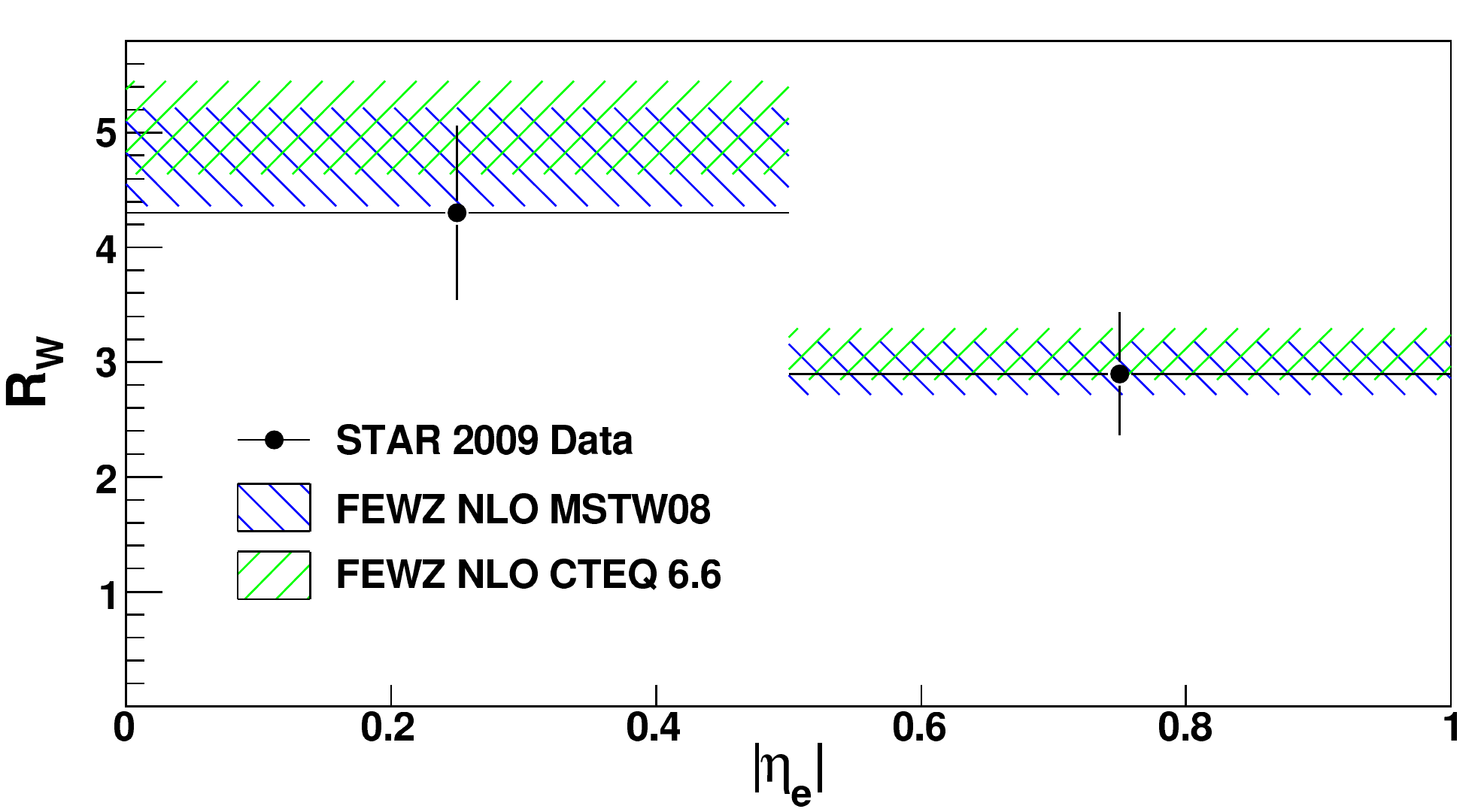}
\caption{Left (a): Prediction of the ratio $R(y)$ for
$p+p$ collision at $\sqrt s$ of 500 GeV using
various parton distribution functions~\cite{yang09}. 
The projected sensitivities
for a run with recorded luminosity of 300 pb$^{-1}$ for the PHENIX detector
are also shown. Right (b): $W$ cross section ratio from
STAR for two pseudorapidity bins~\cite{star12}. Calculations 
at NLO using the MSTW08 and CTEQ 6.6 PDF sets
are shown for comparison.}
\label{fig3.4}
\end{figure}

Another important advantage of $W$ production in $p+p$ collision is the absence of nuclear effects. As noted by several
authors~\cite{wally93,braun94,sawicki93,schmidt01,alekhin05}, the nuclear
modification of parton distributions should be taken into account for
DIS and Drell-Yan process involving deuterium targets. The nuclear shadowing
effect for deuteron could lead to a 4\% to 10\% decrease
in the evaluation of the Gottfried integral by the
NMC~\cite{wally93,schmidt01}. Finally, the $W$ production is sensitive 
to $\bar d/ \bar u$ flavor
asymmetry at a large $Q^2$ scale of $\sim$ 6500 GeV$^2/$c$^2$.
This provides the opportunity to
study the QCD evolution of the $\bar d / \bar u$ flavor asymmetry.

Figure~\ref{fig3.4} (a) shows the prediction of  $R(y)$, defined as

\begin{eqnarray}
R(y)
&=& { d\sigma/dy (p+p \to W^+ x \to l^+ \nu x)
\over d\sigma/dy (p+p \to W^- x \to l^- \bar \nu x)},
\end{eqnarray}

\noindent where $y$ is the rapidity of the charged lepton from the 
$W$ decay. 
The MRS S0$^\prime$ corresponds to the $\bar
d/\bar u$ symmetric parton distributions, while the other three
parton distribution functions are from global fits with asymmetric
$\bar d/\bar u$ sea-quark distributions. Figure~\ref{fig3.4} (a) clearly
shows that $W$ asymmetry
measurements at RHIC could provide an independent determination
of $\bar d / \bar u$~\cite{yang09}.

The STAR collaboration recently reported the measurement of $W^+/W^-$ cross 
section ratio, $R_W$, in $p+p$ collision at $\sqrt{s}$ = 500 GeV, as shown in
Fig.~\ref{fig3.4} (b). Also displayed in Fig.~\ref{fig3.4} (b) are
theoretical calculations of $R_W$ 
at NLO. The data are consistent with predictions using $\bar d / \bar u$
flavor asymmetric PDFs (MSTW08 and CTEQ6.6). The large uncertainty
in the measured $R_W$ is dominated by the statistical precision
of the $W^-$ yield. Future high statistics measurement of the $W$ cross
section ratio at RHIC will provide a sensitive new determination of the
$\bar d / \bar u$ flavor asymmetry.

\subsection{Probe nuclear antishadowing at the LHC}  

With its large mass, $M_Z$, high transverse momentum reach at the LHC, and 
the reliability of QCD factorized production formalism, 
production of $Z^0$ bosons in hadronic collisions is a clean and precise probe for 
short-distance strong interaction physics at a scale as small as attometer.  
On the other hand, when the transverse momentum of $Z^0$-boson is much smaller than
its mass, $p_T\ll M_Z$, the production becomes extremely sensitive to the gluon shower 
of the colliding partons in high energy collisions.  
The physics of gluon shower is very interesting and extremely 
important for understanding the production of large number of particles, and 
various event shapes in high energy collisions, but, it is not easily accessible 
by probes with a single large momentum transfer.  With the dial of transverse momentum, 
the $Z^0$ production is an ideal probe for QCD dynamics at various scales, and 
has become one of many successful examples in perturbative QCD toolkit. 
\begin{figure}[h]
\begin{center}
\includegraphics[width=0.35\textwidth]{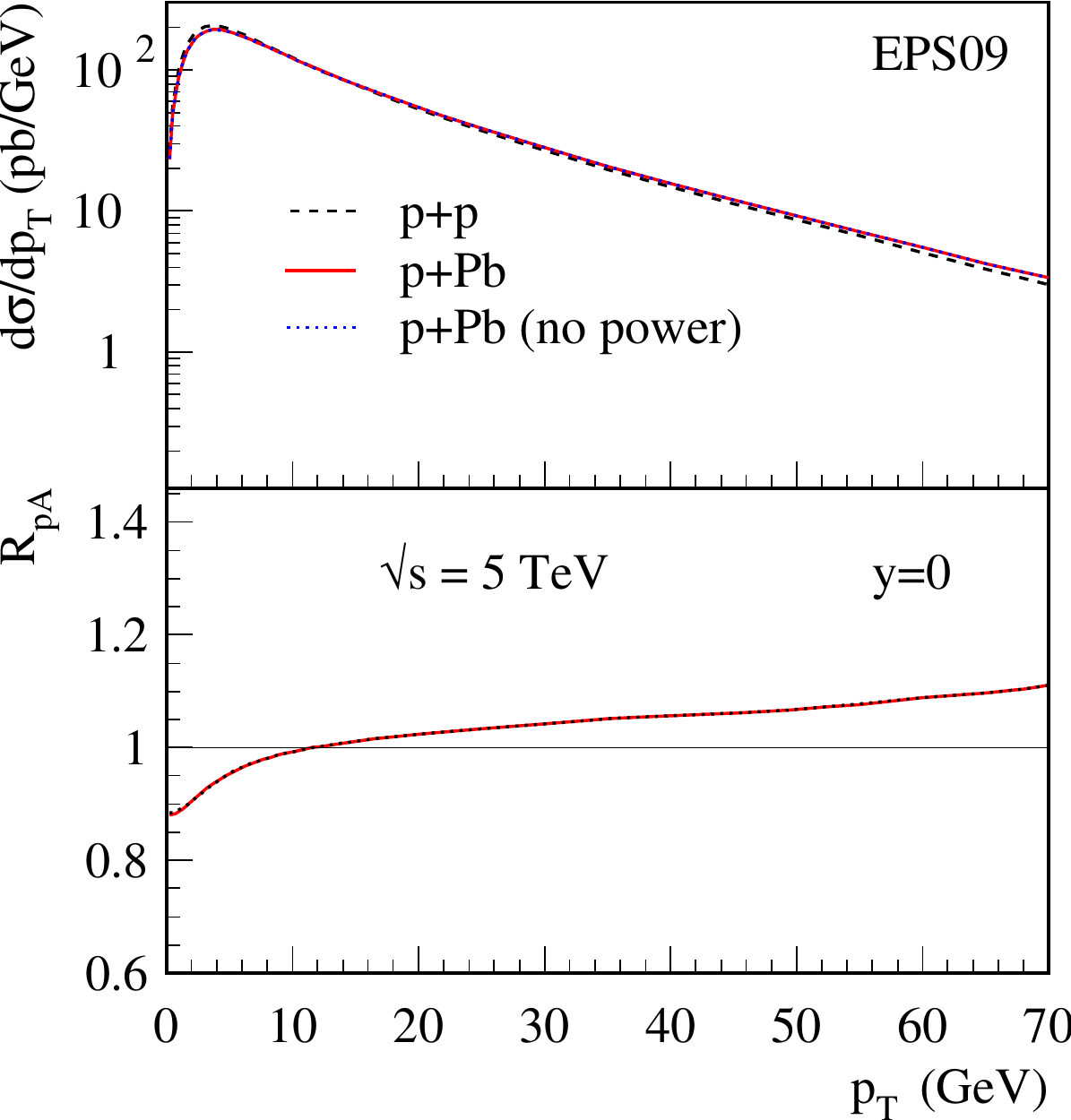}
\hskip 0.3in
\includegraphics[width=0.35\textwidth]{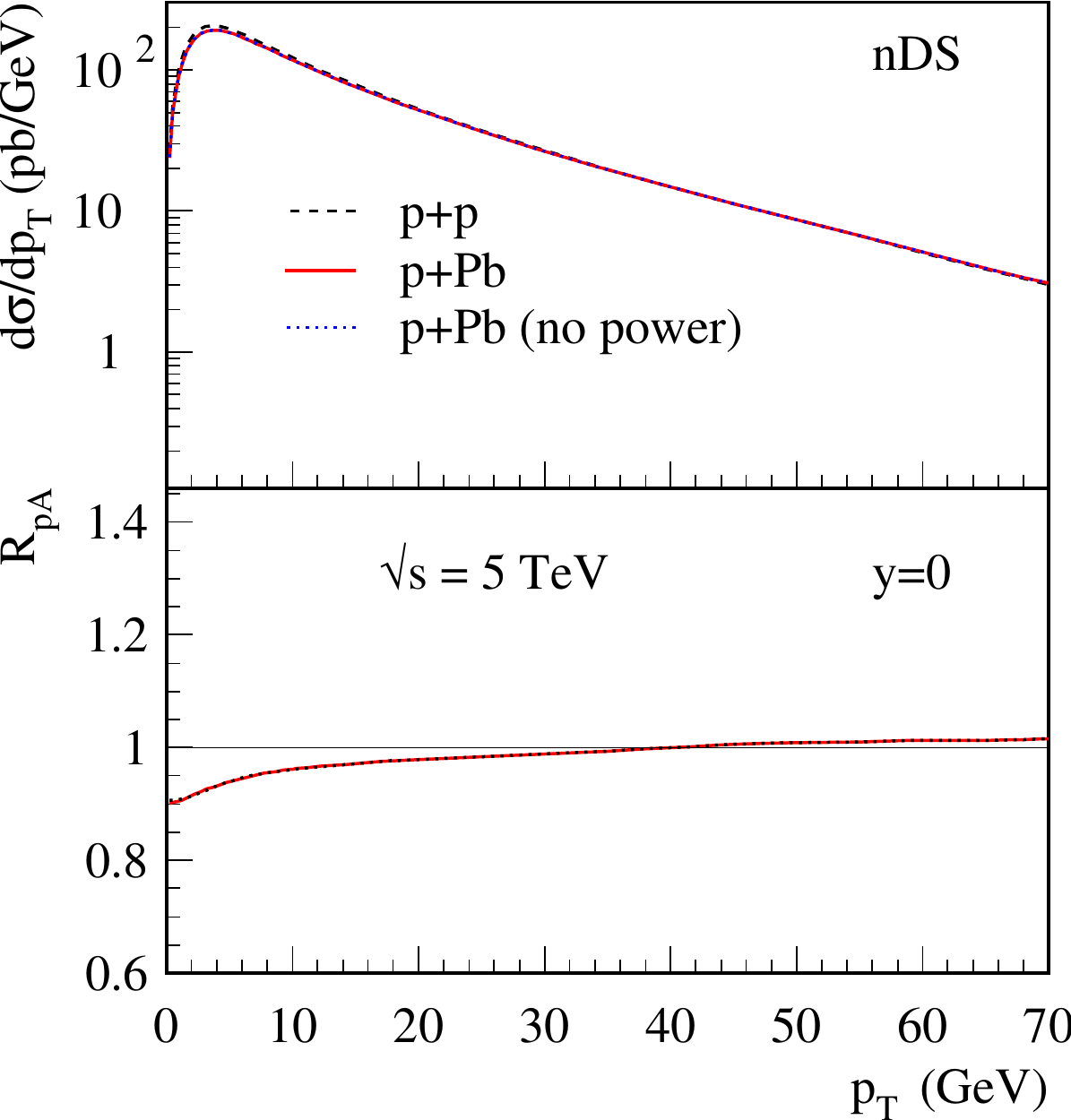}
\caption{$Z^0$ boson production in both p+p and p+Pb collision at center of mass energy 
$\sqrt{S}=5$ TeV and rapidity $y=0$~\cite{Kang:2012am}. The left panel is generated using 
EPS09 nPDFs~\cite{Eskola:2009uj}, while the right panel is by nDS \cite{deFlorian:2003qf}. 
For both panels, the upper plots are the $Z^0$ cross sections per nucleon as a function of transverse momentum $p_T$. The black dashed curve is the p+p baseline, and the red solid curve is for the minimum bias p+Pb collision. Lower panel is for the nuclear modification factor, $R_{pA}$.  }
\label{fig:RpA-z-lhc}
\end{center}
\end{figure}

\begin{figure}[h]
\begin{center}
\includegraphics[width=0.35\textwidth]{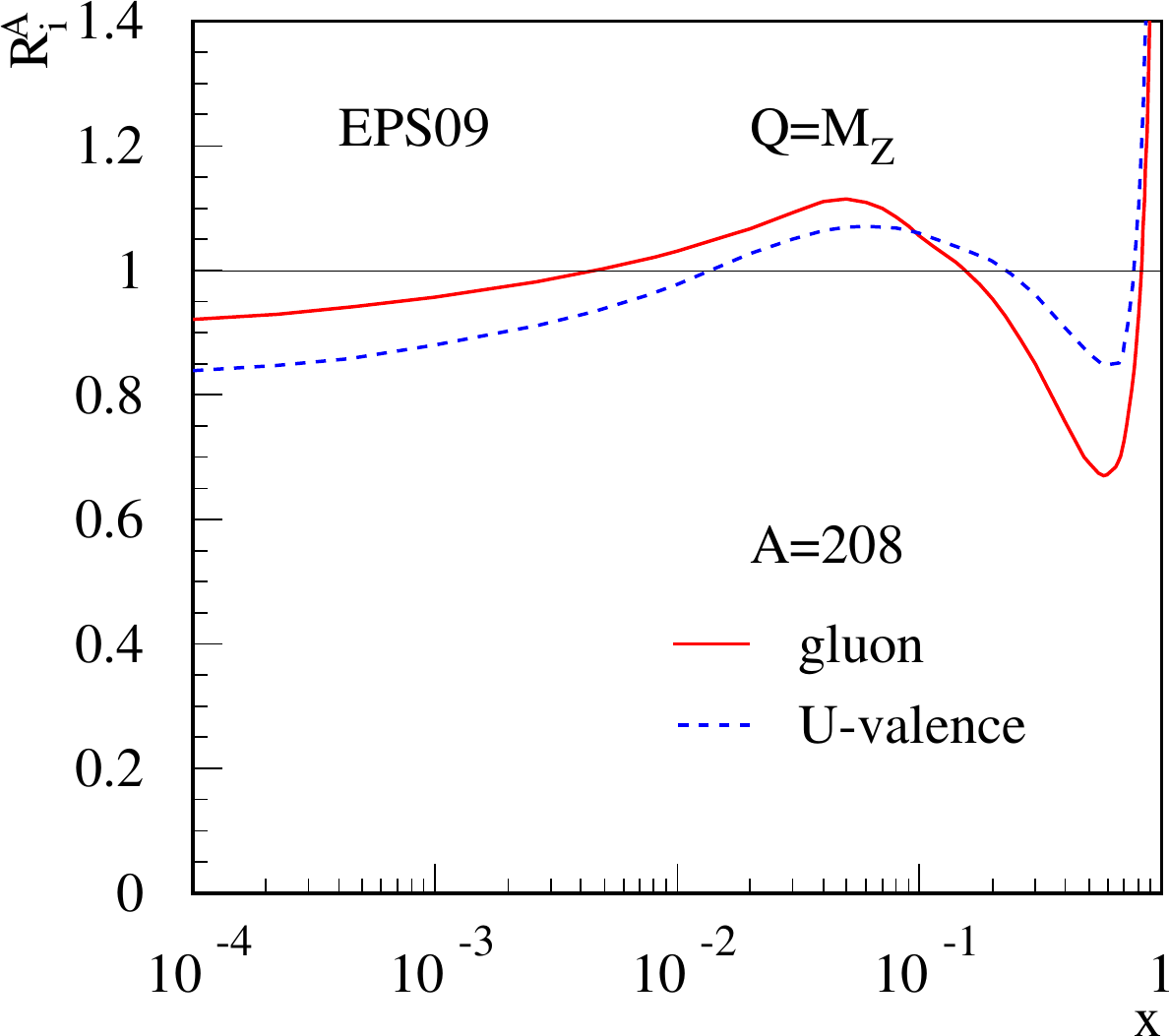}
\hskip 0.3in
\includegraphics[width=0.35\textwidth]{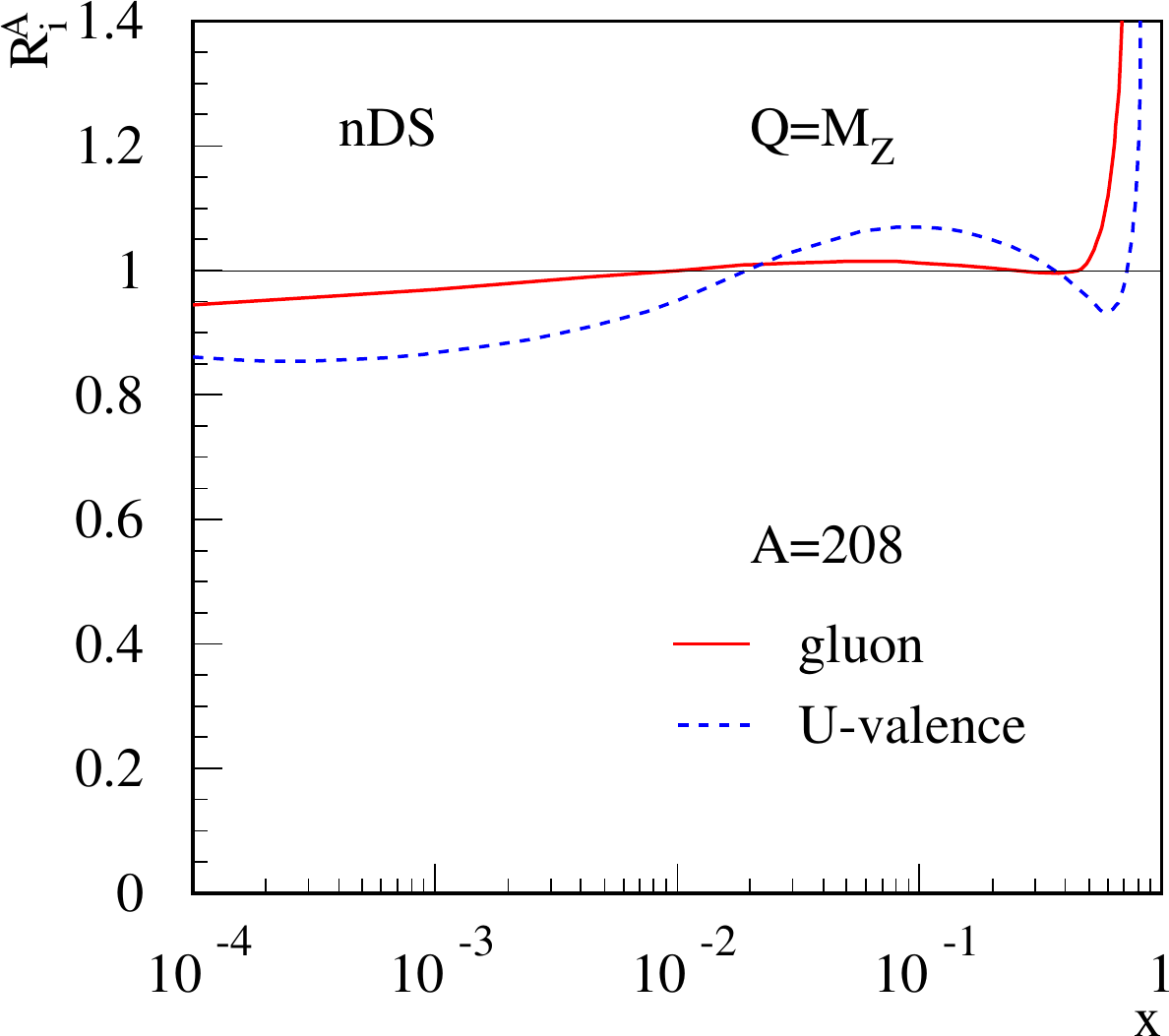}
\caption{
\label{fig:nPDFs}
Ratio of nuclear PDFs over the proton PDFs at scale $Q=M_Z$: 
$R_{i}^A = f_{i/A}(x, Q^2)/f_{i/p}(x, Q^2)$. 
Left panel is for EPS09 nPDFs~\cite{Eskola:2009uj} and 
right panel is for nDS nPDFs~\cite{deFlorian:2003qf}.}
\end{center}
\end{figure}

The study of nuclear modification of $Z^0$ production has finally become available at the LHC 
\cite{Chatrchyan:2011wt,Chatrchyan:2011ua,Aad:2011gj,Aad:2012ew,Aad:2010aa,delaCruz:2012ru}. 
Nuclear modification of the inclusive $Z^0$ production and its implication on nuclear PDFs 
was studied in Ref.~\cite{Paukkunen:2010qg}.   Taking an advantage of the excellent
predictive power of CSS resummation formalism for calculating transverse momentum spectrum
of heavy bosons \cite{Kang:2012am}, recently, Kang and Qiu  
pointed out that nuclear modification of $Z^0$ production as a function of $q_T$
in proton-nucleus collisions at the LHC is directly tied to the nuclear modification of PDFs, 
in particular, nuclear gluon distribution in the antishadowing region.  
In Fig.~\ref{fig:RpA-z-lhc}, the differential cross section of $Z^0$ production 
in proton-nucleus collision, as well as the nuclear modification factor, $R_{pA}$, 
are predicted as a function of $p_T$ by using two commonly used nuclear PDFs: 
EPS09~\cite{Eskola:2009uj} and nDS~\cite{deFlorian:2003qf}.  
Although both parametrizations are based on the global fitting of 
existing experimental data, there is a major difference between these two sets of nuclear PDFs,
as shown in Fig.~\ref{fig:nPDFs},  
EPS09 has a strong anti-shadowing region for gluon distribution inside a large nucleus, 
while nDS does not have antishadowing.  
This difference manifests itself in the nuclear modification factor, $R_{pA}$ in Fig.~\ref{fig:RpA-z-lhc}.  
It was pointed out \cite{Kang:2012am} that the production cross section of $Z^0$ boson at 
this energy, calculated by using the CSS resummation formalism, is dominated by 
gluon initiated subprocess when $p_T > 20$ GeV.  The clear enhancement of $R_{pA}$ 
in large $p_T$ region in Fig.~\ref{fig:RpA-z-lhc} (left) is caused by the large anti-shadowing 
region of nuclear gluon distribution of EPS09.  On the other hand, 
the enhancement of $R_{pA}$ in the large $p_T$ disappears, 
as clearly seen in the lower panel in Fig.~\ref{fig:RpA-z-lhc} (right), 
due to the absence of 
(or much smaller) gluon anti-shadowing in the nDS parametrization of nPDFs. 
It was pointed out that the $Z^0$ production is a direct measurement of 
nuclear gluon distribution, and the measurement of nuclear modification factor $R_{pA}$ 
of $Z^0$ production in p+Pb collisions at the LHC provides a clean and unique test 
of the nuclear gluon anti-shadowing proposed in EPS09 parametrization.

\subsection{Flavor separation of polarized sea}

Polarized Drell-Yan and $W^\pm$ production in polarized $p+p$ collision
were proposed some time ago at RHIC~\cite{bunce92} and they can provide
qualitatively new information on antiquark polarization. At the large
$x_F$ region ($x_F > 0.2$), the longitudinal spin asymmetry $A_{LL}$ in the
$p+p$ Drell-Yan process is given as
\begin{eqnarray}
A^{DY}_{LL}(x_1,x_2) \approx g_1(x_1)/F_1(x_1) \times {\Delta \bar u(x_2)
\over \bar u(x_2)},
\label{eq:pol_dy}
\end{eqnarray}
\noindent where $g_1(x)$ is the proton polarized structure function 
and $\Delta \bar u(x)$ is the polarized $\bar u$ 
distribution function.
Equation~(\ref{eq:pol_dy}) shows that $\bar u$ polarization can be
determined using
polarized Drell-Yan at RHIC. Additional information on 
the sea-quark polarization
can be obtained via $W^\pm$ production~\cite{bs93}.
Parity violation
in $W$ production implies that only singly polarized $p-p$ collision 
is required. At negative $x_F$ (opposite to the direction of the polarized
beam), we have\cite{bs93},
\begin{eqnarray}
A_L^{W^+}\approx {\Delta \bar d(x) \over \bar d(x)},\ \ {\rm and}\ \ \
A_L^{W^-}\approx {\Delta \bar u(x) \over \bar u(x)},  \label{eq:pol_w}
\end{eqnarray}
where $A^W_L$ is the single-spin asymmetry for $W$ production.
Equation~(\ref{eq:pol_w}) shows that the flavor dependence of the sea-quark
polarization can be determined via $W^\pm$ production at RHIC.
A remarkable prediction of the chiral quark soliton model is that
the flavor asymmetry of polarized sea-quark, $\Delta \bar u(x) -
\Delta \bar d(x)$, is large and positive~\cite{dres99a}. This is in 
contrast to the prediction of very small
values for $\Delta \bar u(x) - \Delta \bar d(x)$
in the meson cloud model~\cite{fries98,boreskov99}.

Significant progress in measuring the single-spin asymmetry in $W$ production
has been made at RHIC. Preliminary results~\cite{rhic_spin} of $A_L$ 
measured by the
STAR collaboration at $\sqrt s = 510$ GeV during the 2012 Run
have already made an important impact on determining the $\Delta \bar u$
and $\Delta \bar d$ distributions. In particular, the data
show a strong preference for $\Delta \bar u > \Delta \bar d$
in the range $x>0.05$, consistent with the prediction of the 
chiral quark soliton model. With additional data expected in the near 
future,
the RHIC $W$ production program would greatly improve our
understanding of the helicity distribution of the light quark sea.

\begin{figure}[h]
\begin{center}
\includegraphics[width=0.3\textwidth]{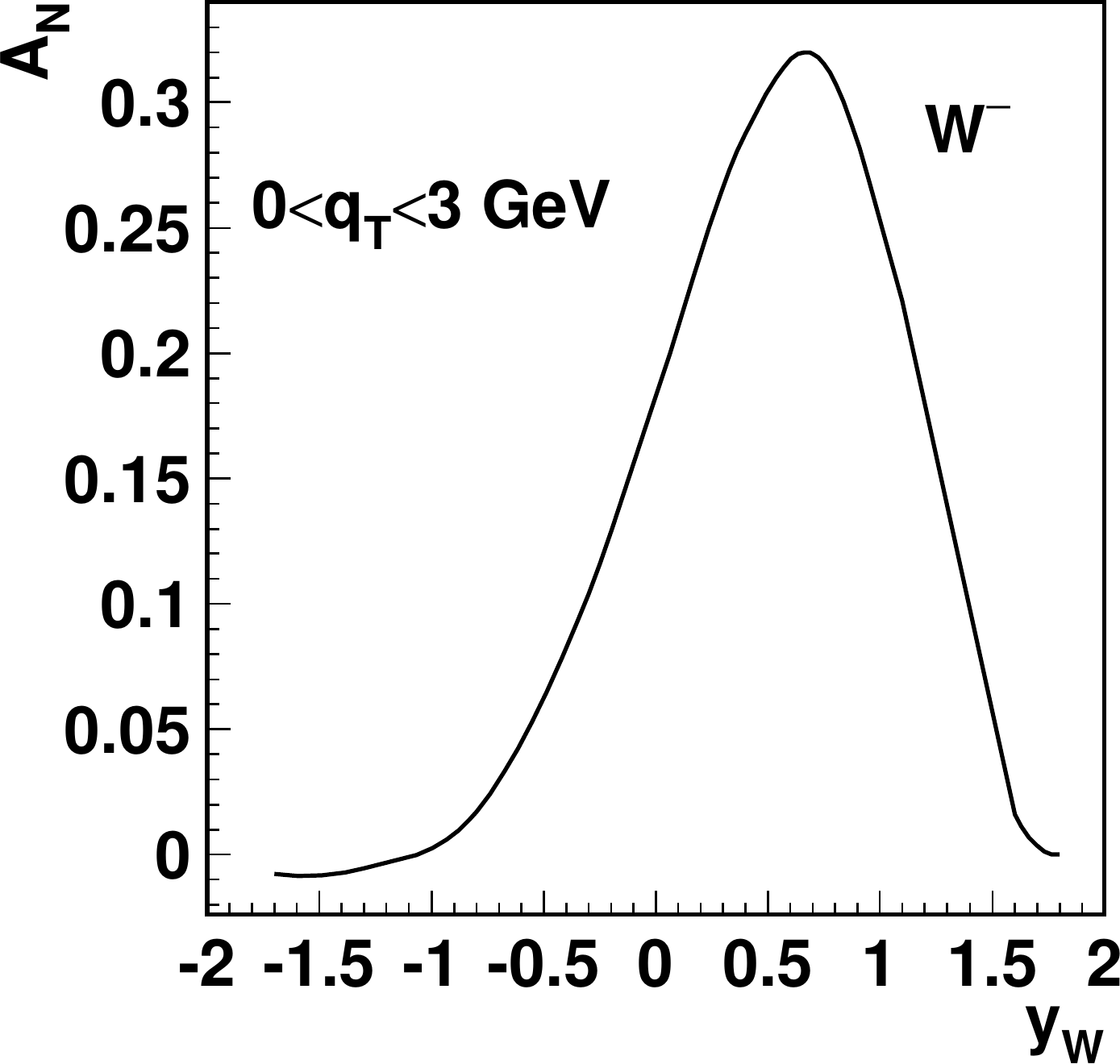}
\hskip 0.1\textwidth
\includegraphics[width=0.3\textwidth]{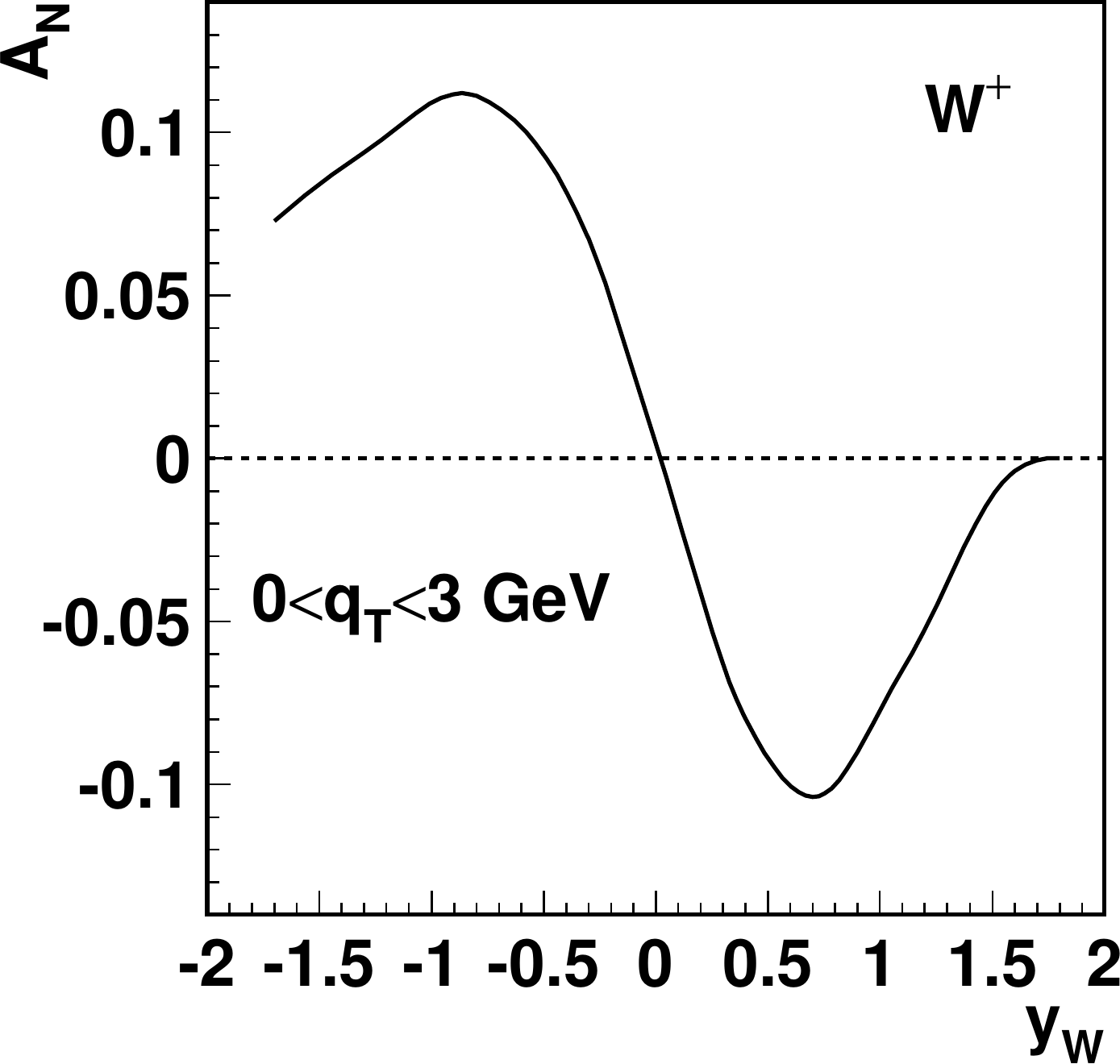}\\
\includegraphics[width=0.3\textwidth]{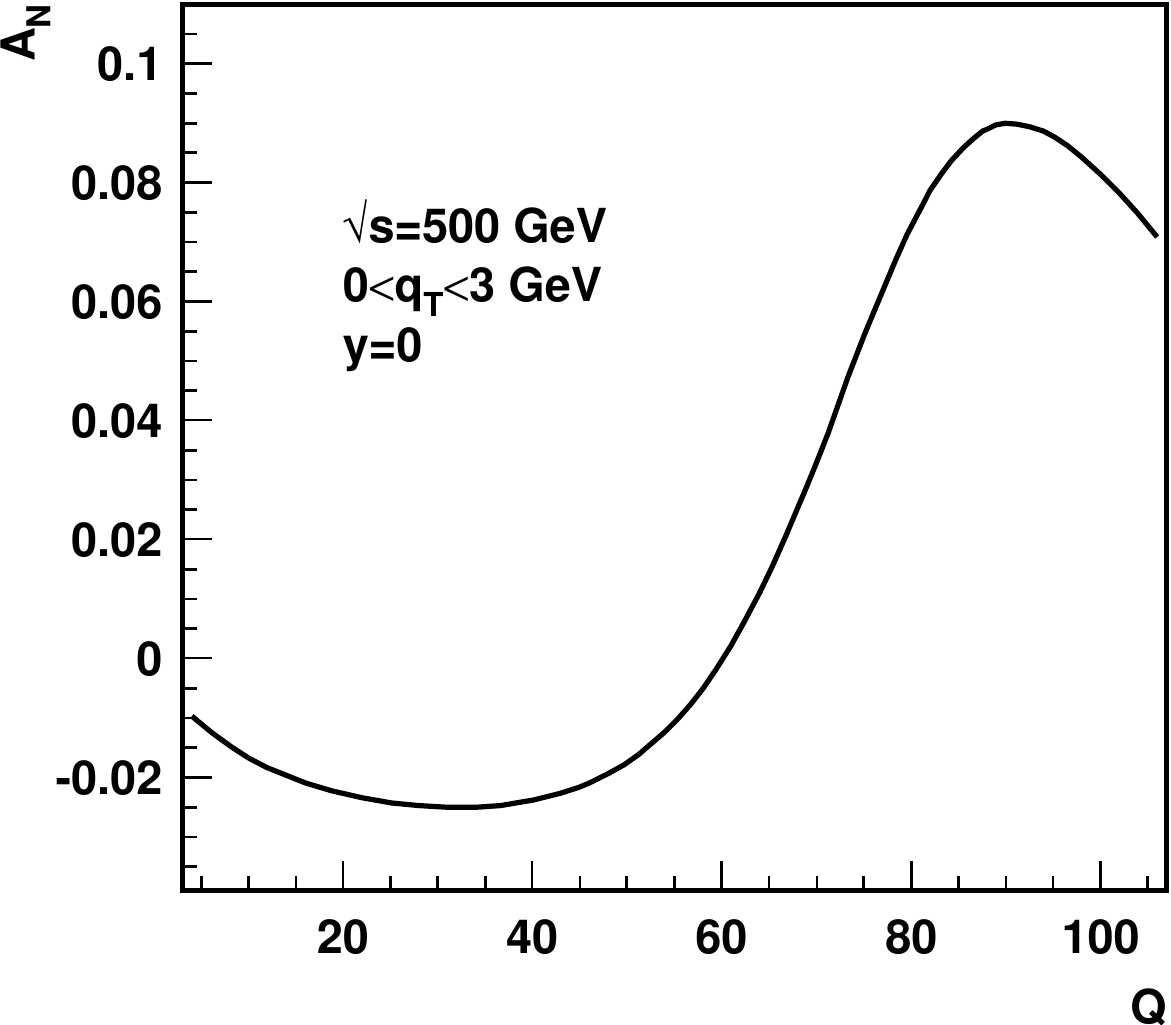}
\hskip 0.1\textwidth
\includegraphics[width=0.3\textwidth]{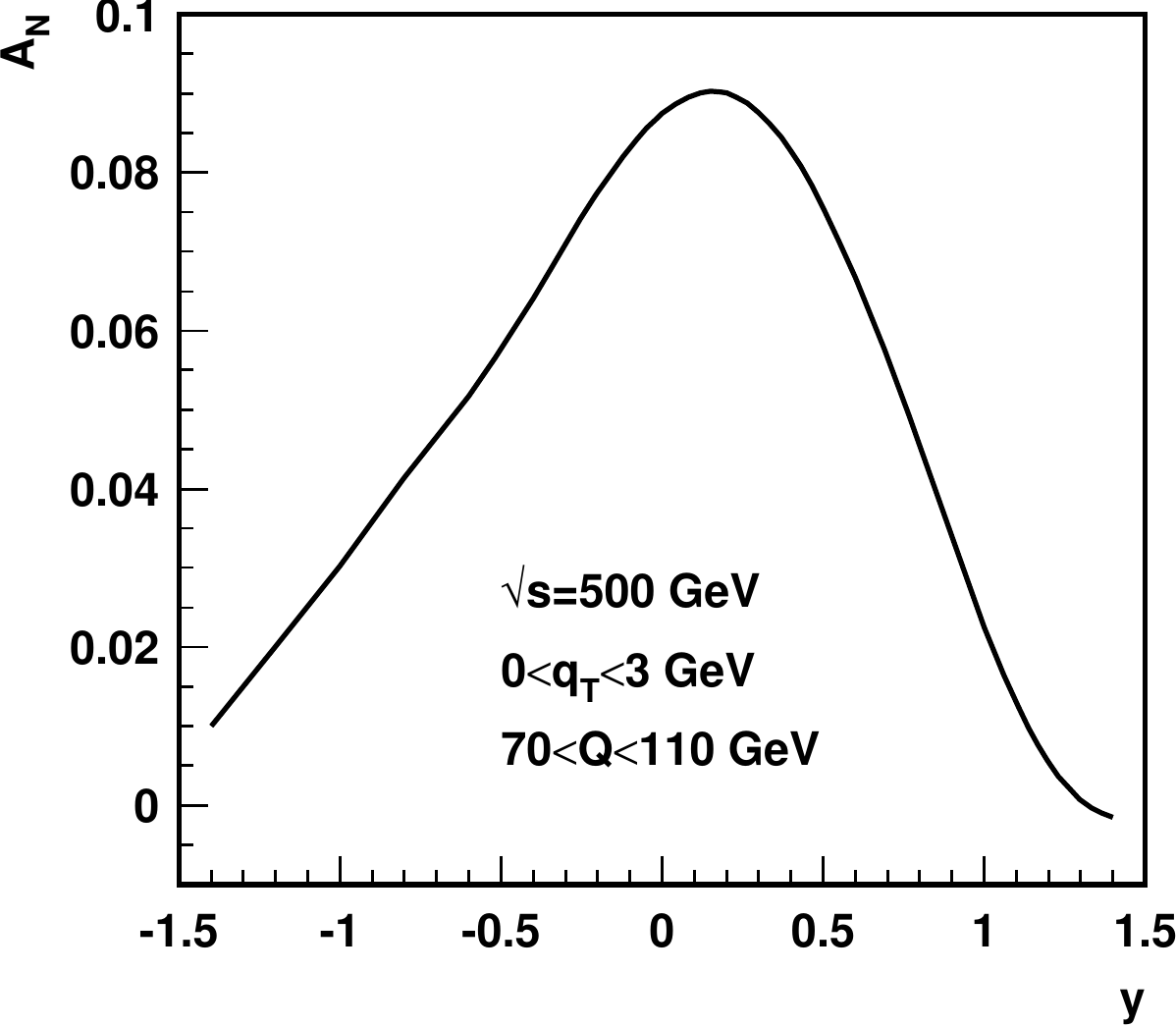}
\caption{\label{fig:an_dy_w_z}
Predictions of SSAs for $W^\pm$ production at RHIC 
as a function of $W$-boson rapidity (top two); and massive lepton pair production as 
a function of rapidity (lower left) and invariant mass of lepton pair (lower right)
 \cite{Kang:2009bp}.
}
\end{center}
\end{figure}

\subsection{Sivers function at $W/Z$ mass}  

With the high energy polarized proton beam, RHIC can also produce $W^\pm$ and $Z^0$
with relatively low transverse momentum $q_T$.  SSAs of $W$ production provide 
excellent tests on QCD evolution of TMD distributions because of the large difference in 
the invariant mass of the lepton pair.  
SSAs of $W^\pm$-boson production at RHIC was calculated at $\sqrt{S}=500$~GeV, 
and numerical predictions for the asymmetries, using 
the Sivers functions extracted by Anselmino {\it et al.} \cite{Anselmino:2008sga}, 
are presented in the top two plots in Fig.~\ref{fig:an_dy_w_z} \cite{Kang:2009bp}.
Because of the uncertainty to reconstruct the $W$ due to the missing neutrino from $W$ decay, 
SSAs of inclusive lepton from $W$ decay were also studied \cite{Kang:2009bp}.  
Although the inclusive lepton SSAs inherited all key features of the 
$W$ asymmetry in Fig.~\ref{fig:an_dy_w_z}, they are diluted in size due to the $W$ decay,
but, still sizable and could be measurable at RHIC for a good range 
of rapidity.  The difference in rapidity dependence
in Fig.~\ref{fig:an_dy_w_z} could provide the excellent information  
on the Sivers functions and their flavor separation. 

In Fig.~\ref{fig:an_dy_w_z}, SSAs of massive lepton pair production near the $Z^0$-pole was also plotted
\cite{Kang:2009sm}.  The figure clearly demonstrates a ``sign change" -- 
the transition from the negative $A_N$ 
at a low invariant mass $Q$ to the positive $A_N$ when $Q$ is near $Z^0$ pole.  This is 
an immediate consequence of the difference in relative strength on how a up and a down quark
couples to a virtual photon or a $Z^0$-boson \cite{Kang:2009sm}, which could be an important test
on how SSAs are generated by Sivers functions of various flavors. 

The SSAs in Fig.~\ref{fig:an_dy_w_z} could be smaller due to the strong scale dependence
of the Sivers function noticed recently \cite{Aybat:2011ge,Aybat:2011ta}.  On the other hand, 
Sun {\it et al.} argue that the decrease of Sivers function as the scale increases may
not be as strong as what was claimed in Ref.~\cite{Aybat:2011ge,Aybat:2011ta}.  In any case,
it is very important to measure the asymmetries experimentally, and test the evolution of the 
SSAs (or TMDs) at different scales, which should be another very important test of the 
QCD TMD-factorization.

%
%

\section{Summary and outlook}
\label{sec:summary}

The massive lepton-pair production (Drell-Yan process) in hadron-hadron collisions 
continues to be a powerful tool for probing the partonic structure of
hadrons and for understanding the dynamics of QCD in
perturbative and nonperturbative regimes. We demonstrated that 
the Drell-Yan process is complementary to the Deep-Inelastic Scattering
in many aspects, and together, they provide interesting and
often surprising information on the flavor structure of the
quark and antiquark contents of the nucleons.  
With the tremendous success of QCD global analysis of all existing high 
energy scattering data from fixed-target experiments to the collider 
experiments at the LHC, the highest energy collider available, 
we are now capable of extracting the universal PDFs to the precision 
that is sufficient for us to question and to explore the hadron's sophisticated 
and rich details of partonic structure, such as the asymmetries in the sea.  

We emphasized that QCD factorization is necessary for extracting 
partonic dynamics and structure from measured hadronic cross sections.
Recent theoretical progress in TMD factorization for Drell-Yan process and
SIDIS provided us with new and much more powerful tools (or ``femtoscopes'') 
to probe (or to ``see'') three-dimensional motion of quarks and gluons 
inside a fast moving hadron, which are quantified in terms of TMDs.  
It is the transverse motion of quarks and gluons that 
carry more direct information on how these 
partons are confined inside a hadron.  
With the transverse momentum dependence, 
TMDs carry more detailed information on hadron's partonic structure, 
and much more than what we have learned from and puzzled by PDFs.  
TMDs provide new and extremely valuable information on 
QCD quantum correlation between the spin and the motion preference of 
quarks and gluons inside a bound hadron, while PDFs, which integrate over
parton's transverse motion, suppress such quantum correlations 
due to their inclusiveness.  With the novel TMDs, we push the investigation 
of hadron's partonic structure to a new frontier.  

With our capability to measure $q_T$ of Drell-Yan production of 
massive lepton pairs, and the fact that the most events have 
$q_T \ll Q$, Drell-Yan process is an ideal observable to probe
TMDs with the large $Q$ to localize the quarks and/or gluons 
of a colliding hadron and the small $q_T$ to explore their transverse 
motion.  Like extracting PDFs, Drell-Yan process is complementary to
SIDIS in extracting TMDs.  In addition, it is also necessary to have both 
Drell-Yan process and SIDIS to study TMDs, 
because of the non-universality of TMDs, and the expected sign change.  
In this respect, Sivers functions and Boer-Mulders functions are
the most interesting TMDs to measure and to explore. 
Drell-Yan production of massive lepton pairs with polarized beam 
and/or target provide the immediate access to these two TMDs.  
The challenge of the Drell-Yan experiment 
is the relatively small cross sections. The great potential of the 
Drell-Yan process in studying TMDs is just beginning to be explored.
As reviewed in this article, many ongoing and future experiments 
at existing or future hadron facilities will explore singly or doubly 
polarized Drell-Yan for the first time.

The advent of the polarized $pp$ collision at RHIC and the high-energy
high-luminosity $pp$ collision at LHC has opened exciting new avenues
for exploring various aspects of partonic structure inside a polarized 
and unpolarized hadron.  
The enormous range in $Q^2$ and $x$, the parton momentum fraction, spanned 
from fixed-target Drell-Yan to $W/Z$ production at collider energies will
undoubtedly reveal exciting and unexpected new results.
The continuous improvement of our knowledge of hadron's partonic structure 
with better measurements in Drell-Yan, DIS, and other hadronic 
observables from current and future facilities will shed some lights on 
how quarks and gluons are bound into color singlet hadrons.

\section*{Acknowledgments}
This work was supported in part by the U.S. Department of Energy under the
contract No. DE-AC02-98CH10886 and the U.S. National Science Foundation
under the contract NSF-PHY-12-05671. 

\appendix

\section{QCD collinear factorization for Drell-Yan process}
\label{sec:appendixa}

In this appendix, we summarize intuitively key arguments for proving QCD collinear factorization 
for Drell-Yan massive lepton-pair production in hadronic collisions.  
The more rigorous and complete arguments for the proof of QCD collinear 
factorization for the leading-power contribution in $1/Q$ expansion of 
the inclusive Drell-Yan cross section can be found in a review article 
by Collins, Soper, and Sterman \cite{Collins:1989gx} and extended r
eferences therein, as well as in a recent book by Collins \cite{collins-book}.  
In the same rigor, complete arguments of QCD collinear factorization for the next-to-leading power contribution to the Drell-Yan cross section can be found in Ref.~\cite{Qiu:1990xy}.  It was known from explicit calculations 
\cite{Doria:1980ak,Di'Lieto:1980dt,Brandt:1988xt} that there is no QCD 
factorization for contributions 
beyond the first leading power corrections \cite{Qiu:1990xy}.

The key for achieving the factorization is to demonstrate the suppression of quantum interference between the short-distance dynamics for producing the massive lepton pair and the nonperturbative physics inside the colliding hadrons, or equivalently, to show that the perturbative and nonperturbative physics are linked by ``long-lived" or approximately on-shell active partons, so that the cross section could be factorized into a product of probabilities: one for the short-distance production, and the other for the probability to find one active parton from each colliding hadron.  However, in QCD, all partons (quarks and gluons) participating in the hadronic collision are off mass-shell, whose virtuality (or four-momentum) needs to be integrated over.  The factorization is possible {\it iff} the dominant contribution to the cross section is from the region of phase space where all active partons (e.g. the quark and antiquark that annihilate into the massive lepton pair) are perturbatively pinched to their respective mass shell to behave as on-shell ``classical'' partons, who are ``long-lived'' compared to the time scale of short-distance hard collision to produce the massive lepton pair. 

Such perturbative pinch of active partons naturally exists in QCD contribution to the cross section, which is proportional to the square of scattering amplitude.  As shown in the lowest order QCD diagram for producing a massive lepton pair in Fig.~\ref{fig:drell-yan} (right), 
the phase space integration of the active quark, $d^4p_a$, 
gets a propagator from the scattering amplitude and another one from its complex conjugate,
so that 
\begin{equation}
\int d^4p_a \,
\frac{1}{p_a^2 + i\varepsilon}\,
\frac{1}{p_a^2 - i\varepsilon}\,
\to \infty\, ,
\label{eq:pinch}
\end{equation}
is perturbatively pinched at $p_a^2\to 0$ and divergent.  
The same is true for the phase space integration of the active antiquark, $d^4p_b$.
That is, the phase space integration of the active quark and antiquark is dominated by
the region of phase space where the virtuality of active quark and antiquark is of 
the order of nonperturbative hadronic scale $\sim 1/{\rm fm}\sim \Lambda_{\rm QCD}$,
which is much smaller than the hard scale of the collision, $Q$, 
the invariant mass of lepton pair.  
If hadron $A$ and hadron $B$ move in $+z$ and $-z$ direction, respectively, 
the factorized Drell-Yan formula in Eq.~(\ref{eq:drell-yan-pm})
can be easily derived from the leading order QCD contribution, the diagram
on the right in Fig.~\ref{fig:drell-yan}, with the following approximation,
\begin{eqnarray}
\mbox{on-shell:} &\  \ & p_a^2, \ p_b^2\ \ll Q^2;
\label{eq:on-shell}\\
\mbox{collinear:} &\ \ & p_{aT}^2, \ p_{bT}^2\ \ll Q^2;
\label{eq:collinear}\\
\mbox{higher-power:}& \ \ & p_a^-\ \ll q^-;\ \ \mbox{and}\ \ p_b^+\ \ll q^+\, ,
\label{eq:power}
\end{eqnarray}
where the notation of light-cone coordinate, $v^{\pm}=(v^0\pm v^3)/\sqrt{2}$, 
for a four-vector $v^\mu$ was used.

In addition to quarks and antiquarks, there are gluons in QCD.
QCD factorization at the leading power means that there should be only one active parton 
(quark, antiquark, or gluon) from each identified hadron.  
The large momentum exchange of the collision, required by
the large invariant mass of the lepton pair, $Q$, keeps the momenta of both active partons
from two colliding hadrons to be of the order of $Q$.  
All relevant nonperturbative contributions to the Drell-Yan 
massive lepton pair production are either suppressed by the powers of $1/Q^2$ 
or factorized into long-distance PDFs, which are not calculable perturbatively.   
But, PDFs are universal or process independent.  
That is, PDFs of hadron $A$ should not be interfered and altered 
by the existence of hadron $B$ or vise versa \cite{collins-qiu-fac}.  

\begin{figure}[h]
\begin{center}
\includegraphics[width=0.38\textwidth]{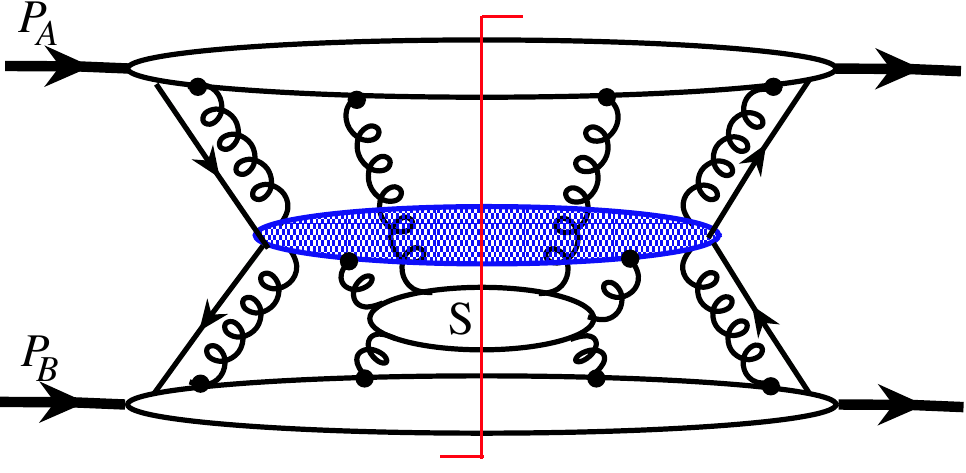}
\hskip 0.05\textwidth
\includegraphics[width=0.38\textwidth]{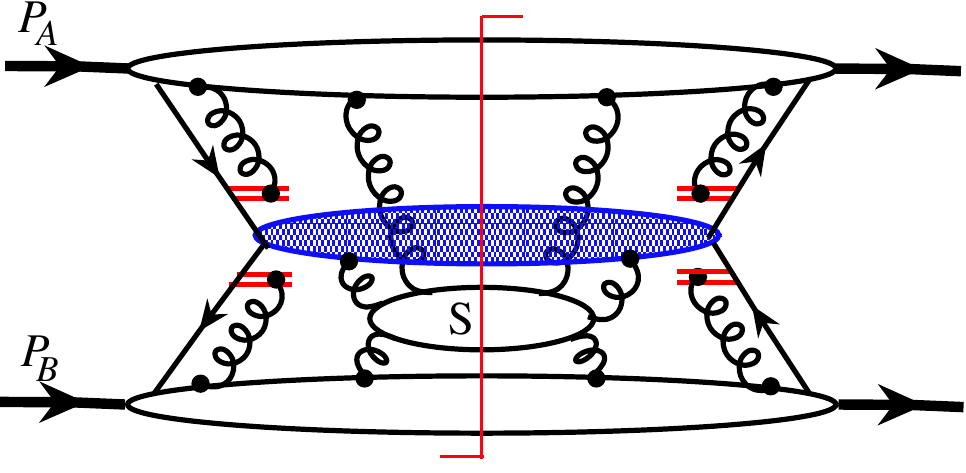}
\caption{Sketch for the leading QCD pinch surface for Drell-Yan process (left),
and QCD contribution to Drell-Yan process with eikonalized gluons.
}
\label{fig:dysurface}
\end{center}
\end{figure}

However, after taking out the active quark from one colliding hadron and an antiquark 
from another colliding hadron, the spectator ``jets'' of both hadrons carry color.  
Gluons from one hadron can interact with the active quark or antiquark of another hadron, 
as well as quarks and gluons inside the spectator jet of another hadron, 
which is the key difference between inclusive DIS and Drell-Yan process. 
In general, gluons can dress the basic Drell-Yan massive 
lepton-pair production process in Fig.~\ref{fig:drell-yan} in many ways.  
The pole structure of the phase space integration of Feynman diagrams in QCD
perturbation theory determines the leading pinch surface that gives the leading power
contribution to the Drell-Yan massive lepton pair production, 
as shown by the diagram on the left in Fig.~\ref{fig:dysurface} \cite{Collins:1989gx}.  
The active quark and antiquark are always accompanied with collinear and longitudinally 
polarized gluons, and soft gluons can attach to both the beam spectator jets, 
as well as the hard part in which all particles are off-shell by the amount of hard scale $Q$.
It is the gluon interaction between hadrons can potentially rotate the color of active parton(s) 
and alter the PDFs, and break the factorization.  

\begin{figure}[h]
\begin{minipage}[c]{0.44\textwidth}
\begin{center}
\includegraphics[width=0.85\textwidth]{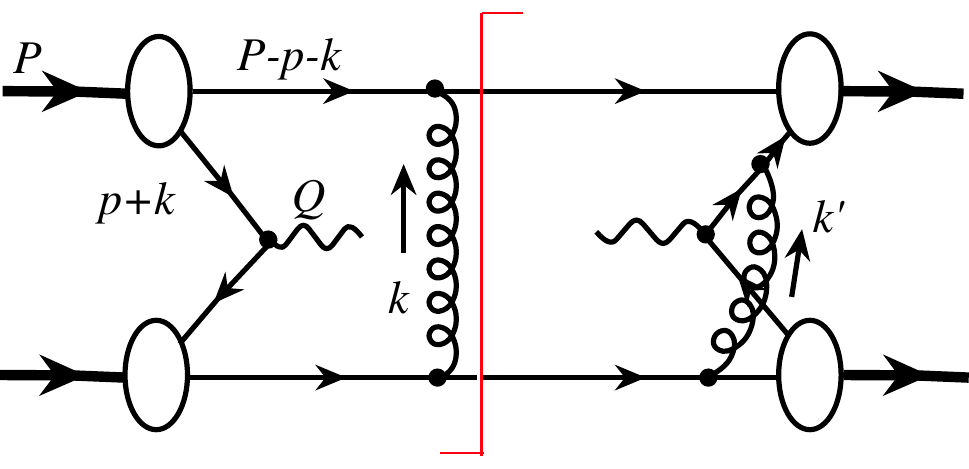}
\end{center}
\end{minipage}
\hskip 0.2in
\begin{minipage}[c]{0.38\textwidth}
\begin{center}
\includegraphics[width=0.99\textwidth]{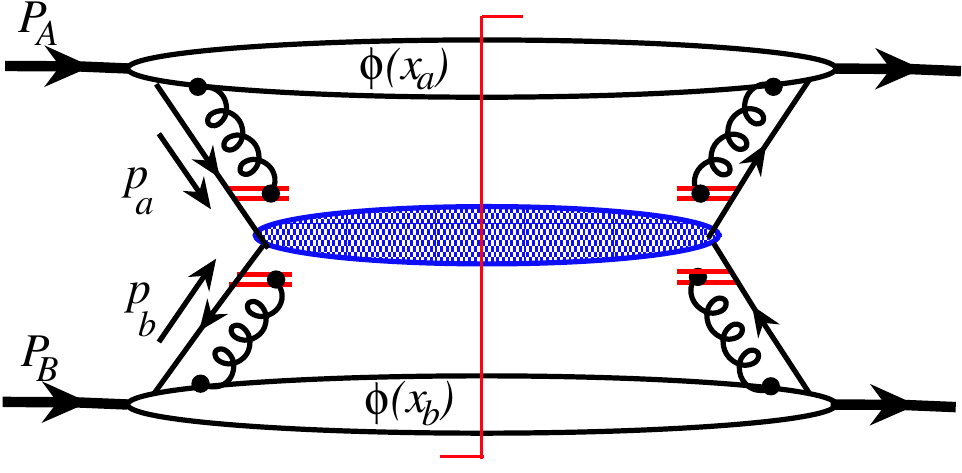}
\end{center}
\end{minipage}
{\hskip -0.3in}$\times$
\begin{minipage}[c]{0.14\textwidth}
\begin{center}
\includegraphics[width=0.95\textwidth]{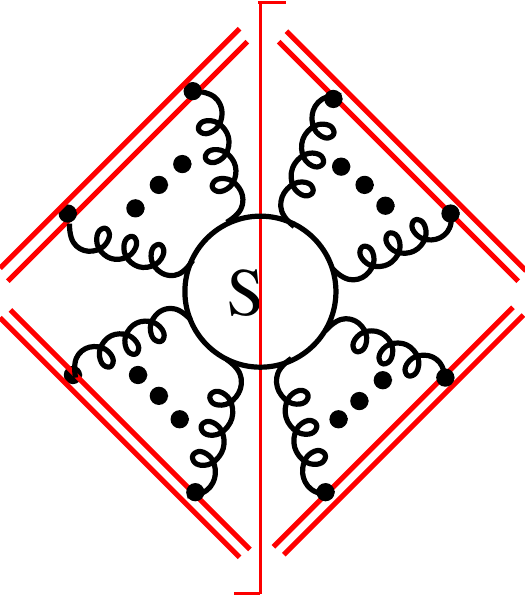}
\end{center}
\end{minipage}
\caption{Sample diagram responsible for pinched Glauber region (left),
and sketch for the factorized Drell-Yan cross section in QCD at leading power in $1/Q^2$ (right).
}
\label{fig:dy-fac}
\end{figure}

For the leading pinch surface of Drell-Yan cross section, in Fig.~\ref{fig:dysurface} (left), 
the collinear gluons are easier to deal with.  
The collinear gluons have polarization vectors proportional to their respective momenta
in a covariant gauge.  The sum of total effect from the collinear gluons can be represented 
by the eikonal lines, as shown by the diagram on the right in Fig.~\ref{fig:dysurface}.  
The eikonal lines, the double lines attached to the active parton, are orthogonal to the 
direction of the active parton that it is attached to.  In order to achieve the factorization, 
we need to get rid of the soft gluons interactions in the diagram in Fig.~\ref{fig:dysurface} (right).  
Soft gluon exchanged between a spectator quark of hadron $B$ and the spectator or 
the active quark of hadron $A$, as shown in the diagram in Fig.~\ref{fig:dy-fac} (left),
could rotate the quark's color and keep it from annihilating with the antiquark of hadron $B$.
The soft gluon approximations (with the eikonal lines) require the $\pm$ light-cone 
momentum components of all soft gluons not too small. 
But, the $k^{\pm}$ of the exchanged gluon of momentum $k$ between the two spectators, 
as shown in the diagram in Fig.~\ref{fig:dy-fac} (left), 
could be trapped in a Òtoo smallÓ region, known as the ``Glauber'' region, 
due to the pinch from the spectator interaction, so that $k^\pm \sim M^2/Q \ll k_T \sim M$, 
where $M\ll Q$ is a hadronic scale of the spectator jet.
Such leading power soft gluon interaction could break the universality of PDFs.  
Without the universality of PDFs, the ``factorized" cross section in Eq.~(\ref{eq:qcd-dy-lp}) 
has no predictive power due to the non-perturbtive nature of the PDFs.

Removal of the trapped Glauber gluons might be the most technical part of the factorization
proof~\cite{Collins:1989gx}.  It was achieved by three key steps: 
1) all poles in one-half plane cancel after summing over 
all final-states (no more pinched poles), 
2) all $k^\pm$-type integrations can be deformed out of the trapped soft region, 
3) all leading power spectator interactions can be factorized and summed into 
an overall unitary soft factor of eikonal lines - gauge links,
as shown in the diagram on the right in Fig.~\ref{fig:dy-fac}.
The soft factor is process independent and made of four gauge links, 
along the light-cone directions conjugated to the directions of two incoming hadrons 
in the scattering amplitude, and the two in the complex conjugate scattering amplitude,
respectively.  For the collinear factorization, the soft factor = 1 due to the unitarity. 
The factorization formalism in Eq.~(\ref{eq:qcd-dy-lp}) 
for the leading power QCD contribution to Drell-Yan cross section
is proved in the sense that all identified sources of leading power contributions  
are either factorizable or canceled in perturbative calculations to all orders in powers of 
$\alpha_s$.  The factorization could be broken if one discovers a new source of 
non-factorizable leading power contribution to Drell-Yan cross section, 
although it might be unlikely.  

\section{TMD parton distribution functions and the sign change}
\label{sec:appendixb}

In this appendix, we briefly review the key arguments of TMD factorization of Drell-Yan process, 
and demonstrate the non-universality of TMD parton distribution functions, which led to the predicted 
sign change of Sivers and Boer-Mulders functions.

With the large invariant mass $Q$, the production of the lepton pair is localized and sensitive to 
the degree of freedom of quarks and gluons.  We have the same QCD Feynman diagrams for
Drell-Yan massive lepton-pair production regardless the value of the pair's transverse momentum 
$q_T$.   The perturbative pinch singularities of these diagrams ensure that partonic contributions
to Drell-Yan cross section come from the region of phase space where both active partons 
of momentum $p_a$ and $p_b$ in Fig.~\ref{fig:dy-fac} are close to their respective 
mass shell and live much longer than $1/Q$ -- the time scale of the hard collision to produce the
lepton pair.   Therefore, the on-shell approximation in Eq.~(\ref{eq:on-shell}) for calculating the 
partonic cross section is valid regardless the value of $q_T$.  All key arguments for the  
factorization given in Appendix~\ref{sec:appendixa} 
(or formally in \cite{Collins:1989gx,collins-book} and references therein)
should carry through.  Drell-Yan differential cross section with
a finite $q_T$ could be factorized as illustrated in Fig.~\ref{fig:dy-fac} (right).  

However, the collinear approximation in Eq.~(\ref{eq:collinear}) for calculating the partonic
hard part should not be valid when the observed $q_T=\sqrt{{\bf q}_T^2}$, where
${\bf q}_T = {\bf p}_{aT} + {\bf p}_{bT} \ll Q$, is of the same size of the transverse momenta
of the colliding active quark and anitquark.  The convolution of active parton momentum $p_a$ 
(and $p_b$) should include both the longitudinal momentum fractions $x_a$ (and $x_b$) and 
transverse momentum ${\bf p}_{aT}$ (and ${\bf p}_{bT}$).  Because of the transverse momentum flow
between active partons, the soft factor in Fig.~\ref{fig:dy-fac} (right) is no longer equal to one, 
and depends on the transverse separation of the gauge links and the rapidity differences of 
active partons \cite{Collins:1984kg}.  Since the soft factor is process independent, 
Collins \cite{collins-book} introduced a new definition of transverse momentum dependent PDFs 
to systematically absorb the soft factor into the TMDs, which leads to the 
TMD factorized Drell-Yan cross section in Eq.~(\ref{eq:dy-tmd}).  

The quark-gluon field operators defining PDFs in QCD collinear factorization formalism 
in Eq.~(\ref{eq:qcd-dy-lp}) are localized to a short-distance of the hard collision, 
$\sim 1/xp^+$ for a parton of momentum $xp^+$.  But, the operators defining TMDs are extended to 
an infinite distance along the light-cone.  For example, the TMD for finding an unpolarized quark 
inside a transverse polarized hadron measured in Drell-Yan process is given by,
\begin{eqnarray}
f_{q/h^\uparrow}^{\rm DY}
(x,\mathbf{p}_\perp,\vec{S})
&=& 
\int \frac{dy^- d^2\mathbf{y}_\perp}{(2\pi)^3}\,
e^{ixp^+ y^- - i\,\mathbf{p}_\perp\cdot \mathbf{y}_\perp}
\langle p,\vec{S}|
\overline{\psi}(0^-,\mathbf{0}_\perp)
\Phi_n^\dagger(\{-\infty,0\},\mathbf{0}_\perp)
\nonumber\\
& \times &
\Phi_{\mathbf{n}_\perp}^\dagger
      (-\infty,\{\mathbf{y}_\perp,\mathbf{0}_\perp\})
\frac{\gamma^+}{2}\,
\Phi_n(\{-\infty,y^-\},\mathbf{y}_\perp)
\psi(y^-,\mathbf{y}_\perp)
|p,\vec{S}\rangle
\label{qkt_dy}
\end{eqnarray}
where the factorization scale and $y^+=0^+$ dependence is suppressed and 
the past pointing gauge links were caused by the initial-state 
interactions of Drell-Yan production \cite{collins02}, and the gauge links are given by
\begin{eqnarray}
\Phi_n(\{\infty,y^-\},\mathbf{y}_\perp) 
&\equiv &
{\cal P}e^{-ig\int_{y^-}^{\infty} dy_1^- 
              n^\mu A_\mu(y_1^-,\,\mathbf{y}_\perp)}\, ,
\nonumber\\
\Phi_{\mathbf{n}_\perp}(\infty,\{\mathbf{y}_\perp,\mathbf{0}_\perp\}) 
&\equiv &
{\cal P}e^{-ig\int_{\mathbf{0}_\perp}^{\mathbf{y}_\perp} 
             d\mathbf{y}'_\perp 
            \mathbf{n}_\perp^\mu A_\mu(\infty,\,\mathbf{y}'_\perp)}\, ,
\label{g_link}
\end{eqnarray}
where ${\cal P}$ indicates the path ordering and 
the direction $\mathbf{n}_\perp$ is pointed from 
$\mathbf{0}_\perp$ to $\mathbf{y}_\perp$. 
Here we define the light-cone vectors, 
$n^\mu=(n^+,n^-,\mathbf{n}_\perp)=(0,1,\mathbf{0}_\perp)$
and $\bar{n}^\mu=(1,0,\mathbf{0}_\perp)$, which project out the 
light-cone components of a four-vector $v^\mu$ as $v\cdot n=v^+$
and $v\cdot\bar{n}=v^-$.

Like PDFs, some of TMDs contribute to unpolarized and some of them contribute to polarized 
Drell-Yan massive lepton pair production \cite{sivers90,boer99,ralston79}.  
But, unlike the PDFs, TMDs are not necessarily universal, 
and could depend on the process from which they are extracted.  
For comparison, the TMD for finding an unpolarized quark inside a transversely 
polarized hadron measured in SIDIS is given by
\begin{eqnarray}
f_{q/h^\uparrow}^{\rm SIDIS}
(x,\mathbf{p}_\perp,\vec{S})
&=& 
\int \frac{dy^- d^2\mathbf{y}_\perp}{(2\pi)^3}\,
e^{ixp^+ y^- - i\,\mathbf{p}_\perp\cdot \mathbf{y}_\perp}
\langle p,\vec{S}|
\overline{\psi}(0^-,\mathbf{0}_\perp)
\Phi_n^\dagger(\{\infty,0\},\mathbf{0}_\perp)
\nonumber\\
& \times &
\Phi_{\mathbf{n}_\perp}^\dagger
     (\infty,\{\mathbf{y}_\perp,\mathbf{0}_\perp\})
\frac{\gamma^+}{2}\,
\Phi_n(\{\infty,y^-\},\mathbf{y}_\perp)
\psi(y^-,\mathbf{y}_\perp)
|p,\vec{S}\rangle ,
\label{qkt_dis}
\end{eqnarray}
where the forward pointing gauge links were caused by the final-state 
interactions of SIDIS.  The definitions of TMDs in Eqs.~(\ref{qkt_dy}) and (\ref{qkt_dis}), 
measured in Drell-Yan and SIDIS, respectively, have different gauge links and 
are in principle different quantities in QCD.  Without the universality of TMDs, 
the predictive power of QCD TMD factorization formalism in Eq.~(\ref{eq:dy-tmd}) 
could be in question.
Fortunately, the process dependence of TMDs is limited to a possible sign change 
due to parity and time-reversal invariance of QCD \cite{collins02,Kang:2009bp}.

From the conservation of parity and time-reversal transformation in QCD,  
hadronic matrix elements of quark and gluon field operators have to satisfy 
\cite{Kang:2008ey,Qiu:1998ia}
\begin{equation}
\langle P,\vec{S}| {\cal O}(\psi,A_\mu) |P,\vec{S}\rangle
= \langle P,-\vec{S}| ({\cal PT}){\cal O}(\psi,A_\mu)^\dagger ({\cal PT})^{-1}
|P,-\vec{S}\rangle \, ,
\label{eq:pt}
\end{equation}
where ${\cal P}$ and ${\cal T}$ are parity and time-reversal operator, respectively. 
Using the relation in Eq.~(\ref{eq:pt}), it is easy to show that \cite{collins02,Kang:2009bp}
\begin{equation}
f_{q/h^\uparrow}^{\rm SIDIS}
(x,\mathbf{p}_\perp,\vec{S})
=
f_{q/h^\uparrow}^{\rm DY}
(x,\mathbf{p}_\perp,-\vec{S})\, ,
\label{eq:pt_inv}
\end{equation}
and conclude that the TMD quark distributions could have processes dependence.
From the identity in Eq.~(\ref{eq:pt_inv}), one finds the quark Sivers functions, 
which are proportional to $[f_{q/h^\uparrow}(x,\mathbf{p}_\perp,\vec{S})
-f_{q/h^\uparrow}(x,\mathbf{p}_\perp,-\vec{S})]/2$, to change sign from SIDIS to Drell-Yan
\cite{collins02,collins-metz-fac,Kang:2009bp}.

Similarily, the conservation of parity and time-reversal transformation in QCD
requires the quark TMDs with a tensor spin projection,
\begin{equation}
f_{h_{1q}/h^\uparrow}^{\rm SIDIS}
(x,\mathbf{p}_\perp,\vec{S})
=
-f_{h_{1q}/h^\uparrow}^{\rm DY}
(x,\mathbf{p}_\perp,-\vec{S})\, ,
\label{eq:pt_inv_tensor}
\end{equation}
which are defined by replacing the vector spin projection $\gamma^+$ in Eqs.~(\ref{qkt_dy}) and 
(\ref{qkt_dis}) by a spin projection proportional to $\sigma^{+\perp}$.
The identity in Eq.~(\ref{eq:pt_inv_tensor}) requires Boer-Mulders functions, $h_{1q}^\perp(x)$, 
which are proportional to $[f_{h_{1q}/h^\uparrow}(x,\mathbf{p}_\perp,\vec{S})
+f_{h_{1q}/h^\uparrow}(x,\mathbf{p}_\perp,-\vec{S})]/2$, to change sign 
from SIDIS to Drell-Yan \cite{collins02,collins-metz-fac,Kang:2009bp}.

The sign change of Sivers and Boer-Mulders functions is the prediction 
of TMD factorization approach, and is one of the most important tests of QCD dynamics 
and factorization approaches to hadronic cross sections.  


\end{document}